\pdfoutput=1
\documentclass[cernpreprint,atlasdraft=true,texlive=2016,UKenglish,texmf]{atlasdoc}
 
\usepackage[full, subfigure, subcaption]{atlaspackage}
\usepackage{atlasbiblatex}
 
\PreprintIdNumber{CERN-EP-2020-031}
 
\usepackage[hepparticle=true, hepprocess=true, hion=true]{atlasphysics}
\usepackage{footnote}
\usepackage{threeparttable}

\usepackage{tablefootnote}
\usepackage{multirow}
 
\graphicspath{{logos/}{figures/}}
 
\usepackage{ANA-TRIG-2018-01-PAPER-defs}

\AtlasTitle{Performance of the ATLAS muon triggers in Run 2}
 
\AtlasJournal{JINST}
\AtlasJournalRef{JINST 15 (2020) P09015}
\AtlasDOI{10.1088/1748-0221/15/09/p09015}
 
\AtlasAbstract{
The performance of the ATLAS muon trigger system is evaluated with proton--proton ($pp$) and heavy-ion (HI) collision data collected in Run 2 during 2015--2018 at the Large Hadron Collider.
It is primarily evaluated using events containing a pair of muons from the decay of \PZ bosons to cover the intermediate momentum range between 26~\GeV\ and 100~\GeV.
Overall, the efficiency of the single-muon triggers is about 68\% in the barrel region and 85\% in the endcap region.
The $p_{\text{T}}$ range for efficiency determination is extended by using muons from decays of \PJgy mesons, \PW bosons, and top quarks.
The performance in HI collision data is measured and shows good agreement with the results obtained in $pp$ collisions.
The muon trigger shows uniform and stable performance in good agreement with the prediction of a detailed simulation.
Dedicated multi-muon triggers with kinematic selections provide the backbone to beauty, quarkonia, and low-mass physics studies.
The design, evolution and performance of these triggers are discussed in detail.
}
 
\author{The ATLAS Collaboration}
 
\AtlasRefCode{TRIG-2018-01}

\hypersetup{pdftitle={ATLAS document},pdfauthor={The ATLAS Collaboration}}
 
\begin{document}
 
\maketitle
 
\tableofcontents

\section{Introduction}
 
The presence of prompt muons in the final state is a distinctive signature for many physics processes studied in collisions at the Large Hadron Collider (LHC).
These studies include measurements of properties of the Higgs boson and Standard Model processes, searches for new phenomena, and a \textit{B}-physics and Light States (BLS) programme.
A high-performance muon trigger is crucial for recording a high-quality data set serving the various physics analyses.
In parallel, a good simulation of the trigger performance is necessary.
 
The ATLAS muon trigger system identifies muons produced in proton--proton (\pp) or heavy-ion (HI) interactions.
It is designed to do so with high efficiency and low muon transverse momentum (\pT) thresholds.
The system employs a two-level, multi-pronged strategy with
\begin{enumerate}
\item fast custom trigger electronics at Level-1 (L1);
\item dedicated algorithms to reconstruct muons and estimate their parameters at the High-Level Trigger (HLT).
\end{enumerate}
 
In order to address a wide variety of physics topologies, ATLAS has developed a suite of triggers designed to select muons.
A single-muon trigger with a \pT threshold of 26~\GeV{} is used by many physics analyses.
In addition, muon triggers in combination with electrons, $\tau$-leptons, jets and missing transverse momentum, as well as multi-muon triggers with lower muon \pT{} thresholds, increase the sensitivity for various physics phenomena which benefit from a lower \pT threshold.
For the BLS programme studying beauty, quarkonia, and low-mass physics, various low-\pT multi-muon triggers are used with a special configuration that allows a high trigger efficiency for non-prompt muons.\footnote{Non-prompt muons are muons which originate from the decay of a secondary particle and are displaced from the primary interaction vertex.}
 
During the LHC \RunTwo (2015--2018), the ATLAS experiment collected \pp collision data at a centre-of-mass energy of 13~\TeV{} with a maximum instantaneous luminosity of $2.1 \times 10^{34}~\mathrm{cm}^{-2} \mathrm{s}^{-1}$.
The average number of interactions occurring in the same bunch crossing, $<\mu>$, was 13 on average in 2015 and increased during the data-taking period to 25 in 2016, 38 in 2017, and 36 in 2018.
Such interactions beyond the interaction of interest, as well as interactions from neighbouring bunch crossings, are called pile-up interactions.
To cope with such harsh conditions, several improvements, on both the hardware side and software side, were deployed before the start of \RunTwo and during the data-taking campaign.
In this paper, the performance of the ATLAS muon trigger is evaluated, primarily using samples containing muon pairs from \PZ boson decays.
The performance of the low-\pT muon triggers ($\pT \lesssim 10$~\GeV{}) is evaluated with samples containing a pair of muons from the decay of \Jpsi mesons.
The performance of the high-\pT muon triggers ($\pT \gtrsim 100$~\GeV{}) is evaluated using events containing top quarks or \PW bosons, where a \PW boson decays into a muon and neutrino.

\section{The ATLAS detector}
 
The ATLAS experiment is a multipurpose particle detector with a forward--backward cylindrically symmetric geometry and almost 4$\pi$ coverage in solid angle,\footnote{ATLAS uses a right-handed coordinate system with its origin at the nominal interaction point (IP) in the centre of the detector and the $z$-axis along the beam pipe.
The $x$-axis points from the IP to the centre of the LHC ring, and the $y$-axis points upward.
Cylindrical coordinates ($r$, $\phi$) are used in the transverse plane, $\phi$ being the azimuthal angle around the beam pipe.
The pseudorapidity is defined in terms of the polar angle $\theta$ as $\eta = -\ln \tan(\theta/2)$.
Angular distance is measured in units of \(\Delta R \equiv \sqrt{(\Delta\eta)^{2} + (\Delta\phi)^{2}}\).} and is composed of four major sub-detectors: an inner tracking detector (ID), a calorimetry system subdivided into an electromagnetic calorimeter and a hadronic calorimeter, and a muon spectrometer (MS).
A detailed description of the ATLAS detector can be found in Ref.~\cite{PERF-2007-01,ATLAS-TDR-2010-19,PIX-2018-001}.
Muons are measured independently in the ID and in the MS.
The ID consists of a silicon pixel detector, a silicon microstrip detector and a transition radiation straw tube tracker, and is embedded in a solenoid, providing a magnetic field of 2~T.
The ID measures charged-particle tracks up to $|\eta|=2.5$.
\begin{figure}[htbp]
\centering
\includegraphics[width=0.85\textwidth]{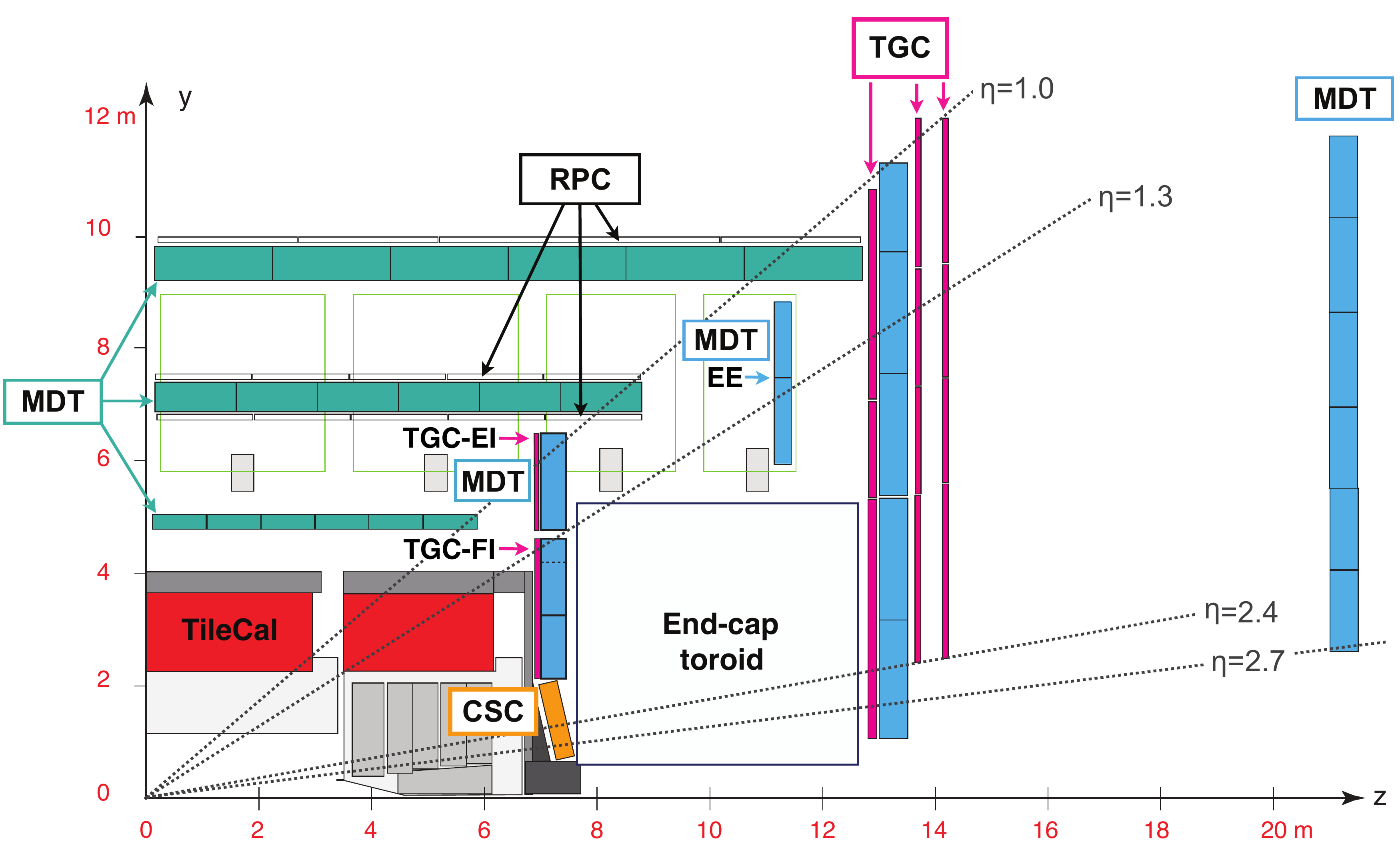}
\caption{A schematic picture showing a quarter-section of the muon system in a plane containing the beam axis, with monitored drift tube (MDT) and cathode strip (CSC) chambers for momentum determination and resistive plate (RPC) and thin gap (TGC) chambers for triggering.
The Forward Inner and Endcap Inner TGC chambers are marked as TGC-FI and TGC-EI, respectively. The Extended Endcap MDT chambers are referred to as EE.}
\label{fig:muon_spectrometer}
\end{figure}
 
The MS is based on three large air-core superconducting toroidal magnet systems (two endcaps and one barrel) with eight coils each, providing a field integral between 2.0~Tm and 6.0~Tm across most of the detector acceptance.
Figure~\ref{fig:muon_spectrometer} shows a quarter-section of the muon system in a plane containing the beam axis.
In the central region, the detectors compose a barrel that is arranged in three concentric cylindrical shells around the beam axis.
In the endcap region, the muon chambers form large wheels, perpendicular to the $z$-axis.
Several detector technologies are utilised to provide both precision tracking and triggering.
The deflection of the muon trajectory in the magnetic field is detected using hits in three layers of precision monitored drift tube (MDT) chambers for $|\eta|$ < 2.
In the region 2.0 < $|\eta|$ < 2.7, two layers of MDT chambers in combination with one layer of cathode strip chambers (CSC) are used.
Three layers of resistive plate chambers (RPC) in the barrel region ($|\eta|$ < 1.05), and three layers of thin gap chambers (TGC) in the endcap regions (1.05 < $|\eta|$ < 2.4) provide the L1 muon trigger and the read-out of the coordinate in the $r$--$\phi$ projection.
 
\section{Data and simulation samples}
\label{sec:data_selection}
 
Several data samples collected by the ATLAS detector are used to measure the muon trigger efficiency.
In the following, the data samples used in the analysis are summarised.
The data used in the measurements to derive the \pp collisions trigger performance rates and efficiencies were collected during \RunTwo in 2015--2018 with \pp collisions at a centre-of-mass energy of 13~\TeV{}, amounting to a total integrated luminosity of 139~\ifb~\cite{ATLAS-CONF-2019-021,LUCID2}.
Only data recorded with stable beams and with all relevant sub-detector systems fully operational are considered and accounted for in the integrated luminosity calculation.
The trigger performance measured in \pp collision data is compared with predictions of Monte Carlo (MC) simulation.
Generated samples were processed by the detector simulation of the ATLAS experiment based on $\GEANT4$~\cite{Agostinelli:2002hh}.
 
The intermediate-\pT (low-\pT) analysis uses \Zmm (\Jpsi$\rightarrow\mu\mu$) samples for the performance measurements.
Samples of prompt \Jpsi$\rightarrow\mu\mu$ decays were generated using \pythia.186~\cite{Sjostrand:2007gs} complemented with Photos++ (v3.52)~\cite{Golonka:2005pn,Davidson:2010ew} to simulate the effects of final-state radiation.
A requirement on the minimum transverse momentum of each muon ($\pT > 4$~\GeV{}) was applied at the generator level.
The \POWHEGBOX~v1 MC generator~\cite{Nason:2004rx,Frixione:2007vw,Alioli:2010xd,Alioli:2008gx} was used for the simulation of the hard-scattering processes of $Z$ boson production and decay in the muon channel.
It was interfaced to \pythia.186 for the modelling of the parton shower, hadronisation, and underlying event, with parameter values set according to the AZNLO tune~\cite{STDM-2012-23}.
The CT10 parton distribution function~(PDF) set~\cite{Lai:2010vv} was used for the hard-scattering processes, whereas the CTEQ6L1 PDF set~\cite{Pumplin:2002vw} was used for the parton shower.
The effect of quantum electrodynamics final-state radiation was simulated with Photos++ (v3.52).
The \evtgen~v1.2.0 program~\cite{EvtGen} was used to decay bottom and charm hadrons.
 
The high-\pT analysis focuses on two event topologies: events with muons from top-quark pair production (\ttbar), and events containing a leptonically decaying \PW boson and jets (\Wjets).
Events from $V+$jets production ($V = W$ or $Z$) and diboson processes were simulated with the \SHERPA~v2.2~\cite{Bothmann:2019yzt} generator at next-to-leading order (NLO) in QCD.
Samples were generated using the \nnpdfnnlo PDF set~\cite{Ball:2014uwa}, along with the dedicated set of tuned parton shower parameters developed by the \SHERPA authors.
The production of \ttbar events was modelled using the \POWHEGBOX~v2~\cite{Frixione:2007nw} generator at NLO in QCD.
Background processes are also estimated from simulation, using several generation set-ups depending on the process.
Single-top production in the $s$- and $t$-channels was simulated with the same set-up using \POWHEGBOX~v2~\cite{Alioli:2009je,Frederix:2012dh} at NLO in QCD. 
The $Wt$ process was modelled using the \POWHEGBOX~v2~\cite{Re:2010bp,Nason:2004rx,Frixione:2007vw,Alioli:2010xd} generator at NLO in QCD in the five-flavour scheme.
The $tZ$ process was modelled with \MGMCatNLO at leading order in QCD. 
The production of $tWZ$, $t\bar{t}W$, $t\bar{t}Z$ and $t\bar{t}t\bar{t}$ events was modelled using the \MGMCatNLO~v2.3.3~\cite{Alwall:2014hca} generator at NLO in QCD.
All top-quark samples were produced with the NNPDF3.0 PDF set and were interfaced with \pythia~\cite{Sjostrand:2014zea} using the A14 tune~\cite{ATL-PHYS-PUB-2014-021} and the \nnpdflo PDF set~\cite{Ball:2012cx}.
 
In addition, the trigger efficiencies are also derived for data from low-pile-up \pp collisions with an average pile-up of 1.1 at a centre-of-mass energy of 5.02~\TeV{}.
This data set was collected in November 2015, and amounts to a total integrated luminosity of 25~\ipb.
The corresponding sample of simulated \Zmm events was produced with a set-up similar to the one described above.
 
Trigger efficiency measurements for HI collisions described in this document are based on the data collected during \RunTwo for \PbPb and \pPb collisions at centre-of-mass energies of $\sqn=5.02$~\TeV{} and $\sqn=8.16$~\TeV{} per nucleon pair.
The \PbPb data were collected in November--December 2015 and at the end of \RunTwo, amounting to total integrated luminosities of 0.49~\inb{} and 1.42~\inb, respectively.
The \pPb data were collected in November--December 2016, amounting to a total integrated luminosity of 165~\inb.
 
The \pPb collisions were divided into two periods corresponding to different beam orientations of the protons and lead nuclei.
The two beam orientations are defined as follows:
\begin{itemize}
\item The `\pPb{}' configuration: protons circulate in beam 2 and lead ions circulate in beam 1;
protons go in the negative-$\eta$ direction.
The total integrated luminosity amounts to 56.8 \inb.
\item The `\Pbp{}' configuration: protons circulate in beam 1 and lead ions circulate in beam 2;
protons go in the positive-$\eta$ direction.
The total integrated luminosity amounts to 107.8 \inb.
\end{itemize}
 
\section{Offline object reconstruction and identification}
 
Reconstructed muons and jets are used to measure the muon trigger performance.
The detailed selections of the reconstructed objects differ depending on the measurement and are outlined in Section~\ref{sec:perf}, while this section provides an overview of the object reconstruction itself.
 
Muons are reconstructed~\cite{PERF-2015-10} from combined tracks in the MS and the ID.
Their transverse momentum is calibrated~\cite{PERF-2015-10}, and they are required to fulfil certain identification criteria~\cite{PERF-2015-10} which may vary between different measurements.
To be selected as prompt muons, their tracks must point to the primary vertex~\footnote{The primary vertex is defined as the reconstructed vertex with the highest sum of the squared transverse momenta of the associated tracks.} (PV), which is ensured by requiring that the track's transverse impact parameter significance, $|d_{0}/\sigma(d_0)|$, is less than 3, and that the distance of closest approach to the PV along the $z$-axis satisfies $|\zzsth| < 0.5$~mm.
In order to suppress background from non-prompt muons, an isolation criterion can be applied.
The scalar sum of the \pT of tracks within a variable-size cone around the muon (excluding its own track) must be less than 6\% of the muon \pT.
The track isolation cone size for muons, $\Delta R$, is given by the smaller of $\Delta R = \SI{10}{~\GeV{}}/\pT$ and $\Delta R = 0.3$.
 
Jets are reconstructed from topological clusters of energy deposits in calorimeter cells~\cite{PERF-2014-07} with the \antikt algorithm~\cite{Cacciari:2008gp} with a radius parameter of 0.4.
Jets are calibrated to the jet energy scale at particle level~\cite{PERF-2016-04} and are required to be within the fiducial volume $|\eta| < 2.5$.
For jets with $|\eta| < 2.4$ and $\pT < 60$~\GeV{}, pile-up contributions are suppressed by the use of the jet vertex tagger~\cite{PERF-2014-03}.
Jets containing $b$-hadrons are identified as `$b$-tagged' using the MV2c10 algorithm, a multivariate discriminant based on the track impact parameters and displaced vertices~\cite{FTAG-2018-01_published}.
These $b$-tagged jets are reconstructed in the region $|\eta| < 2.5$ and have $\pT > 20$~\GeV{}.
The $b$-tagging working point (WP) with 77\% efficiency for jets containing $b$-hadrons in simulated \ttbar events provides rejection rates of 110 and 4.9 for light-flavour jets and jets containing $c$-hadrons, respectively.
 
The missing transverse momentum $\mathbf{p}_{\text{T}}^{\text{miss}}$, with magnitude \met, is calculated as the negative vectorial sum of the transverse momenta of all reconstructed objects and the soft term.
The soft term includes all tracks associated with the PV but not matched to any reconstructed physics object.
Tracks not associated with the PV are not considered in the \met calculation, improving the \met resolution by suppressing the effect of pile-up~\cite{PERF-2016-07,ATLAS-CONF-2018-023}.
 
\section{The ATLAS muon trigger}
\label{sec:MuonTrigger}
 
The muon trigger system is a part of the ATLAS trigger system, allowing event triggering based on muons in a wide muon momentum range with high efficiency.
The ATLAS trigger, including the muon trigger system, conceptually consists of two levels: the hardware-based L1 trigger and the software-based HLT.
The L1 decision is formed by a Central Trigger Processor~\cite{Ask_2008}, based on information received from the calorimeters and muon trigger chambers.
For select multi-object triggers, the L1 topological trigger processor~\cite{ATL-COM-DAQ-2014-019,Artz_2015}, commissioned in 2016, combines information about several objects into topological information.
After the L1 trigger accepts the event, it is processed by the HLT.
If also accepted at the HLT level, the event is transferred to local storage and exported to a Tier-0 facility at the CERN computing centre for offline reconstruction and finally moved to permanent storage.
Recorded events are gathered in data streams, depending on their primary use case and their specific offline reconstruction needs.
The event selection in the HLT is referred to as a trigger, and the collection of all L1 and HLT triggers and their prescales\footnote{The term prescale means that only one out of $N$ events passing the trigger decision is accepted.} is called the trigger menu.
The trigger menu defines several types of triggers~\cite{ ATL-DAQ-PUB-2019-001}, but this paper focuses on primary triggers which are used for physics measurements and typically have no prescale applied.
 
In the L1 processing the degree of deviation from the hit pattern expected for a muon with infinite momentum is used to estimate the \pT of the muon with six programmable thresholds.
The number of muon candidates passing each threshold is used in the conditions for the global L1 trigger.
Following an L1 accept decision, the \pT thresholds and the corresponding detector regions, called regions of interest (RoIs), are sent to the HLT for further consideration~\cite{PERF-2007-01,TRIG-2016-01}.
The typical dimensions of the RoIs are $0.1 \times 0.1$ ($0.03 \times 0.03$) in $\Delta\eta\times\Delta\phi$ in the RPCs (TGCs)~\cite{PERF-2007-01}.
The geometric coverage of the L1 trigger is $\approx$99\% in the endcap regions and $\approx$80\% in the barrel region.
The limited geometric coverage in the barrel region is due to gaps around $\eta = 0$ (to provide space for services of the ID and calorimeters), the feet and rib support structures of the ATLAS detector, and two small elevator shafts in the bottom part of the MS.
The barrel region is equipped with three concentric layers of RPCs.
The L1 trigger decision in the barrel region is based on the coincidence of hits from three (two) concentric RPC stations~\cite{1748-0221-4-04-P04010} for the three high-\pT (low-\pT) thresholds.
During \RunTwo, only the high-\pT triggers were used for single-muon signatures, while the low-$p_\mathrm{T}$ triggers were used in coincidence with other trigger objects to select multi-object signatures, including muon pairs.
During the shutdown of the LHC in 2013--2014, a fourth layer of RPC chambers was added to the trigger system in the detector feet region ($-2.16 < \phi < -1.77$ and $-1.37 <\phi < -0.98$) to cover holes in the geometrical acceptance caused by the ATLAS support structures.
These RPC chambers were installed during the construction of ATLAS but not equipped with trigger electronics at that time.
Figure~\ref{fig:rpc_eff_feet} (left) shows the position of these chambers in the outer part of the muon spectrometer.
For these new RPC chambers a two-layer coincidence is used also for the highest \pT thresholds.
This resulted in a 20\% increase of acceptance in those $\phi$ regions.
These additional `feet RPC' chambers were commissioned during the data-taking in 2015 and enabled for physics in 2016.
Figure~\ref{fig:rpc_eff_feet} (right) shows the impact of the feet RPC chambers on the trigger efficiency in one of the two $\phi$ regions.
Additionally, new RPC detectors were installed in 2014 in the elevator regions at $\phi \simeq 1.57$, $|\eta| \simeq 0.7$.
Part of these chambers use small-gap RPCs with a 1~mm gas gap, rather than 2~mm as in standard ATLAS RPCs, similar to those that will be used for the High Luminosity LHC upgrade.
\begin{figure}[htbp]
\centering
\includegraphics[width=0.49\textwidth]{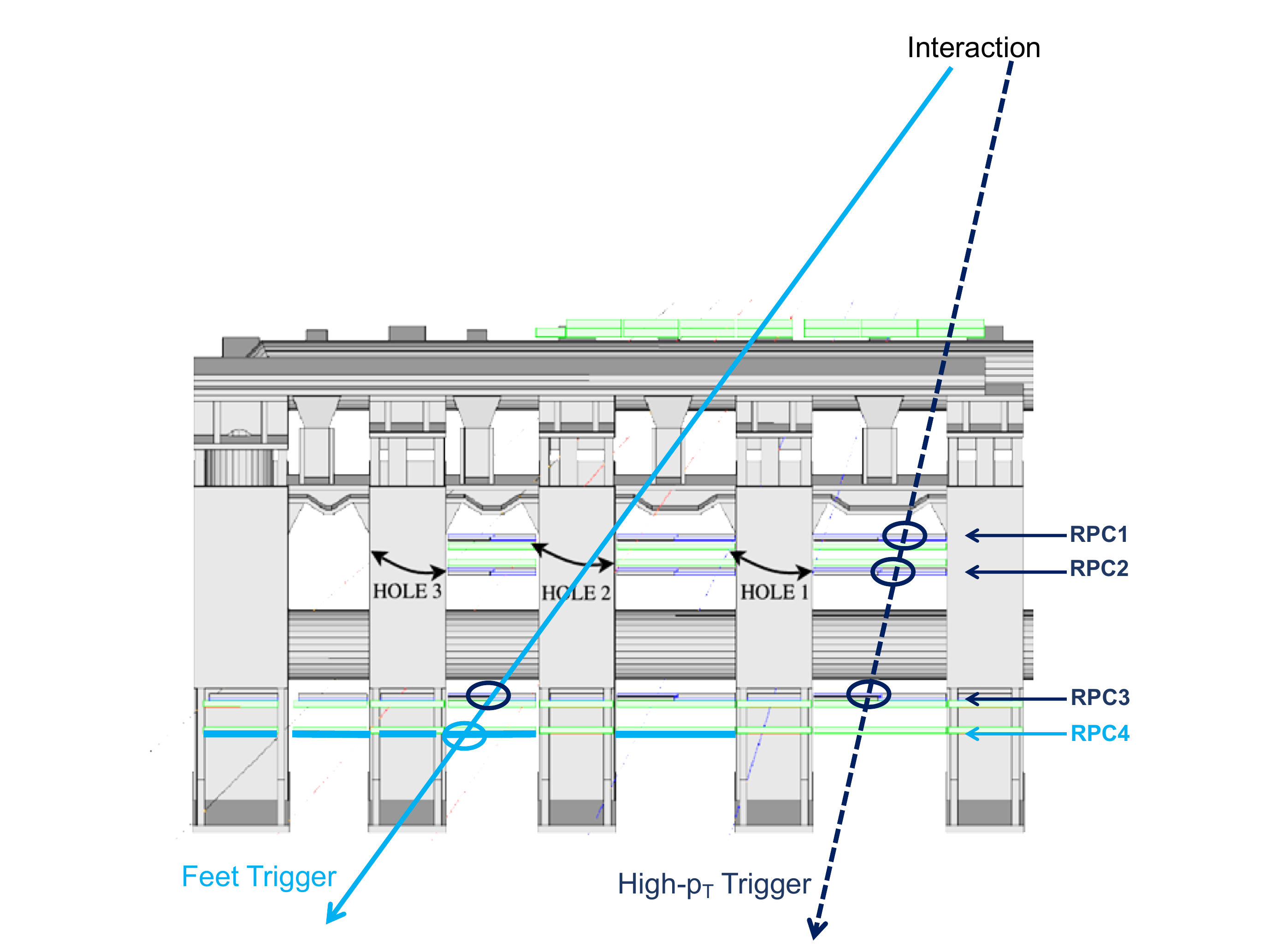}
\includegraphics[width=0.49\textwidth]{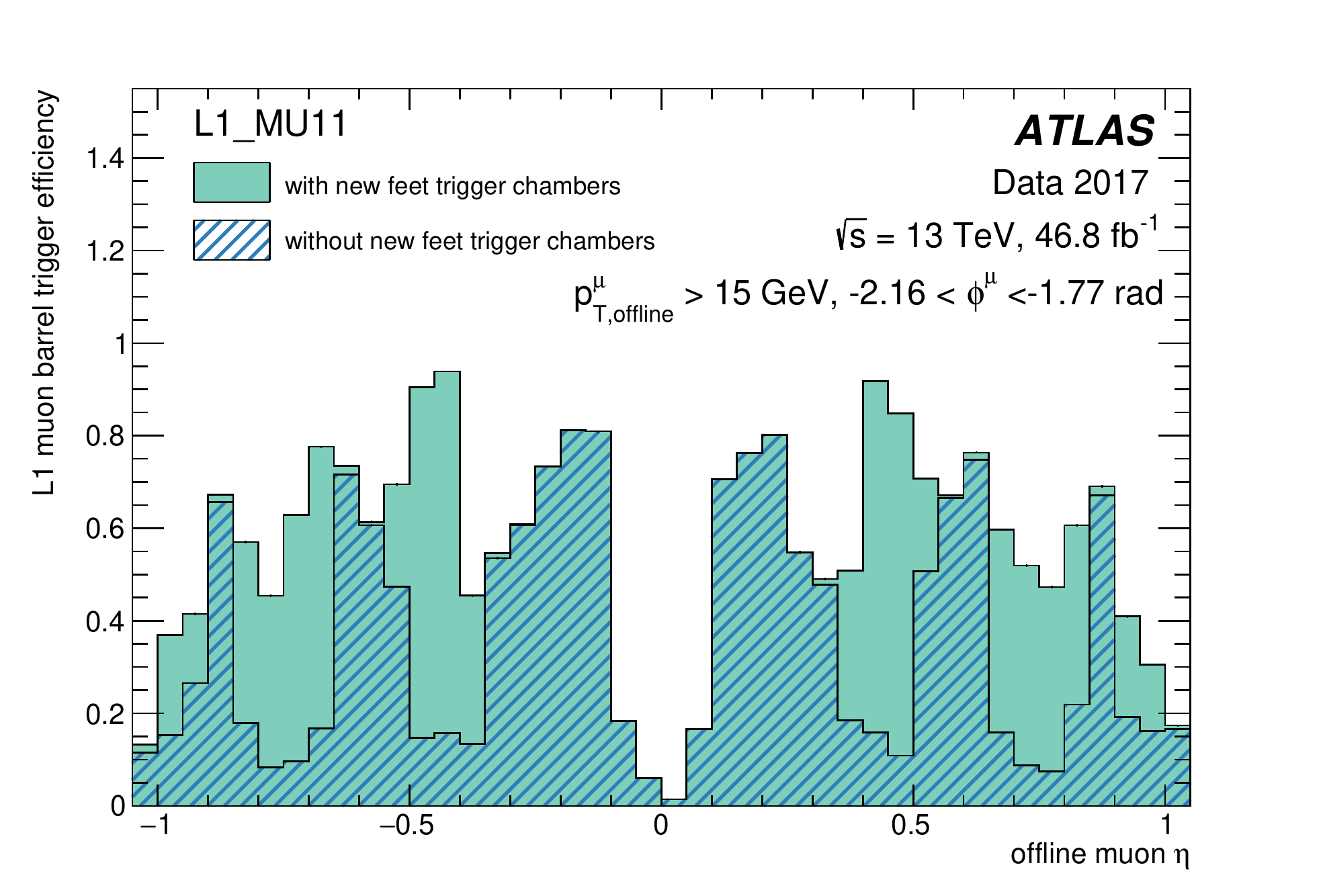}
\caption{Left: Sketch showing the position of the new RPC chambers (RPC4) in the feet sectors.
The muon trajectory depicted by the solid line is accepted by a two-layer coincidence (RPC3-RPC4) using the new feet chambers.
The dashed-line trajectory shows a muon accepted by the standard three-layer (RPC1-RPC2-RPC3) high-\pT trigger.
Right: L1 trigger efficiency as a function of the pseudorapidity of offline reconstructed muon candidates in the barrel detector region for $-2.16 < \phi < -1.77$ evaluated for a trigger applying a 3-station coincidence requirement and a \pT threshold of 10~\GeV{} (L1\_MU11).
The efficiency with the newly installed RPC chambers is shown as the solid filled histogram, while the efficiency without these chambers is overlaid as the hatched histogram.
Muons are required to pass \textit{Medium} quality requirements~\cite{PERF-2015-10} and have a transverse momentum of at least 15~\GeV{}.}
\label{fig:rpc_eff_feet}
\end{figure}
 
The L1 trigger decision in the endcap region ($1.05<|\eta|<2.4$) is based on the coincidence of hits in the TGC stations of the middle layer, called the Big Wheel.
Many upgrades were deployed in \RunTwo in order to reduce the L1 trigger rate while keeping the efficiency high.
The main source of trigger background in the L1 muon endcap system is low-momentum charged particles emerging from the endcap toroid magnets and beam shielding.
To suppress these backgrounds, a coincidence requirement between the Big Wheel and TGC Forward Inner (TGC-FI) chambers was introduced in 2015.
The optimisation of this coincidence and a new coincidence between the Big Wheel and TGC Endcap Inner (TGC-EI) chambers was performed in 2016.
The effect of these coincidence requirements is shown in Figure~\ref{fig:L1:endcaps:FI}.
The asymmetry is due to the different acceptance of charged fake muons given the magnetic field configuration of the MS.
\begin{figure}[htbp]
\centering
\includegraphics[width=0.60\textwidth]{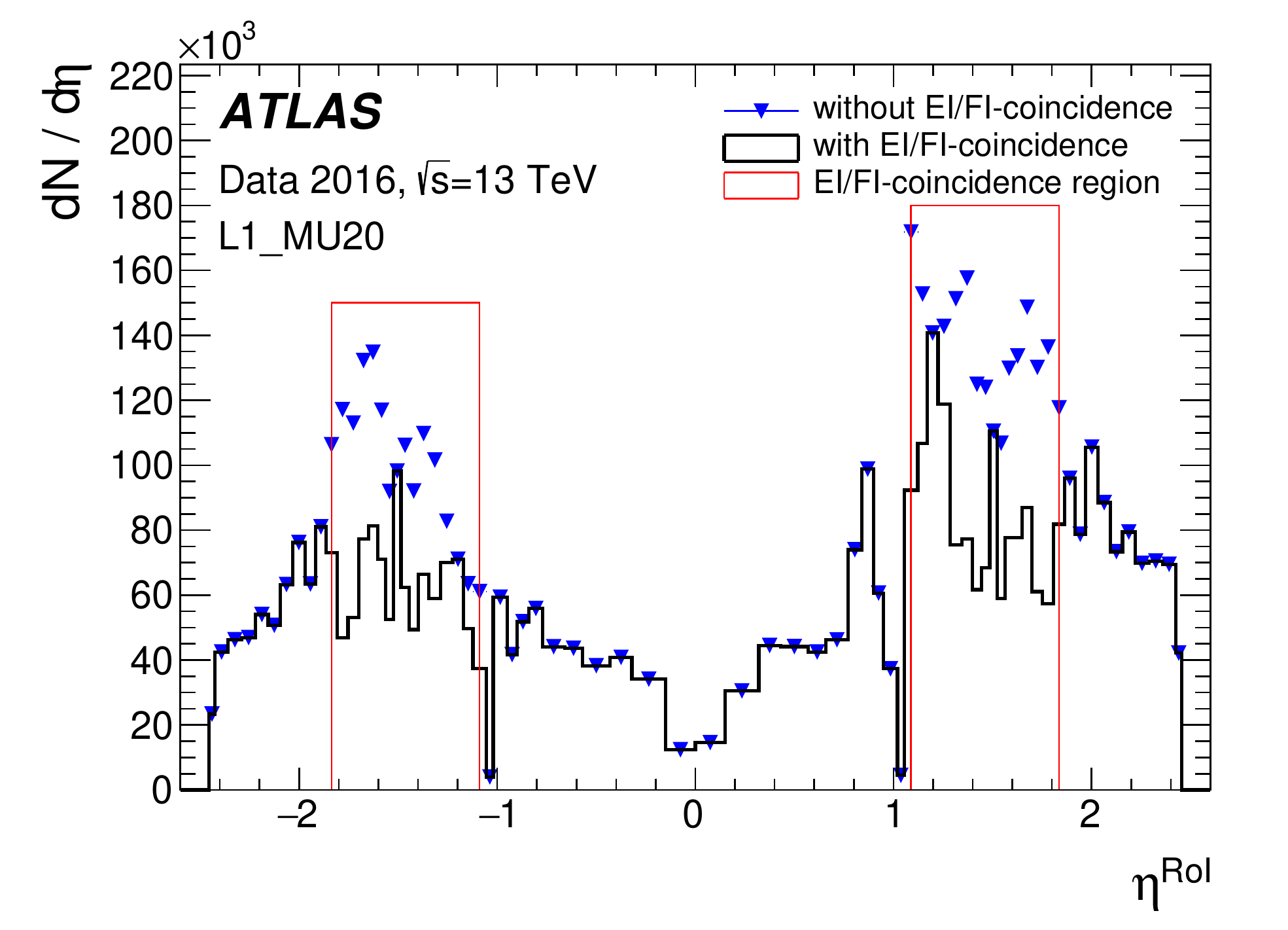}
\caption{Pseudorapidity distributions of the L1 muon trigger with a \pT threshold of 20~\GeV{} (L1\_MU20) and the rate reduction due to requiring a coincidence with TGCs consisting of Endcap Inner (EI) and Forward Inner (FI) chambers are shown, using events taken by a lower-threshold L1 trigger (L1\_MU11) in 2016. The $\eta^\mathrm{RoI}$ distribution after the inclusion of the EI/FI-coincidence is shown as a solid black line. The $\eta^\mathrm{RoI}$ distribution before the inclusion of the EI/FI-coincidence is also shown as a reference (blue triangles) to examine the reduction of the L1\_MU20 trigger rate at  $1.05<|\eta|<1.8$, which is highlighted by the red rectangles.
No coincidence was required for $|\eta|\sim 1.5$ because the FI chambers in this region were inactive, which explains the lack of rate reduction.
Moreover, the coincidence around $|\eta|\sim 1.2$ was intentionally not applied, to allow commissioning of a coincidence with the TileCal.}
\label{fig:L1:endcaps:FI}
\end{figure}
 
A coincidence between TGC chambers and the tile hadronic calorimeter (TileCal) assists in the rejection of fake muon triggers in the region $1.05<|\eta|<1.3$.
The coincidence mitigates the effect of the limited $\phi$ coverage~($\sim$50\%) of the inner layer of the muon detector due to the toroidal magnets.
Figure~\ref{fig:TileMuon_L1MuE}~(left) shows the pseudorapidity distribution of the single-muon trigger with \pT requirement above 20~\GeV{} (L1\_MU20).
A rate reduction is observed in $1.05<|\eta|<1.3$.
Figure~\ref{fig:TileMuon_L1MuE}~(right) shows the L1\_MU20 trigger rate as a function of instantaneous luminosity.
A reduction of 6\% in the L1\_MU20 trigger rate is observed for the entire coverage of the MS when requiring the TileCal coincidence, at a cost of at most 2.5\% inefficiency.
This is compatible with the geometrical gaps between TileCal modules.
\begin{figure}[!h]
\begin{minipage}[c][1\width]{0.5\textwidth}
\includegraphics[clip,width=1\textwidth]{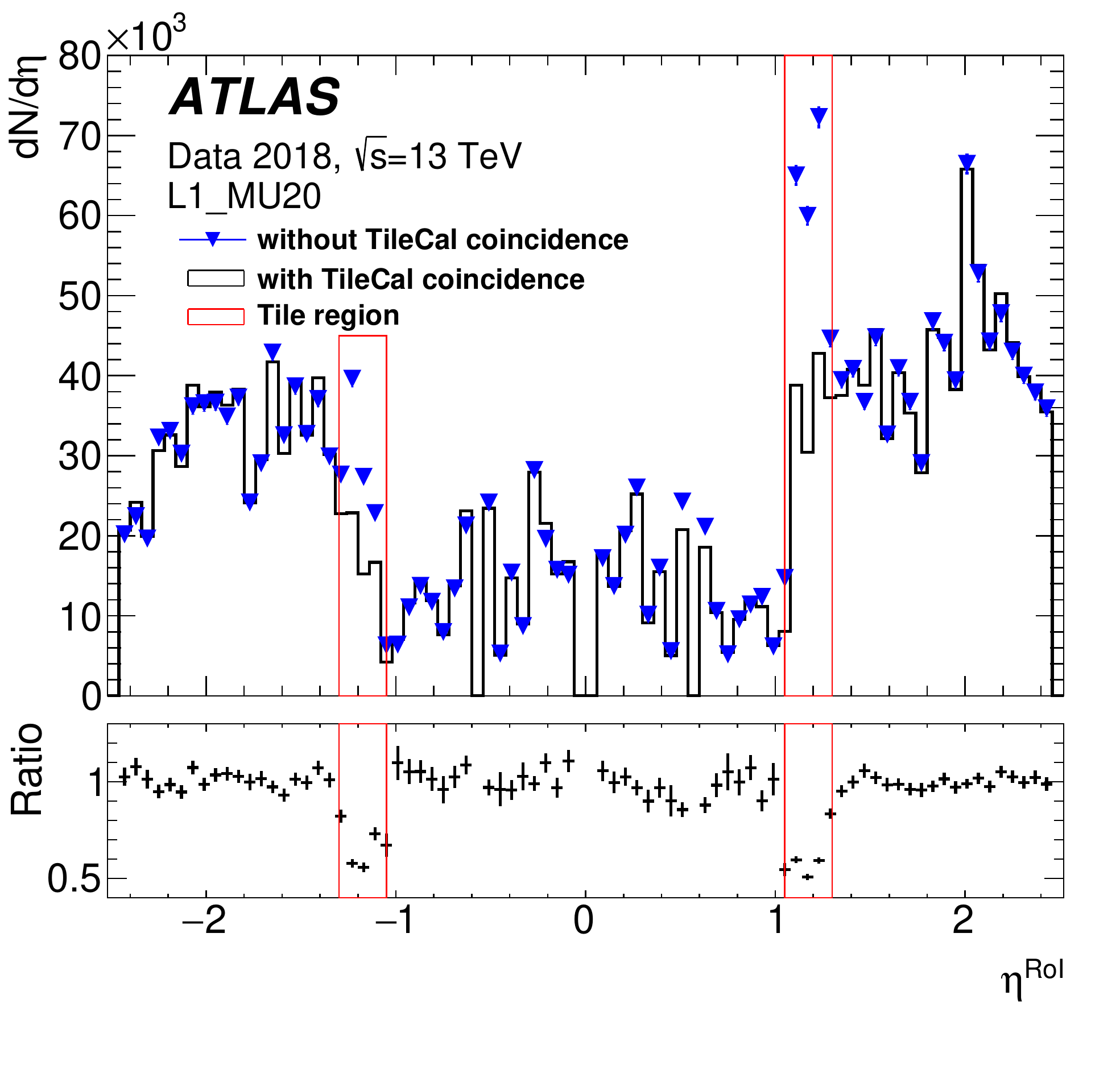}
\end{minipage}{\centering{}
\begin{minipage}[c][1\width]{0.5\textwidth}
\begin{center}
\includegraphics[clip,width=1.\textwidth]{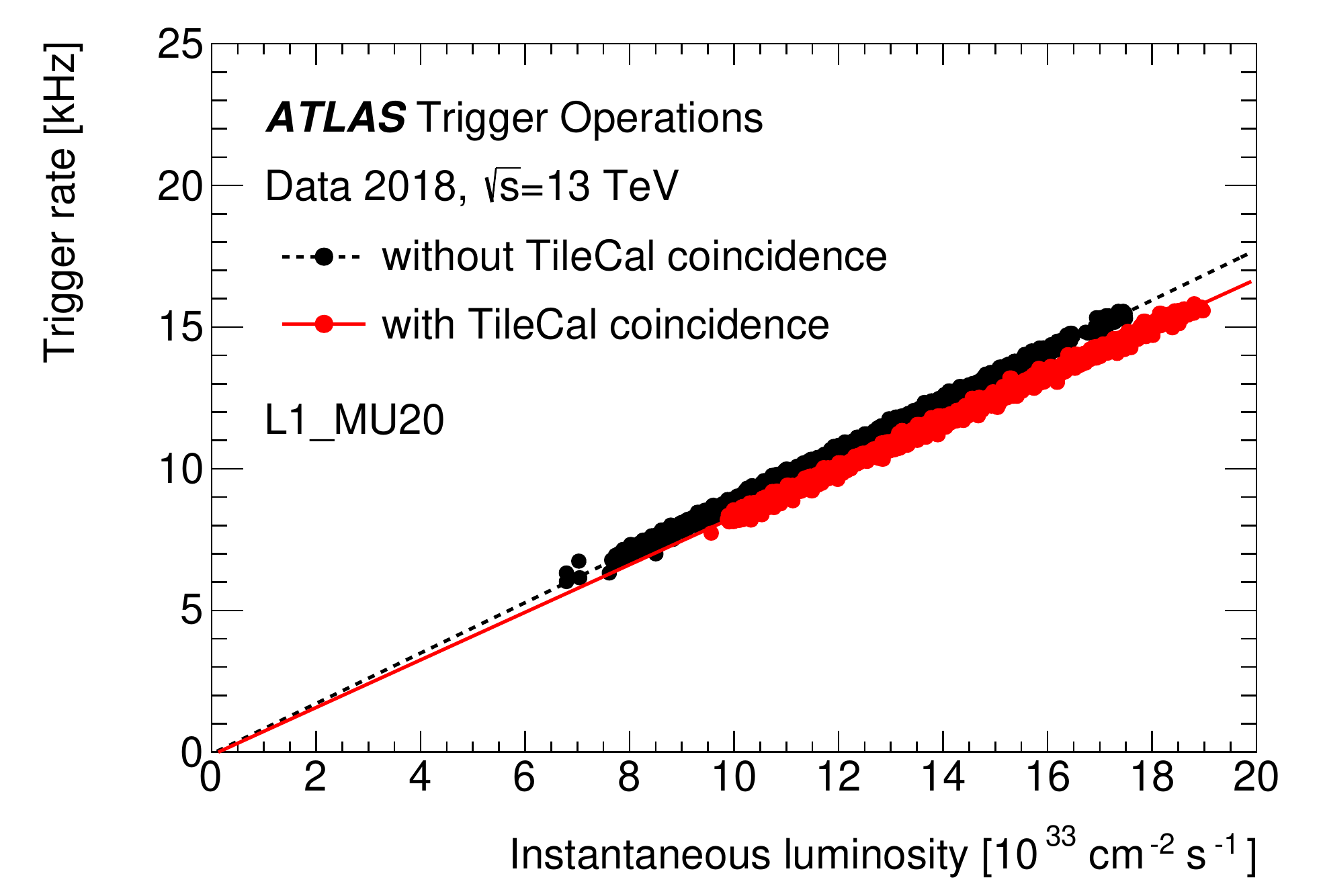}
\par\end{center}
\end{minipage}}
\caption{Left: Pseudorapidity distribution of the L1 RoIs~($\eta^\mathrm{RoI}$) which fulfil the 20~\GeV{} requirement~(L1\_MU20) after the new TileCal coincidence in the L1 trigger decisions~(solid black line).
The $\eta^\mathrm{RoI}$ distribution before the deployment of the TileCal coincidence is also shown as a reference~(blue triangles) to examine the reduction of the L1\_MU20 trigger rate at $1.05<|\eta|<1.3$, which is highlighted by the red rectangles.
The reference histogram is normalised so that the entries out of the acceptance of the TileCal coincidence (1.05 < $|\eta^\mathrm{RoI}|$ < 1.3) are compatible between the two distributions for the comparison.
The ratio of after to before deployment of the TileCal coincidence is also shown.
The error bars show the statistical uncertainties only.
Right: L1 trigger rate for the L1\_MU20 trigger as a function of instantaneous luminosity.
The black~(red) points correspond to data recorded without~(with) the TileCal coincidence requirement.}
\label{fig:TileMuon_L1MuE}
\end{figure}
In 2016, the 2-station-strip coincidence in the Big Wheel was changed to a 3-station-strip coincidence for the single-muon trigger with \pT requirement above 4~\GeV{} (L1\_MU4).
A chamber-by-chamber Coincidence Window (CW) optimisation procedure was introduced to take the detector alignment into account by varying the \pT depending on the bending magnitude.
Originally the CWs were extracted from MC simulation with perfect alignment.
During the 2017 data-taking the CWs were optimised for most triggers based on the data taken in 2015 and 2016.
CW optimisations for lower-\pT triggers were performed at the beginning of 2018.
The rate reduction for the L1 trimuon trigger with a 4~\GeV{} threshold is shown in Figure~\ref{fig:L1:endcaps:CWoptimization}.
\begin{figure}[htbp]
\centering
\includegraphics[width=0.46\textwidth]{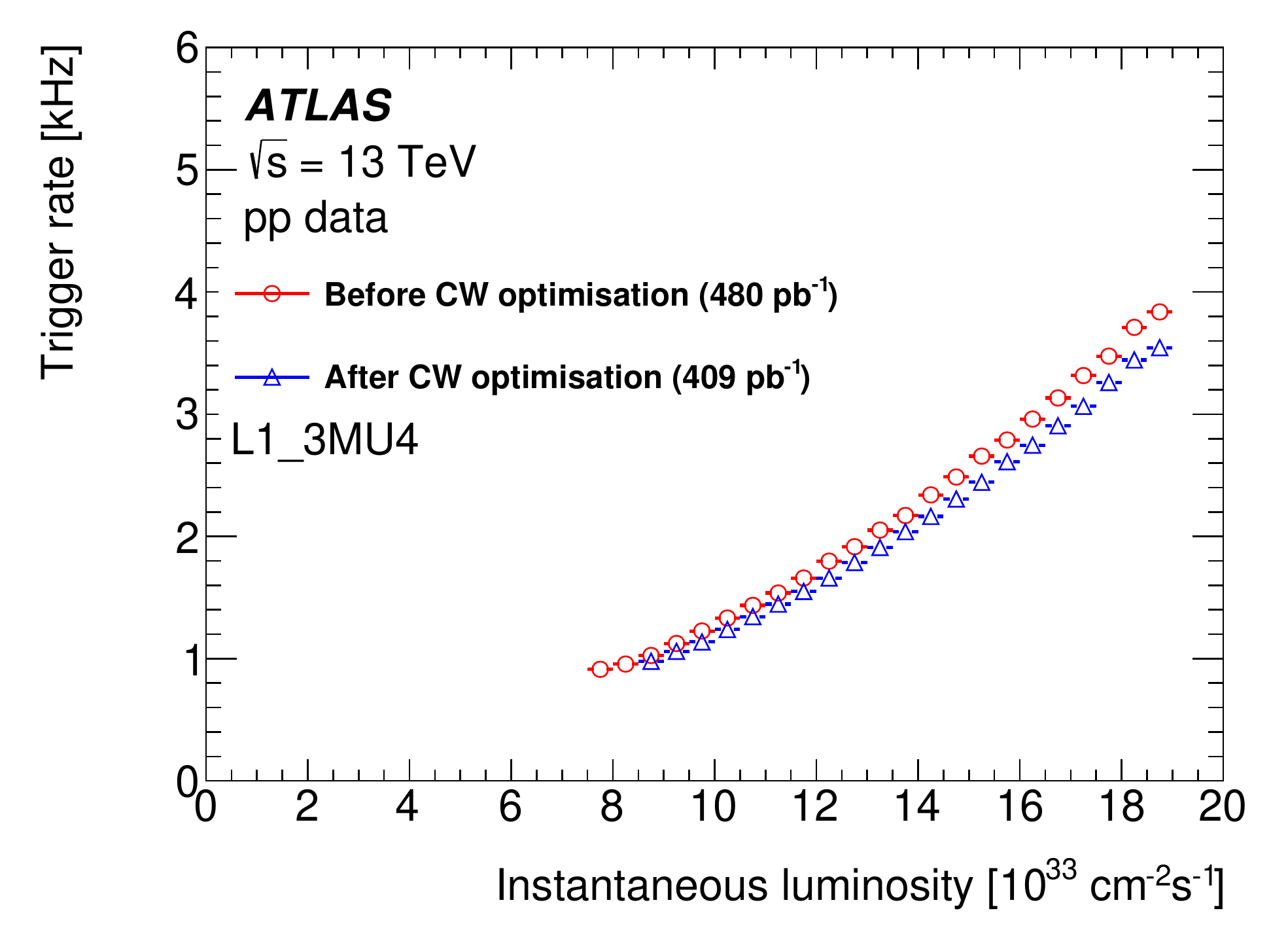}
\caption{Trigger rate of the trimuon L1 trigger with a 4~\GeV{} threshold as a function of the instantaneous luminosity before (open circles) and after (open triangles) the coincidence window optimisation.}
\label{fig:L1:endcaps:CWoptimization}
\end{figure}
 
The HLT selects events in two stages, executing fast reconstruction algorithms first, followed by muon algorithms similar to the ones used for offline muon reconstruction~\cite{PERF-2015-10}.
The RoI identified by the L1 trigger enables the fast algorithms to select precisely the region of the detector in which the interesting features reside, therefore reducing the amount of data to be transferred and processed.
The muon stand-alone (SA) algorithm constructs a track using the MDT hits within the RoI, refining the L1 candidate.
To achieve the needed resolution in sufficiently short time, the \pT of the SA muon is reconstructed with simple parameterised functions.
Several changes were deployed during \RunTwo in order to improve the efficiency and \pT resolution of the SA algorithm.
The efficiency is optimised by refining the fitting algorithm such that noise hits in the MDTs are removed.
Further improvements in the \pT resolution are obtained in the regions $1.05 < |\eta| < 1.35$ and $2.0 < |\eta| < 2.4$ by including additional hit information from the Extended Endcap (EE) and CSC chambers respectively.
Figure~\ref{fig:hlt:l2:improvements} summarises the improvements in the efficiency (left) and \pT resolution (right) which have been measured with the \tap method discussed in Section~\ref{sec:eff}.
The raise in efficiency at \pT values below the threshold is caused by two effects: the L1 trigger mis-reconstructs a low-\pT muon as a high-\pT object leading to a wrong reconstruction by the fast reconstruction algorithms, and an overestimation of the muon \pT if it traverses regions with a weak magnetic field, leading to an almost straight trajectory.
\begin{figure}[htbp]
\centering
\includegraphics[width=0.45\textwidth]{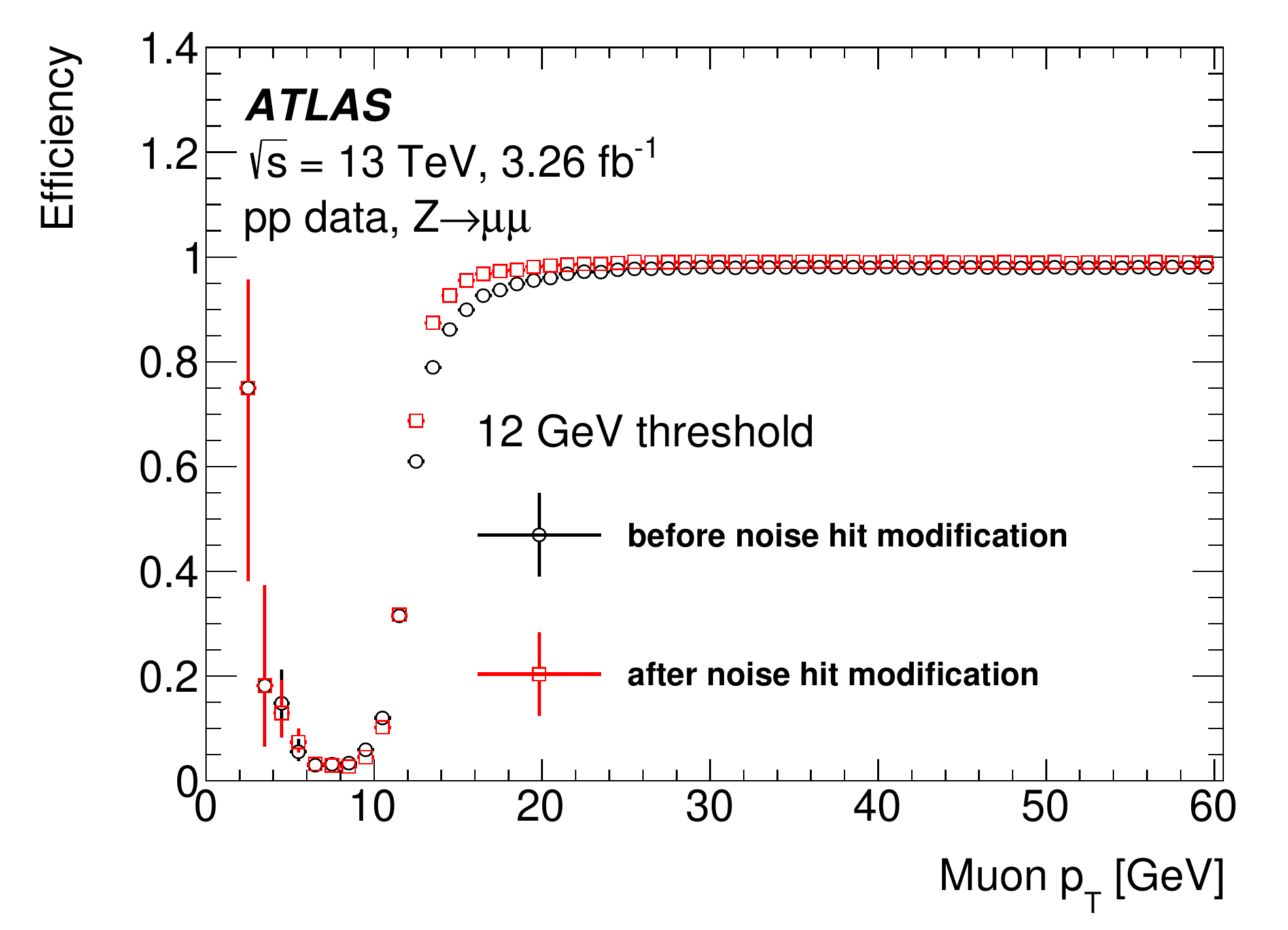}
\includegraphics[width=0.45\textwidth]{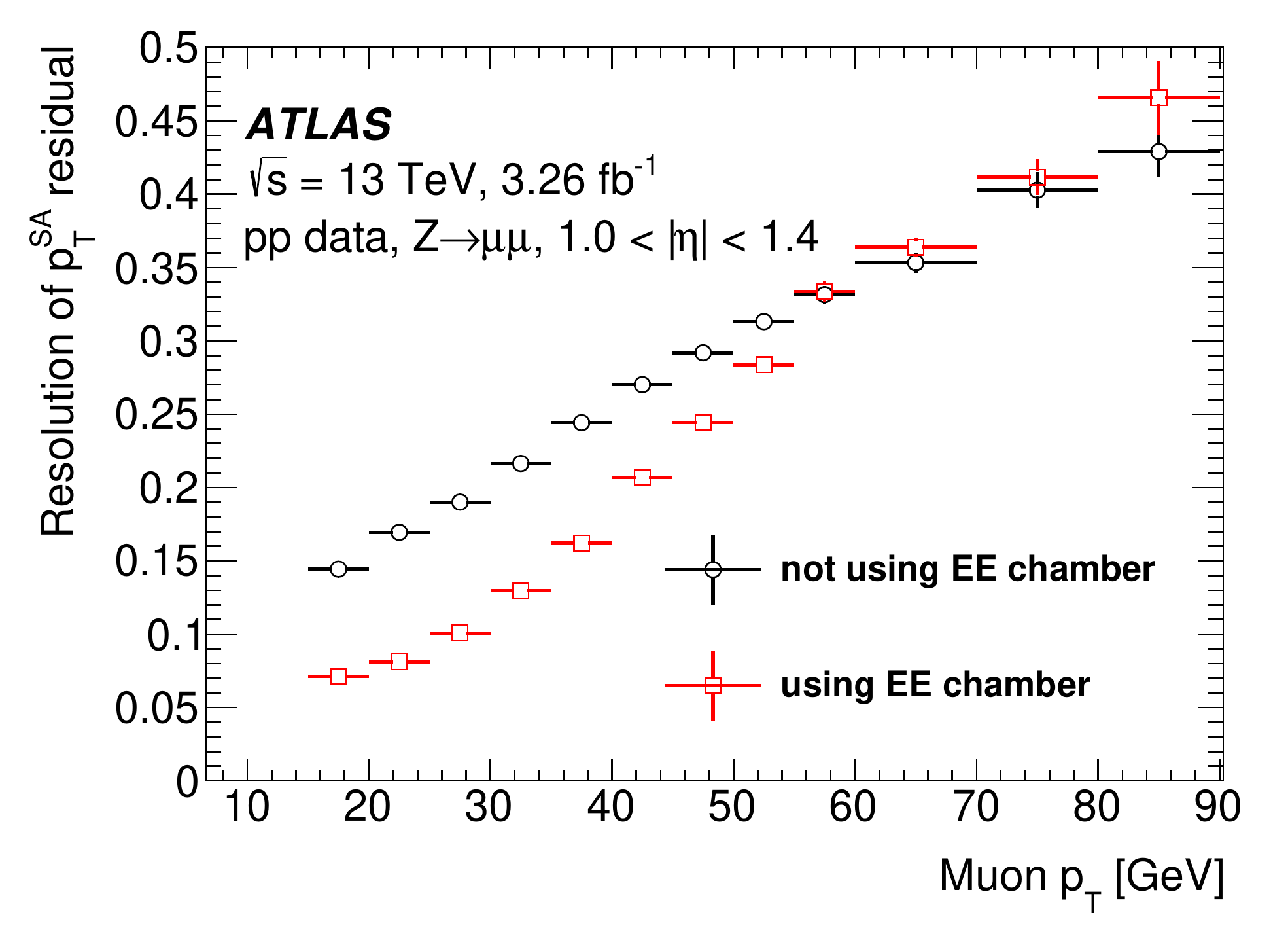}
\caption{Improvements of the fast reconstruction algorithms measured in \pp collision data taken in 2015 evaluated in \Zmm events using a \tap approach outlined in Section~\ref{sec:eff}.
The error bars show the statistical uncertainties only.
Left:~Efficiency of the fast SA reconstruction before (black circles) and after (red squares) the modification of the noise hits requirement for a \pT threshold of 12~\GeV{}.
Right:~\pT resolution of the fast SA reconstruction obtained by including additional hit information from the EE chambers (red squares) and not including (black circles) this information.
The resolution is extracted by a Gaussian fit to the distribution of the fractional residual of inverse-\pT, $\left(1/p_{\mathrm{T}}^{\mathrm{SA}} - 1/p_{\mathrm{T}}^{\mathrm{offline}}\right)/\left(1/p_{\mathrm{T}}^{\mathrm{offline}}\right)$.
}
\label{fig:hlt:l2:improvements}
\end{figure}
 
Following the SA algorithm step, the MS-only muon track is combined with reconstructed tracks in the ID.
For this the muon track is back-extrapolated to the interaction point using the offline track extrapolator and statistically combined with tracks reconstructed in the ID to form a combined muon candidate with refined track parameter resolution.
 
The final step in the HLT selection is the precision stage, which can be operated in two modes: the RoI-based mode and the full-scan (FS) mode.
The first mode focuses on RoIs defined by the L1 and fast reconstruction steps, while the second searches the full detector without using information from any of the previous steps.
Using the FS mode allows circumvention of the L1 inefficiencies, but given the high CPU demand it cannot be executed for every event.
 
In the RoI-based method, muon candidates, called precision stand-alone muons, are first formed by using the muon detectors, and are subsequently combined with ID tracks, by means of a fit of the hits from both, leading to precision combined muons.
If no combined muon is formed, muon candidates are searched for by extrapolating ID tracks to the muon detectors.
If there are corresponding track segments, combined muons are formed.
Additionally, the degree of isolation for the combined muon is quantified by summing the \pT of ID tracks with $p_{\mathrm{T}}^{\mathrm{trk}} > 1~\GeV{}$ found near the combined muon candidate.
The `near'  criterion is defined by a cut requiring $\Delta z < 6~\mathrm{mm}$, with $\Delta z$ being the distance of the track from the primary vertex in the $z$-direction.
This cut was found to be slightly inefficient in events with high pile-up in 2017 and thus was tightened to $\Delta z < 2~\mathrm{mm}$, which allowed the loosening of the isolation criterion for data-taking in 2018.
The isolation criterion is defined by a cut on the ratio of the summed transverse momentum of the additional nearby tracks to the \pT of the muon candidate, $\sum_{\Delta z < 6\,(2)\,\mathrm{mm}} p_{\mathrm{T}}^{\mathrm{trk}} / p_{\mathrm{T}}^{\mu} < \mathrm{cut}$.
Several cut values are defined, corresponding to different WPs.
Commonly, the \textit{Medium} WP is used with a cut at 7\%, while in 2015 a cut at 6\% was applied.
A second procedure extends the combined precision algorithm by searching for muon candidates in the MS only.
This can be used, for example, in searches for long-lived particles leaving no signature in the ID.
 
The FS mode is used to find additional muons that are not found by the RoI-based method mainly due to L1 inefficiencies.
In the FS mode, muon candidates are first sought in all muon detectors.
Then, RoIs are constructed around the found MS tracks and ID tracks are reconstructed within these RoIs.
The same combination procedure described for the RoI-based method is used to construct combined FS muons.
Given the high CPU demand of the FS procedure, it is only executed in multi-object triggers with at least one of the trigger objects found by an RoI-based algorithm.

\section{Muon trigger menu}
\label{sec:menu}
 
The trigger system is configured via the trigger menu, which defines the set of trigger selection criteria used for data-taking or simulation.
A sequence of reconstruction and selection steps for specific muon objects in the trigger system is specified by a trigger chain which is often referred to simply as a trigger.
Due to changing LHC conditions and continuous improvements in the trigger algorithms, the menu is subject to frequent changes.
Details of the menu configuration for each year of data-taking can be found in Refs.~\cite{ATL-DAQ-PUB-2016-001, ATL-DAQ-PUB-2017-001, ATL-DAQ-PUB-2018-002, ATL-DAQ-PUB-2019-001}.
In the following, the most common selections used by the majority of physics analyses are summarised.
 
In 2015 and 2016, the six programmable \pT thresholds of the L1 trigger were set as L1\_MU4, L1\_MU6, L1\_MU10, L1\_MU11, L1\_MU15 and L1\_MU20.
The numbers in the L1 trigger names correspond to the \pT threshold (in \GeV{}), except for L1\_MU11, which applies a 10~\GeV{} threshold, but contrary to L1\_MU10 a 3-station coincidence is required for the RPCs.
After the deployment of the additional RPC chambers in the feet region, L1\_MU15 was replaced by an L1\_MU20 version disabling those chambers to be used as backup in case of rate increases at high instantaneous luminosities.
The sequence of the single-muon trigger chains which ran without any prescale in 2016--2018 are shown in Table~\ref{tab:menu}.
In 2015, lower thresholds were supported.
Details of the trigger menu in 2015 are documented in Ref.~\cite{ATL-DAQ-PUB-2016-001}.
 
The single-muon trigger chain HLT\_mu26\_ivarmedium is designed to select isolated muons with $\pT > 26$~\GeV{} by requiring a relatively loose isolation criterion in order to control trigger rates.
The isolation criterion is chosen such that the efficiency for well-isolated muons from \PZ boson decays is very close to 100\%, while about half of the muons from heavy-flavour decays are rejected.
The HLT\_mu50 trigger is designed to collect muons with large \pT without efficiency loss due to any isolation requirement.
The HLT\_mu60\_0eta105\_msonly trigger decision is based only on MS reconstruction, is active only in the barrel region, and is inactive in the endcap regions due to the high rate.
The HLT\_mu26\_ivarmedium, HLT\_mu50, and HLT\_mu60\_0eta105\_msonly chains are called \textit{primary} single-muon triggers.
The single-muon triggers cover the needs of many physics analyses.
Physics analyses which benefit from muon triggers with lower \pT thresholds use either dimuon triggers or muon triggers in combination with other triggers.
The multi-muon triggers are made either by requiring multiple muon candidates, each of which independently fires a single-muon trigger, or by finding multiple muons using the FS strategy after the leading muon candidate has been confirmed by a single-muon trigger.
The sequence of the main multi-muon trigger chains which were used unprescaled in 2015--2018 are also summarised in Table~\ref{tab:menu}.
The HLT\_2mu14 chain requires two or more muon candidates, each of which passes a single-muon trigger HLT\_mu14 chain.
\begin{table}[htbp]
\centering
\caption{Sequence for the muon trigger chains at L1 and HLT for \pp collision data.
The \pT and isolation cuts applied at each step of the chain are also shown.
The isolation requirement was updated in 2018 by tightening the $\Delta z$ selection on additional tracks to reduce inefficiencies observed in high-pile-up conditions during data-taking in 2017.
The threshold on the tri-muon trigger was increased from 4~\GeV{} to 6~\GeV{} in 2018.}
\resizebox{\columnwidth}{!}{
\begin{tabular}{lll}
\toprule
Trigger chain & Level-1 & HLT\\
\midrule
Single-muon triggers & & \\
\midrule
HLT\_mu26\_ivarmedium     & L1\_MU20  & $\geq$ 1 CB muon with \pT > 26 \GeV{} and $\Sigma_{\Delta z<6\,(2)\,\mathrm{mm}} \pT^\mathrm{trk} / \pT $ < 0.07\\
HLT\_mu50                  & L1\_MU20  & $\geq$ 1 CB muon with \pT > 50 \GeV{} \\
HLT\_mu60\_0eta105\_msonly & L1\_MU20  & $\geq$ 1 SA muon with \pT > 60 \GeV{} in $|\eta|$ < 1.05\\
\midrule
Multi-muon triggers & & \\
\midrule
HLT\_2mu14                 & L1\_2MU10 & $\geq$ 2 CB muons with \pT > 14 \GeV{}\\
HLT\_mu22\_mu8noL1         & L1\_MU20  & $\geq$ 1 CB muon with \pT > 22 \GeV{} (mu22 trigger)\\
&           & and $\geq$ 2 FS muons with \pT > 22 and > 8 \GeV{}\\
HLT\_3mu4(6)                  & L1\_3MU4(6)  & $\geq$ 3 CB muons with \pT > 4 (6) \GeV{}\\
HLT\_3mu6\_msonly          & L1\_3MU6  & $\geq$ 3 SA muons with \pT > 6 \GeV{}\\
HLT\_mu20\_2mu4noL1        & L1\_MU20  & $\geq$ 1 CB muon with \pT > 20 \GeV{} (mu20 trigger)\\
&           & and $\geq$ 1 FS muons with \pT > 20 \GeV{} and  $\geq$ 2 FS muons with \pT > 4 \GeV{}\\
\bottomrule
\end{tabular}
}
\label{tab:menu}
\end{table}
The HLT\_mu22\_mu8noL1 chain requires at least one muon candidate which passes a single-muon trigger HLT\_mu22, and subsequently employs the FS algorithm to find two or more muon candidates with $\pT > 22$ and 8~\GeV{} for leading and subleading muons.
The choice of a leading \pT cut of 22~\GeV{} is driven by computing resource limitations when invoking the FS muon finding at the HLT level.
 
As mentioned in Section~\ref{sec:data_selection}, the LHC also provides HI collisions and \pp collisions with low pile-up.
To achieve the optimal data-taking efficiency, the trigger menu is adapted accordingly.
 
For the low-pile-up \pp collisions at a centre-of-mass energy $\sqrt{s} = 5.02~\TeV{}$, the thresholds of the muon triggers were relaxed due to the lower rates.
The lowest unprescaled single-muon trigger was operated at a \pT threshold of 14~\GeV{}.
In the 2016 \pPb run, only five programmable \pT thresholds of the L1 trigger were used.
These are the same as for \pp collision data-taking, except for L1\_MU11.
The primary trigger chain HLT\_mu15\_L1MU6 was designed to select muons with $\pT > 15$~\GeV{} and was used in the analysis to select muons originating from \PZ or \PW boson decays.
For the \PbPb data-taking, the lowest unprescaled single-muon trigger was HLT\_mu8.
The primary dimuon trigger chain in the menu was HLT\_2mu4.
Table~\ref{tab:menu_HI} summarises the sequences of the muon trigger chains that ran without any prescale during the low-pile-up \pp and HI data-taking.
\begin{table}[htbp]
\centering
\caption{Sequence for the muon trigger chains at L1 and HLT for HI and low-pile-up \pp collision data.
The \pT cut applied at each step of the chain is also shown.}
\begin{tabular}{llll}
\toprule
Trigger chain & Level-1 & HLT & Data-taking campaign\\
\midrule
Single-muon triggers & & & \\
\midrule
HLT\_mu15\_L1MU6  & L1\_MU6  & $\geq$ 1 CB muon with \pT > 15 \GeV{} & \pPb\\
HLT\_mu14 & L1\_MU10  & $\geq$ 1 CB muon with \pT > 14 \GeV{} & low-pile-up \pp\\
HLT\_mu8 & L1\_MU6  & $\geq$ 1 CB muon with \pT > 8 \GeV{} & \PbPb\\
\midrule
HLT\_2mu4         & L1\_2MU4 & $\geq$ 2 CB muons with \pT > 4 \GeV{} & \pPb; low-pile-up \pp \\
HLT\_2mu3         & L1\_2MU4 & $\geq$ 2 CB muons with \pT > 3 \GeV{} &  \PbPb \\
\bottomrule
\end{tabular}
\label{tab:menu_HI}
\end{table}

\section{CPU timing studies}
 
The total computation time used by triggers is important to monitor, as available resources for running online are limited.
The times for the fast reconstruction and precision algorithms are evaluated by rerunning a representative trigger configuration on a single run using an environment similar to the online HLT computing farm set-up and are shown in Figure~\ref{fig:timing}.
In general, the fast and precision reconstruction algorithms described in Section~\ref{sec:MuonTrigger} take up about 30\% of the total trigger processing time, while the fast algorithms take a very small fraction of the time.
The latter did not undergo any particular CPU optimisation, but show a strong menu dependence which caused changes during the data-taking campaign.
During \RunTwo, several improvements in the software chain were implemented to reduce the needed processing time.
Notably, the CPU needs are significantly reduced in 2016 relative to 2015 by introducing a caching algorithm deployed towards the end of 2015.
Furthermore, the design of the most time-consuming chains were revisited.
In 2016, an additional speed-up was achieved by deploying the Eigen~\cite{eigenweb} library to perform the combined fit, which is more efficient than the previously used MA27~\cite{Duff:1982:MSF} implementation.
Before data-taking in 2017 started, the call sequence of the algorithms was optimised to avoid running algorithms multiple times in a single RoI.
After the changes with significant impact on the average processing time, the developments during 2017 and 2018 focused on the tail of the time distribution, addressing special event topologies requiring a long processing time.
Due to several reconfigurations of the reconstruction algorithms, such as disabling precision steps not required by the trigger demands, these contributions were strongly reduced.
\begin{figure}[htbp]
\centering
\includegraphics[width=0.49\textwidth]{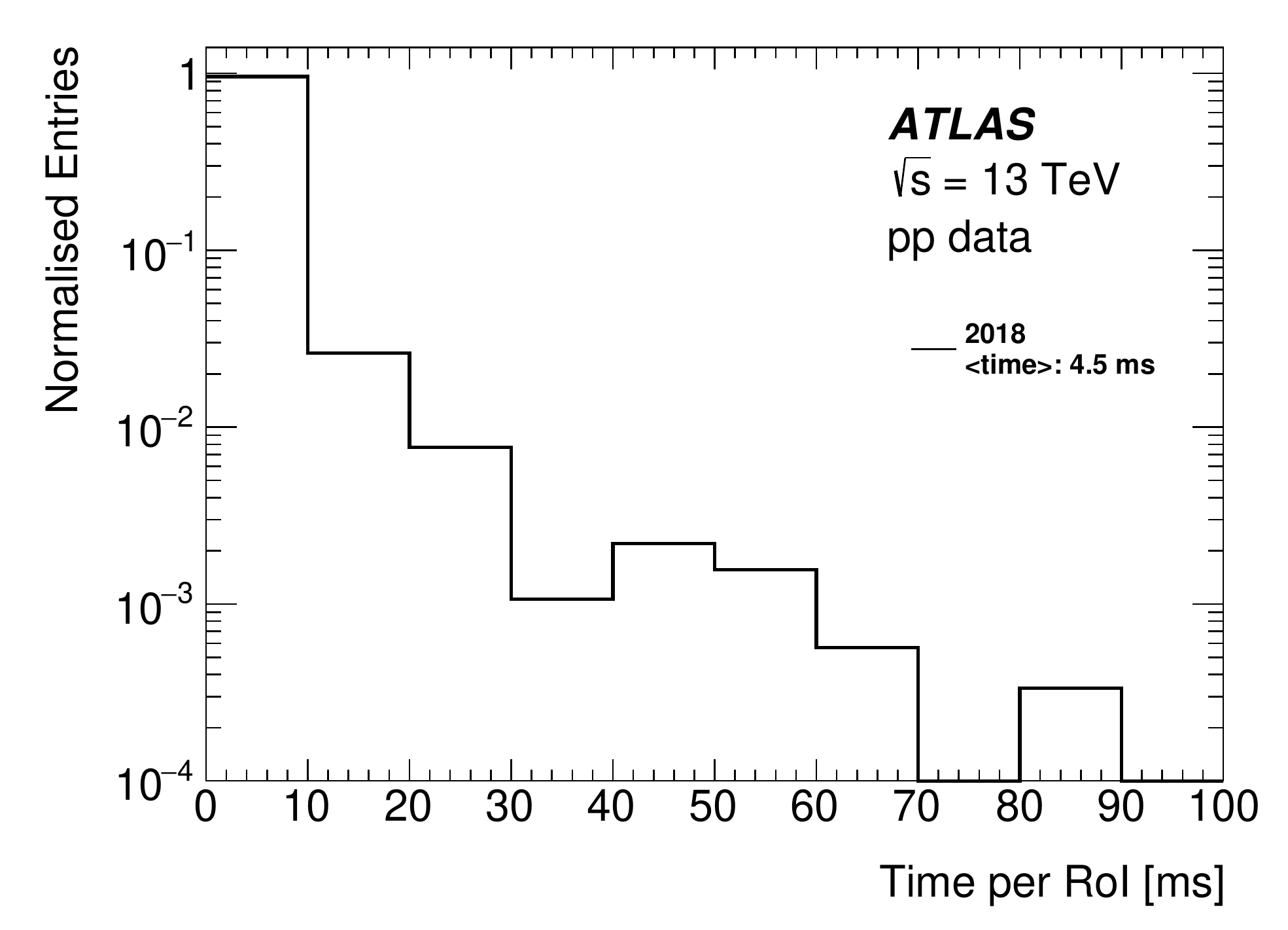}
\includegraphics[width=0.49\textwidth]{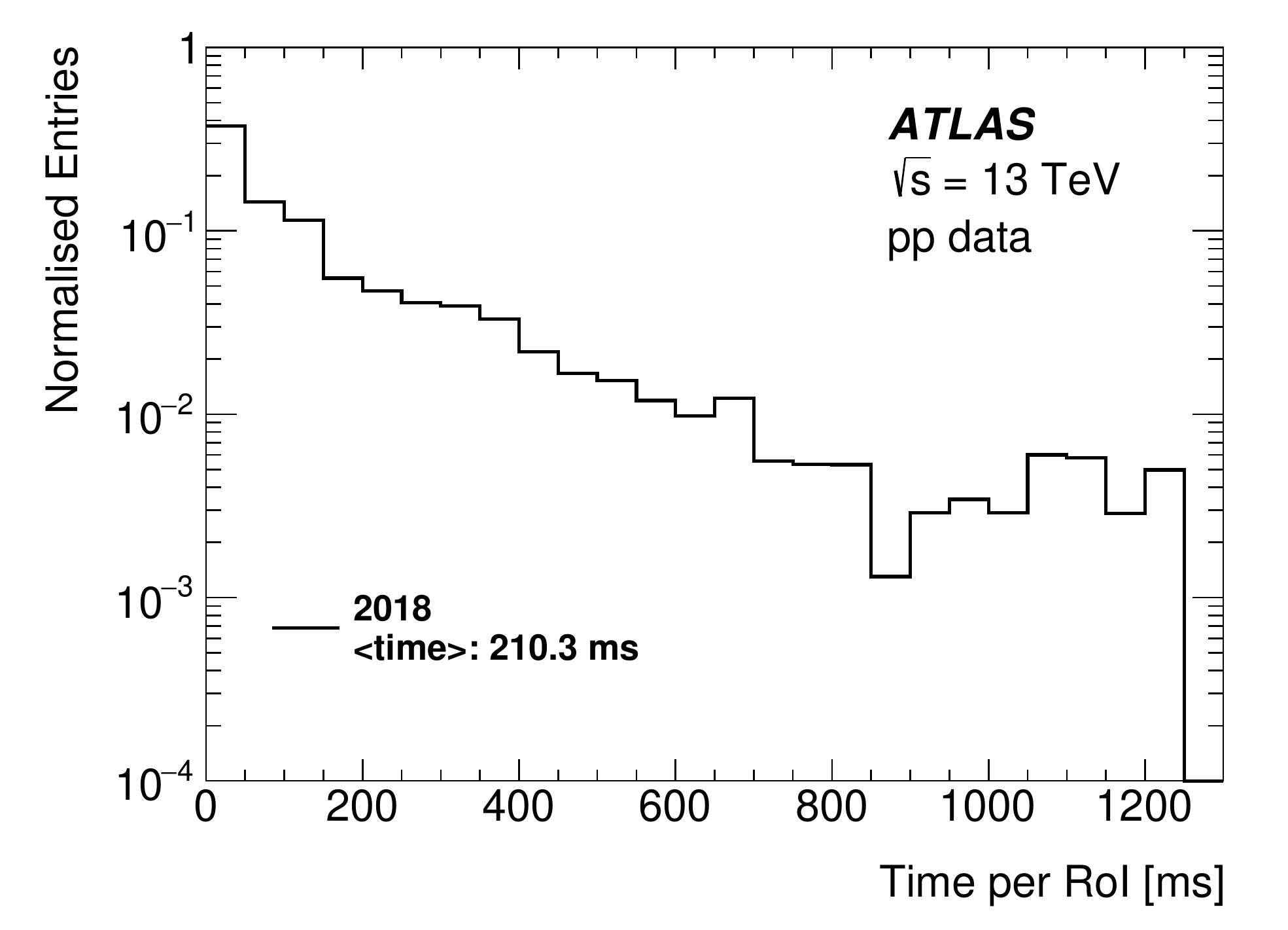}
\caption{HLT processing time per RoI for the fast (left) and precision (right) algorithms for a representative configuration of the muon trigger chains at the end of \RunTwo.}
\label{fig:timing}
\end{figure}

\section{Trigger performance measurements}
\label{sec:perf}
 
\subsection{Resolution studies}
 
The tag-and-probe method applied to \Zmm{} events, described below, was used to evaluate the quality of the \pT, $\eta$, and $\phi$ determination for muon tracks at the HLT compared with the offline reconstruction.
Online algorithms are almost identical to offline reconstruction algorithms, but several simplifications are required in order to cope with the limitations of the trigger system.
Additionally, offline reconstruction can feature updated calibration and alignment corrections not available at the time the data were taken.
Therefore, finite differences are expected in the resolution.
The offline relative momentum resolution is measured to be 2.3\% (2.9\%) in the barrel (endcap) region using \Zmm{} events~\cite{PERF-2015-10}.
The residual of the trigger-reconstructed \pT is defined relative to the offline reconstructed value as $\delta_{\pT} = (1/{\pT^{\mathrm{trigger}}} - 1/{\pT^{\mathrm{offline}}})/(1/{\pT^{\mathrm{offline}}})$, where $\pT^{\mathrm{trigger}}$ and $\pT^{\mathrm{offline}}$ are the transverse momenta determined by the HLT and offline algorithm, respectively.
The resolution of the trigger reconstruction with respect to the offline reconstruction is defined as the standard deviation of a Gaussian function fitted to the $\delta_{\pT}$ distribution.
Figure~\ref{fig:resolution:pt} shows the \pT resolution as a function of the offline muon \pT for CB and SA HLT muons in the barrel and endcap regions.
The \pT resolution is about 1\% (2\%) and 3\% (5\%) in the barrel (endcap) region for CB and SA muons, respectively.
 
\begin{figure}[htbp]
\centering
\includegraphics[width= 0.49\textwidth]{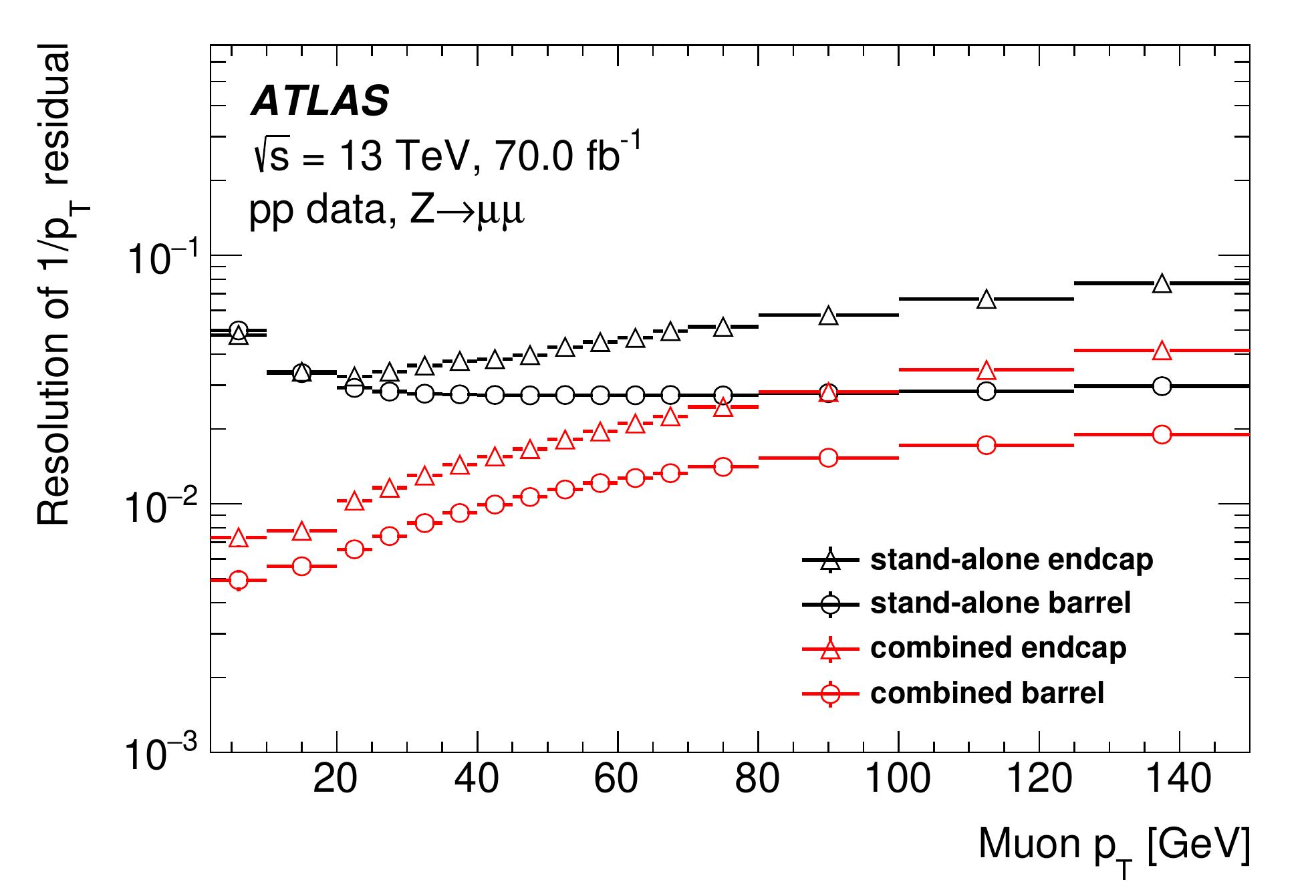}
\caption{Resolution with respect to offline reconstruction of the inverse-\pT residual as a function of offline muon \pT for CB (red) and SA (black) muons in the barrel (circles) and endcap (triangles) regions.}
\label{fig:resolution:pt}
\end{figure}
 
The resolution with respect to offline reconstruction in the $\eta$ and $\phi$ determination were examined accordingly by defining the residual as the absolute value of the difference between the trigger and offline reconstructed quantities.
Figure~\ref{fig:resolution:angular} summarises the resolution of the $\eta$ and $\phi$ residuals as a function of the offline muon~\pT.
The angular residual resolutions are typically below $10^{-4}$ and $10^{-2}$ for CB and SA muons, respectively.
 
\begin{figure}[htbp]
\centering
\includegraphics[width= 0.49\textwidth]{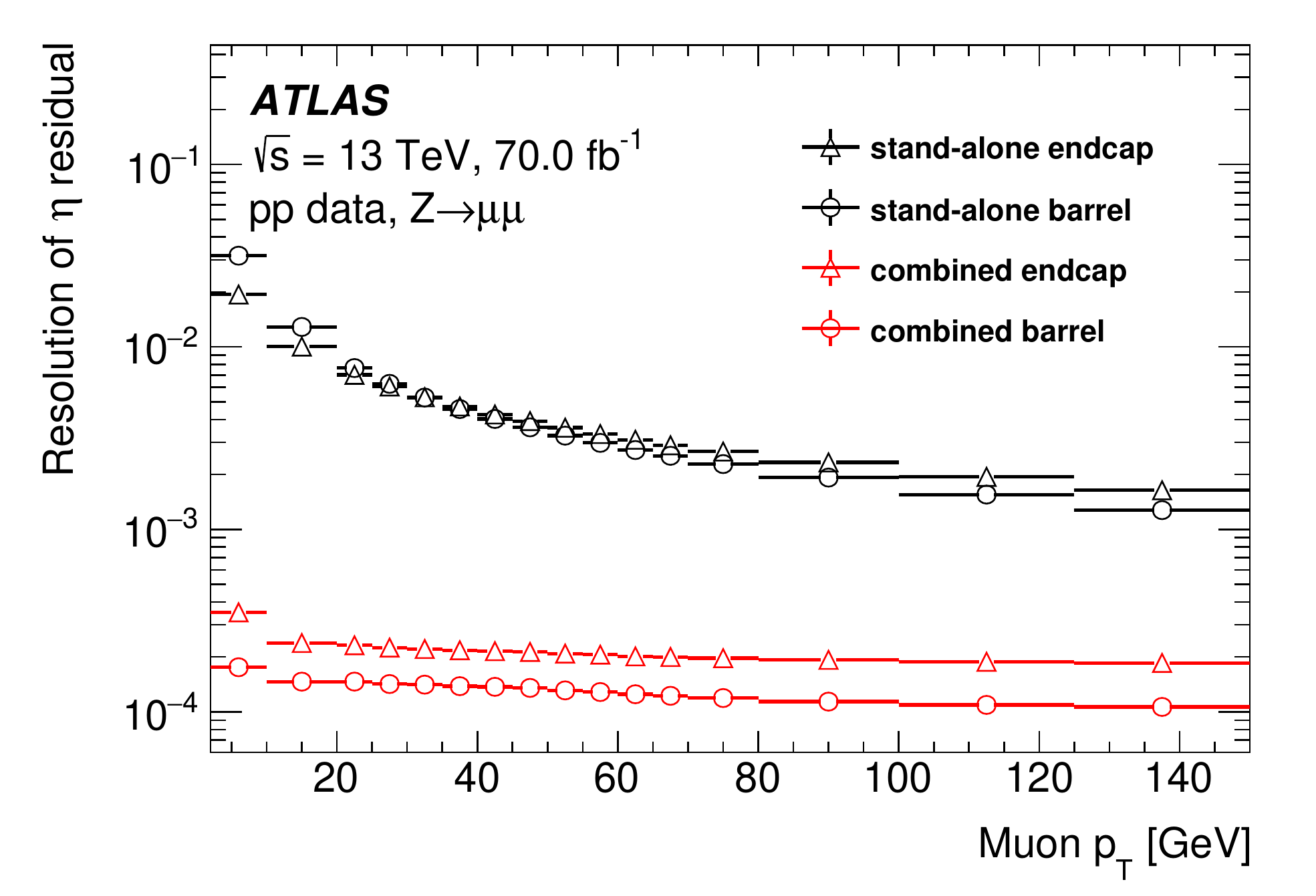}
\includegraphics[width= 0.49\textwidth]{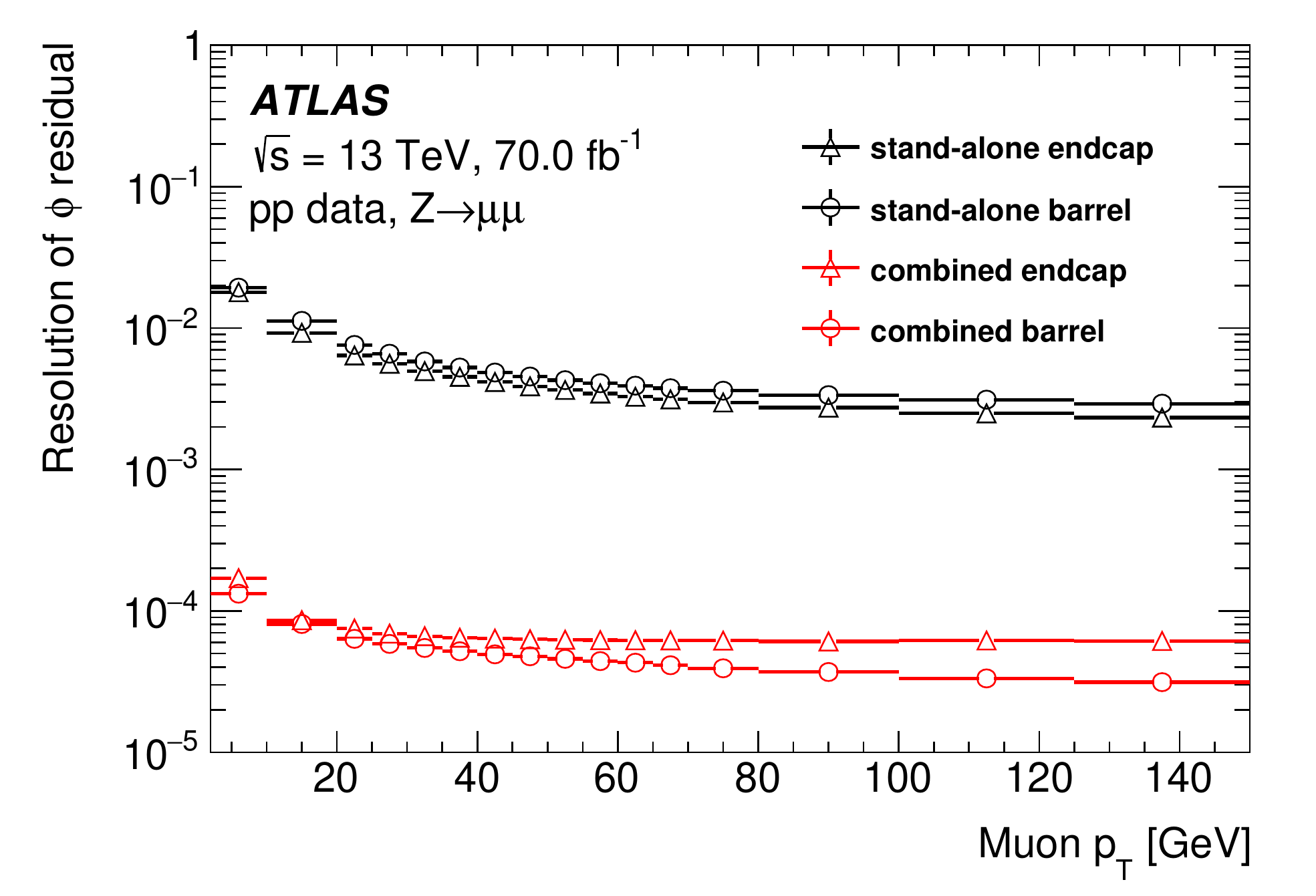}
\caption{Resolution with respect to offline reconstruction in the pseudorapidity (left) and azimuthal angle (right) determination as a function of offline muon \pT for the combined (red) and stand-alone (black) muons in the barrel (circles) and endcap (triangles) regions.}
\label{fig:resolution:angular}
\end{figure}
 
\subsection{Rate measurements}
The ATLAS data-taking conditions are archived with a time interval of about one minute, which defines a luminosity block.
In order to obtain the rate of a given trigger as a function of the instantaneous luminosity~\cite{LUCID2}, individual rate measurements on different luminosity blocks from all data collected in a given year are used.
Any rate measurement for which the ratio of the trigger rate to instantaneous luminosity varies by more than 20\% from the average is discarded as an estimator of the rate for that trigger.
This avoids averaging rate measurements that fluctuate because of unpredictable and temporary changes of LHC conditions.
 
\subsection{Trigger efficiency measurement}
\label{sec:eff}
The muon trigger efficiency, $\epsilon$, is measured with respect to muons reconstructed offline using the data set collected during 2015--2018, and considering only periods of data-taking with all sub-detectors functioning nominally.
The measurement is based on a \tap method which selects muons from known decays such as \Zmm.
An unbiased sample of probe muons is selected by a stringent selection of a second tag muon.
The trigger efficiency is defined as the ratio of the number of probe muons matched to at least one trigger object, $N_{\text{match}}$, to the total number of probe muons, $N_{\text{probe}}$,
\begin{equation}
\epsilon = \frac{N_{\text{match}}}{N_{\text{probe}}} \, .\notag
\end{equation}
 
If both muons satisfy the selection criteria of the tag muon, the probe muons are considered as tag candidates in turn.
The efficiency is measured with respect to a specific offline selection of reconstructed muons and may vary between different topologies.
The efficiency measurements are performed for a range of different topologies; \Jpsi and $Z$ boson candidates decaying into a dimuon system are used to measure the trigger efficiencies of low-\pT and moderate-\pT muons, respectively.
For high-\pT muons, two topologies targeting semileptonic \ttbar candidates and \Wjets events are employed.
Finally, a dedicated measurement is performed for HI collisions, again targeting \Jpsi and $Z$ decays into dimuon systems.
Except where stated otherwise, the effects of background are negligible and thus no background subtraction is applied.
 
In each analysis, the event quality is checked to remove events which contain noise bursts or coherent noise in the calorimeters.
Further, at least one reconstructed \pp interaction vertex is required for each event, with the primary vertex defined as the vertex with the highest sum of track $\pT^2$.
 
Trigger efficiencies for low-\pT muons (\pT $\lesssim$ 10~\GeV{}) are evaluated using the \tap method with \Jmumu events.
Due to varying and non-negligible contributions of the background under the \Jpsi signal mass peak across the muon kinematic range, a fit to the dimuon mass distribution is used to extract the \Jpsi yields.
Pairs of oppositely charged muons successfully fitted to a common vertex within the invariant mass range $2.7~\GeV{} < m_{\mu\mu} < 3.5~\GeV{}$ are selected as offline \Jmumu candidates.
The quality of the vertex fit must satisfy $\chi^2 < 20$ for one degree of freedom.
 
Events for which a trigger muon candidate and an HLT ID track have an invariant mass compatible with the \Jpsi mass are accepted; various trigger muon \pT thresholds are used.
If one of the two muons satisfies the \textit{Tight} quality requirement~\cite{PERF-2015-10} and has a \pT above the tag trigger threshold, it is considered as a tag candidate.
The other muon is then taken as a probe candidate.
The tag candidate is required to be within a cone of size $\Delta R = 0.01$ around an HLT muon object that fired the tag trigger.
A probe muon is identified as triggered if it lies within a cone of size $\Delta R = 0.01$ around an object that caused the trigger to fire.
In order to avoid correlation effects between the tag muons and the probe muons, events with $\Delta R(\mu^+,\mu^-) < 0.2$ are rejected from the study.
Additionally, both tag and probe candidates with  \pT $\lesssim$ 10~\GeV{} must satisfy a \pT-dependent matching to the L1 RoI following the relation $\Delta R(\text{L1 RoI},\mu) < -0.01 * \pT^{\mu} + 0.18$, while tag muons with  \pT > 10~\GeV{} are required to satisfy $\Delta R(\text{L1 RoI},\mu) < 0.08$.
To ensure the tag and probe HLT triggers have different L1 RoIs, the angular distance between the two offline muons is required to be larger than the sum of the angular distances between the L1 RoI and the offline muon that correspond to the tag muon and the probe muon: $\Delta R(\mu^+,\mu^-) > \Delta R(\text{L1 RoI},\mu_\text{tag}) + \Delta R(\text{L1 RoI},\mu_\text{probe})$.
To extract the yield of \Jpsi candidates for which the probe muon is triggered or not triggered, an extended unbinned maximum-likelihood fit to the dimuon invariant mass spectrum is performed.
In the fit, the background probability density function is described with an exponential function, while the signal is modelled with a sum of a Gaussian function and a Crystal Ball function with the parameters of the latter fixed to the values obtained from a large sample of simulated events.
A simultaneous fit is performed on the candidates with the probe muon either triggered or not triggered.
Common parameters of the signal and background shapes are used in the simultaneous fit of triggered and not triggered muons, while the signal and background yields are left to float independently.
 
A very clean sample of $\Zmm$ events is selected to measure the trigger efficiency for muons in the \pT range from $\sim$10 to 100~\GeV{}.
A pair of oppositely charged muons consistent with the \PZ boson mass, $|\mZ - m_{\mu\mu}| < 10~\GeV{}$, with $m_Z=91.2~\GeV{}$, is selected.
The two muons are required to originate from the same interaction vertex by imposing impact parameter requirements.
If one of the two muons has $\pT > 28~\GeV{}$ and satisfies the \textit{Loose} isolation WP requirements~\cite{PERF-2015-10}, it is considered as a tag muon.
The other muon is then taken as a probe.
The tag muon must further have an angular distance of $\Delta R < 0.1$ from an object that has fired the HLT\_mu26\_ivarmedium trigger.\footnote{In 2015, HLT\_mu20\_iloose\_L1MU15 was used as the lowest unprescaled single-muon trigger.
See Section~\ref{sec:menu} for further details.}
The probe muon must satisfy several impact parameter requirements and is subject to selections applied in the offline identification corresponding to the identification WP of interest.
A probe muon is identified as triggered if it lies a distance $\Delta R < 0.1$ from an object that caused the trigger to fire.
The trigger efficiency is measured with respect to several offline selection criteria as defined in the identification WPs.
 
In order to evaluate trigger efficiencies for high-\pT muons, namely muons with \pT$\gtrsim$ 100~\GeV{}, the \tap method is used with \ttbar and \Wjets event topologies.
The tag trigger used in this study is a trigger based on selecting events with large missing transverse momentum, \met.
In addition to this, events need to contain exactly one muon with $\pT > 27~\GeV{}$ and no isolation requirement.
Events are rejected if they contain an electron, and are required to have $\met > 200~\GeV{}$.
For \Wjets events, between one and four jets with $\pT > 25~\GeV{}$ are required, and events are rejected if they contain any \bjets.
At least four jets with $\pT > 25~\GeV{}$ are required for \ttbar events, and at least one among them has to be a \bjet defined using the MV2c10 77\% WP.
The \bjet requirement ensures the orthogonality of the \Wjets and \ttbar measurements.
 
In the HI analysis a similar selection is applied using \Jmumu (\Zmm) events to evaluate low-\pT (high-\pT) muon trigger performance, but adapting for the different conditions.
A pair of oppositely charged muons with an invariant mass $2.7 < m_{\mu\mu} < 3.5~\GeV{}$ ($81 < m_{\mu\mu} < 101~\GeV{}$) is required for \Jpsi (\PZ) events.
In addition, both muons must be within the pseudorapidity range $|\eta|<2.5$.
In the \Jmumu analysis, the same matching criteria as mentioned above are applied to the tag and probe muons.
Different requirements are applied for the measurement using \Zmm events in the \pPb (\PbPb) analysis:
both muons must be matched to the appropriate trigger object within $\Delta R < 0.01$ ($\Delta R < 0.1$),
the tag muon must have $\pT > 17~\GeV{}$ ($\pT > 24~\GeV{}$), and the probe muon must have  $\pT > 4~\GeV{}$ ($\pT > 8~\GeV{}$).
 
In order to improve the accuracy of the modelling of data, ATLAS physics analyses make use of the ratio, called the scale factor (SF), of measured and simulated efficiencies to correct simulated samples.
Effects due to the choice of event selection for the efficiency measurement are assessed by varying the requirements in the selection of both the tag and probe muon candidates and are quantified as systematic uncertainties.
Several sources of systematic uncertainty are considered, which, for the \Zmm analysis, are:
\begin{itemize}
\item Pile-up dependence: \\
Estimated by splitting the data set into a low- and high-pile-up sample at the approximate peak in the number of reconstructed vertices for each year, and computing the trigger efficiency for events above and below this cut.
The cut was set at 11 reconstructed vertices for 2015 and 2016, and at 19 reconstructed vertices for 2017 and 2018 except for data-taking periods in 2017 with exceptionally high pile-up, for which the cut was set at 25 reconstructed vertices.
 
\item Correlation between the tag and probe muons: \\
Muon pairs from the $Z\rightarrow \mu\mu$ decay tend to be back-to-back in $\phi$.
Since the barrel and endcaps have 16-fold and 12-fold symmetry respectively, this means that if the tag muon traverses a highly efficient region of the detector, the probe muon is likely to do so also.
To estimate this effect, the trigger efficiency is calculated using an extra cut to remove back-to-back muons, $\Delta\phi(\mu_{\text{tag}},\mu_{\text{probe}})< \pi-0.1$, where $\Delta\phi(\mu_{\text{tag}},\mu_{\text{probe}})$ gives the azimuthal separation of the tag and probe muons.
 
\item Background contribution:\\
Estimated by measuring the efficiency when the \PZ boson mass window is enlarged by $\pm 5~\GeV{}$.
 
\item Probe selection criteria:\\
The effects of various probe selection cuts on the trigger efficiency were investigated.
\begin{itemize}
\item Charge: Since the charge affects the behaviour of muons in the magnetic field present in the MS, it could conceivably affect the muon trigger efficiency.
This effect is evaluated by calculating the trigger efficiency using only positively or negatively charged probe muons.
\item Impact parameter: Estimated by calculating the efficiency without impact parameter requirements.
\item Isolation: Estimated by applying different isolation WP requirements similar to the ones described in Ref.~\cite{PERF-2015-10} to the probe muons and measuring the efficiency.
The isolation systematic uncertainty is taken as the largest of the resulting deviations from the nominal (no isolation) case.
\item \pT: Estimated by splitting the probe muons into two groups according to a \pT cut.
The value of the cut depends on the \pT threshold of the trigger of interest; for trigger thresholds of $26~\GeV{}$, the cut is $40~\GeV{}$, while for thresholds of $50~\GeV{}$, a cut at $70~\GeV{}$ is used.
Trigger efficiency is then calculated using only the subset of probe muons with \pT above or below that cut.
\end{itemize}
\end{itemize}
 
Due to the different topology and selection in the \Wjets and \ttbar analysis, a different set of systematic uncertainties is considered:
\begin{itemize}
\item Muon quality WP:\\
Estimated by changing the muon quality WP from the \textit{Medium} WP to the \textit{high-}\pT WP.
 
\item \met reconstruction:\\
Estimated by measuring the efficiency while changing the \met event-level selection from 200~\GeV{} to 150~\GeV{}.
 
\item Cut on jet \pT:\\
Estimated by measuring the efficiency while changing the object-level selection for jets from 25~\GeV{} to 30~\GeV{}.
 
\item Identification of \bjets:\\
Estimated by measuring the trigger efficiency while changing the object-level selection for \bjets.
The selection requirement is tightened from the MV2c10 77\% WP to the 70\% WP for the \ttbar analysis, while for the \Wjets analysis the \bjet veto uses the 85\% WP instead of the 77\% WP.
 
\item Muon isolation:\\
Estimated by applying different isolation WP requirements to the probe muon and measuring the efficiency.
The isolation systematic uncertainty is taken as the largest of the resulting deviations from the nominal (no isolation) case.
\end{itemize}
 
All contributions are assumed to be independent and are added in quadrature to obtain the total systematic uncertainties.
The impact of the systematic uncertainties on the muon momentum scale and resolution~\cite{PERF-2015-10} is found to be negligible.
For the low-\pT analysis with \Jpsi events, only statistical uncertainties are considered.
 
\section{L1 muon trigger performance}
 
The L1 muon trigger thresholds remained unchanged throughout the data-taking in \RunTwo, except for the replacement of L1\_MU15 in 2017 after the new RPC chambers in the feet region were commissioned.
The typical maximum L1 rate is 90~kHz.
Of this rate, the L1 seed of the single-isolated-muon trigger HLT\_mu26\_ivarmedium, which is L1\_MU20, is fired at about 15~kHz for an instantaneous luminosity of $1.7\times10^{34}~\mathrm{cm}^{-2}\mathrm{s}^{-1}$.
Figure~\ref{fig:rates:L1} shows the rates of the single-muon and dimuon triggers as a function of the instantaneous luminosity.
A clear linear dependence is visible, with year-to-year slope changes due to the improvements discussed above.
This indicates a negligible contribution from effects not related to \pp collisions.
\begin{figure}[htbp]
\centering
\includegraphics[width=0.49\textwidth]{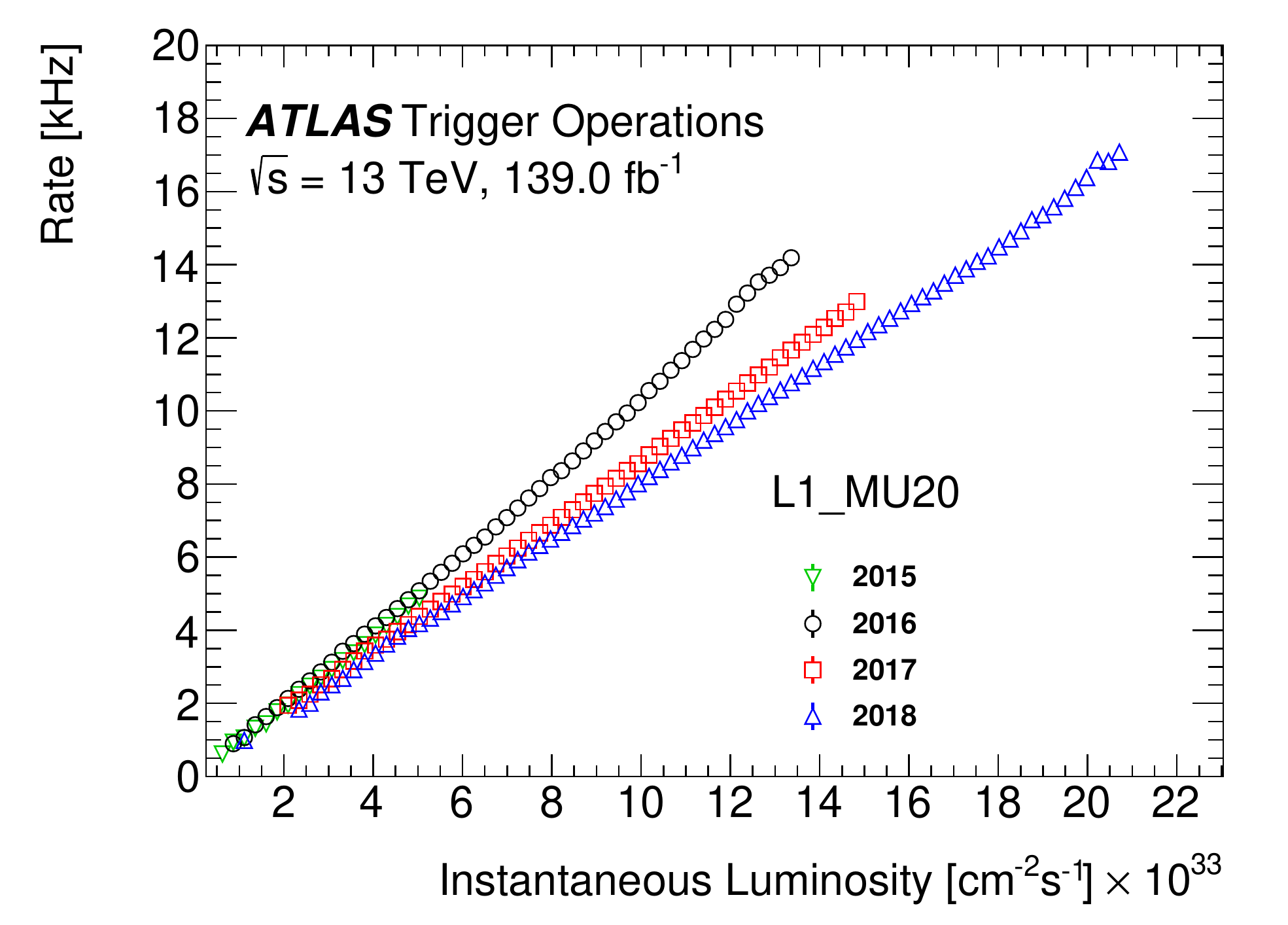}
\includegraphics[width=0.49\textwidth]{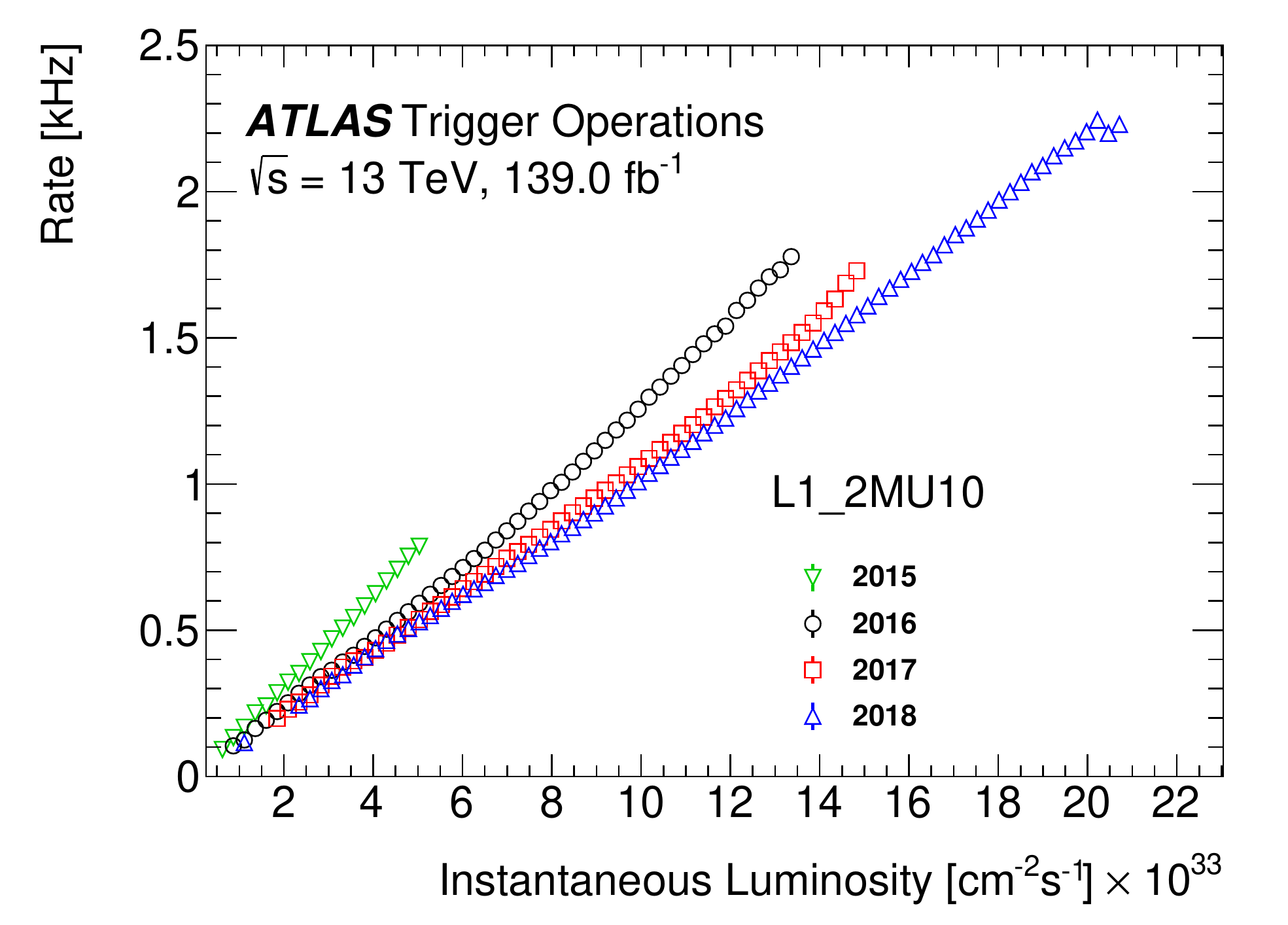}
\caption{L1 trigger rates as a function of the instantaneous luminosity for the lowest unprescaled single-muon and dimuon triggers, L1\_MU20 (left) and L1\_2MU10 (right), in different years in Run 2.}
\label{fig:rates:L1}
\end{figure}
 
The efficiencies for single-muon L1 triggers with different \pT thresholds for each year of data-taking during \RunTwo are measured using the \Zmm \tap method described above.
Figures~\ref{fig:eff:l1:mu4} and~\ref{fig:eff:l1:mu20} summarise the efficiencies as a function of the offline muon \pT for the lowest and highest threshold in the barrel and endcap regions, respectively.
In the barrel region the plateau efficiency is about 75\% for the low-\pT trigger and 70\% for the high-\pT trigger, the latter being lower due to the applied 2-/3-station requirement.
The steeper turn-on behaviour in the barrel region is due to the better \pT{} resolution.
In 2016, the coincidence requirements for the low-\pT triggers were tightened, leading to an improved turn-on behaviour.
A slightly increased plateau efficiency in 2016 was achieved by enabling the feet RPC chambers, while in 2017 the feet RPC thresholds had to be tightened for rate reduction, causing a slight decrease of the plateau efficiency.
The increase in the efficiency in 2016 for the high-\pT triggers is due to the inclusion of the feet RPC triggers.
Operation of the RPC chambers in 2017 and 2018 suffered from gas leaks in some chambers, which could partially be mitigated but cause variations of the efficiency.
The efficiencies in the endcap are higher as they do not suffer from the reduced geometrical coverage affecting the barrel triggers.
For L1\_MU4, the efficiency reaches 97\%, while L1\_MU20 reaches 90\%, again caused by the different requirements applied for low- and high-\pT L1 triggers.
The additional coincidence requirements deployed during \RunTwo to mitigate rate increases due to higher luminosities cause a small degradation of the plateau efficiency for the low-\pT triggers in the endcap region.
For 2017 and 2018, a clear improvement in the turn-on region is observed for high-\pT L1 triggers due to the CW optimisation discussed in Section~\ref{sec:MuonTrigger}. A similar CW optimisation for low-\pT L1 triggers was implemented in 2018 that reduced the trigger rates but had a negligible impact on the turn-on.
\begin{figure}[htbp]
\centering
\includegraphics[width=0.49\textwidth]{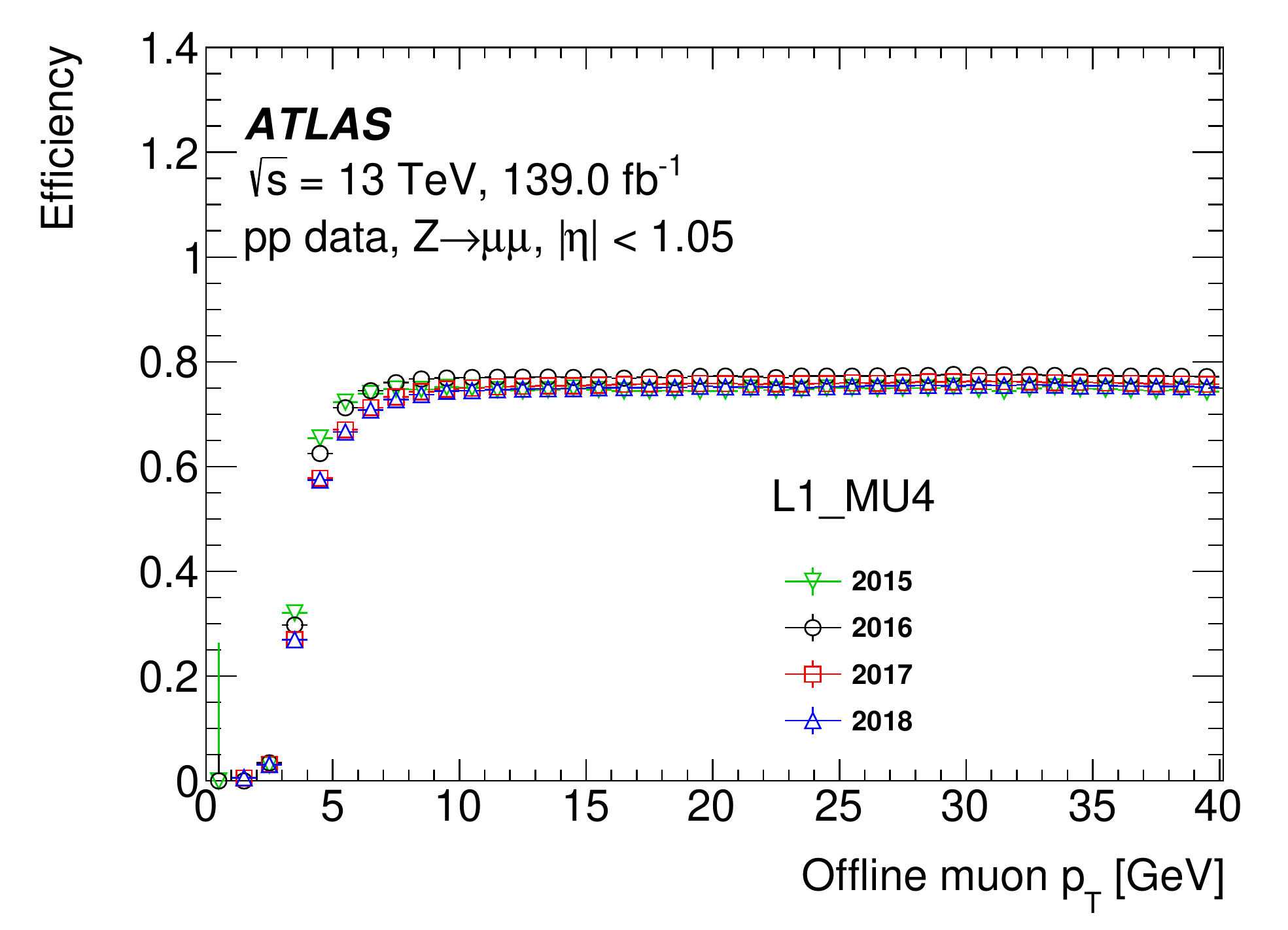}
\includegraphics[width=0.49\textwidth]{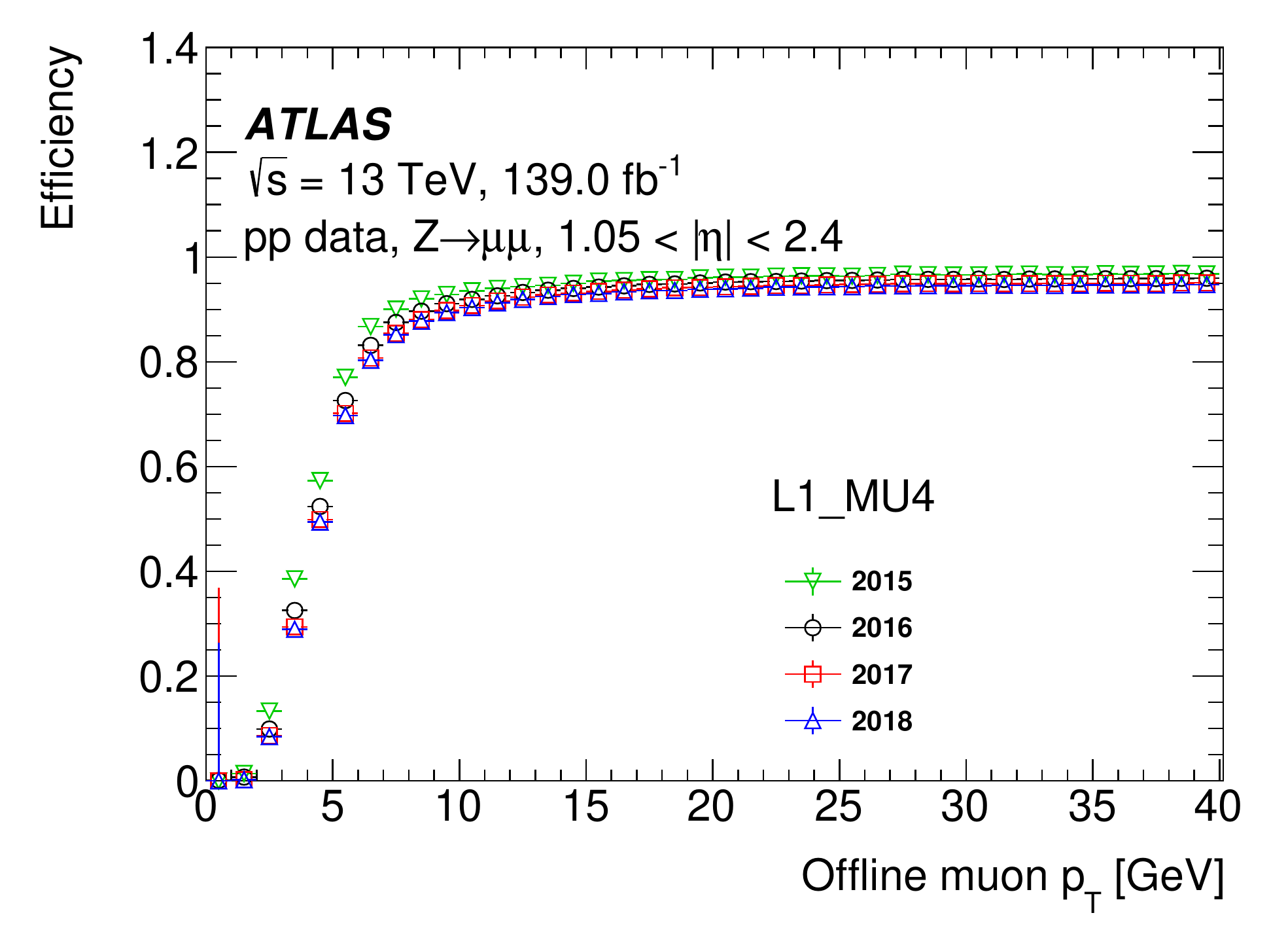}
\caption{Trigger efficiencies of L1 single-muon triggers for a threshold of 4~\GeV{} as a function of the transverse momentum of the reconstructed muon in the barrel (left) and endcap (right) region in different years in Run 2.
The efficiency is measured using the \tap method in \Zmm events. The error bars show the statistical uncertainties only.}
\label{fig:eff:l1:mu4}
\end{figure}
\begin{figure}[htbp]
\centering
\includegraphics[width=0.49\textwidth]{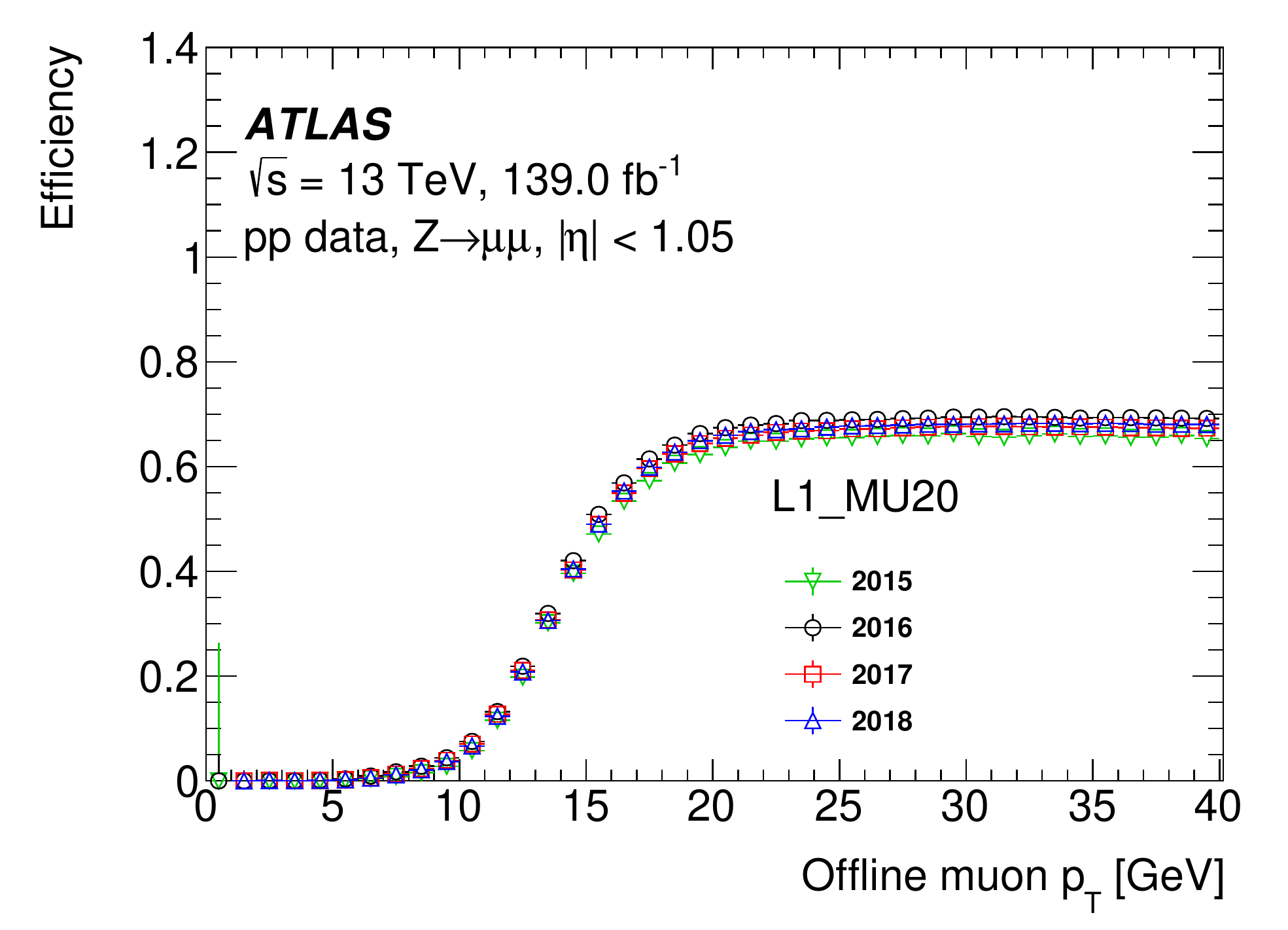}
\includegraphics[width=0.49\textwidth]{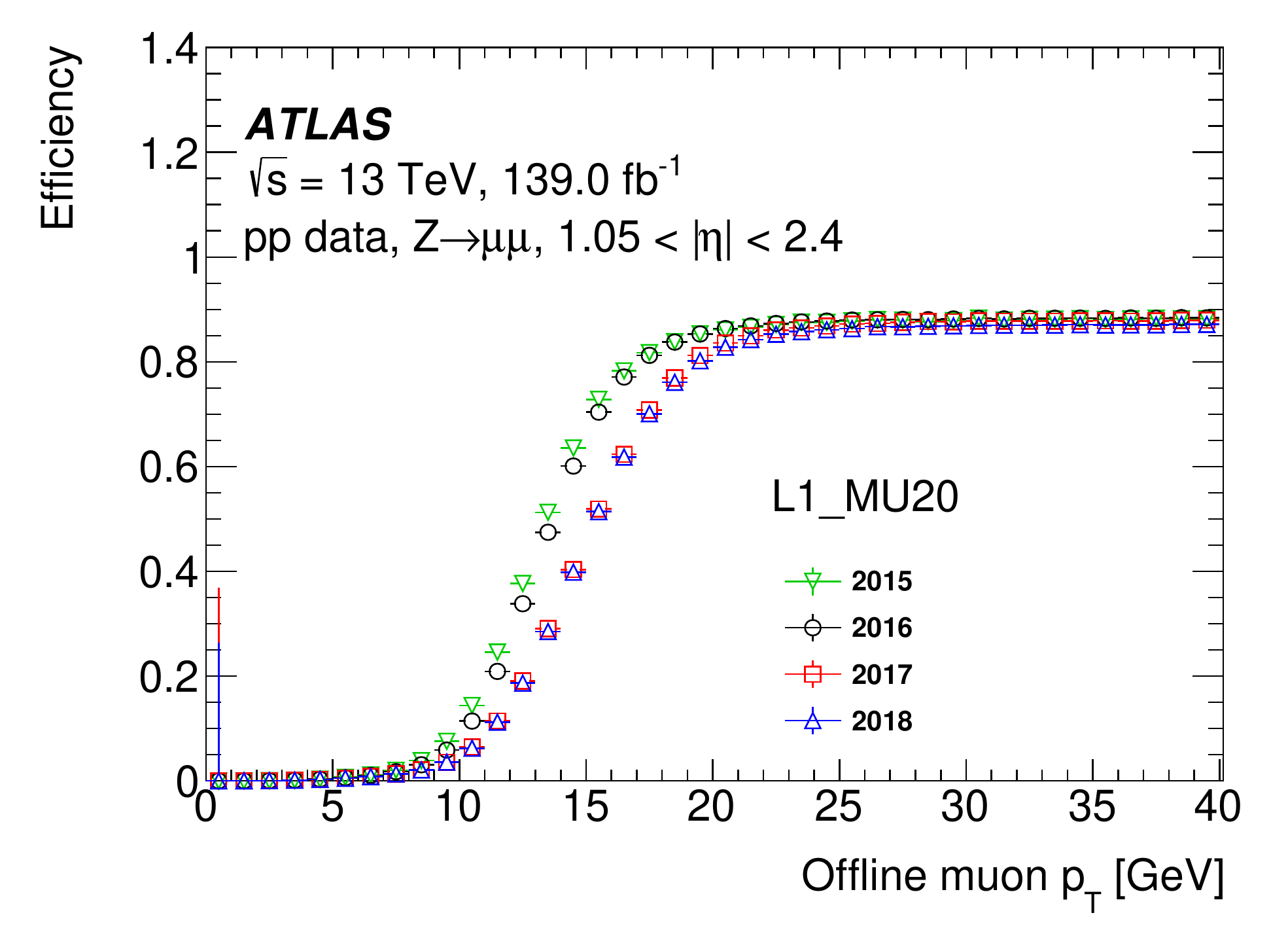}
\caption{Efficiencies of L1 single-muon triggers for a threshold of 20~\GeV{} as a function of the transverse momentum of the reconstructed muon in the barrel (left) and endcap (right) region in different years in Run 2.
The efficiency is measured using the \tap method in \Zmm events. The error bars show the statistical uncertainties only.}
\label{fig:eff:l1:mu20}
\end{figure}

\section{Muon trigger efficiency in \texorpdfstring{\pp}{pp} data-taking}
 
\subsection{Single-muon trigger efficiency}
Requiring events to pass either of the lowest unprescaled single-muon triggers, HLT\_mu26\_ivarmedium (HLT\_mu20\_iloose\_L1MU15 in 2015) or HLT\_mu50, serves as a general-purpose single-muon trigger for many physics analyses.
The rate reduction relative to L1 is about two orders of magnitude, leading to a typical rate of 180~Hz for an instantaneous luminosity of $1.7 \times 10^{34}$ cm$^{-2}$s$^{-1}$, as shown in Figure~\ref{fig:rates:HLT}.
\begin{figure}[htbp]
\centering
\includegraphics[width=0.49\textwidth]{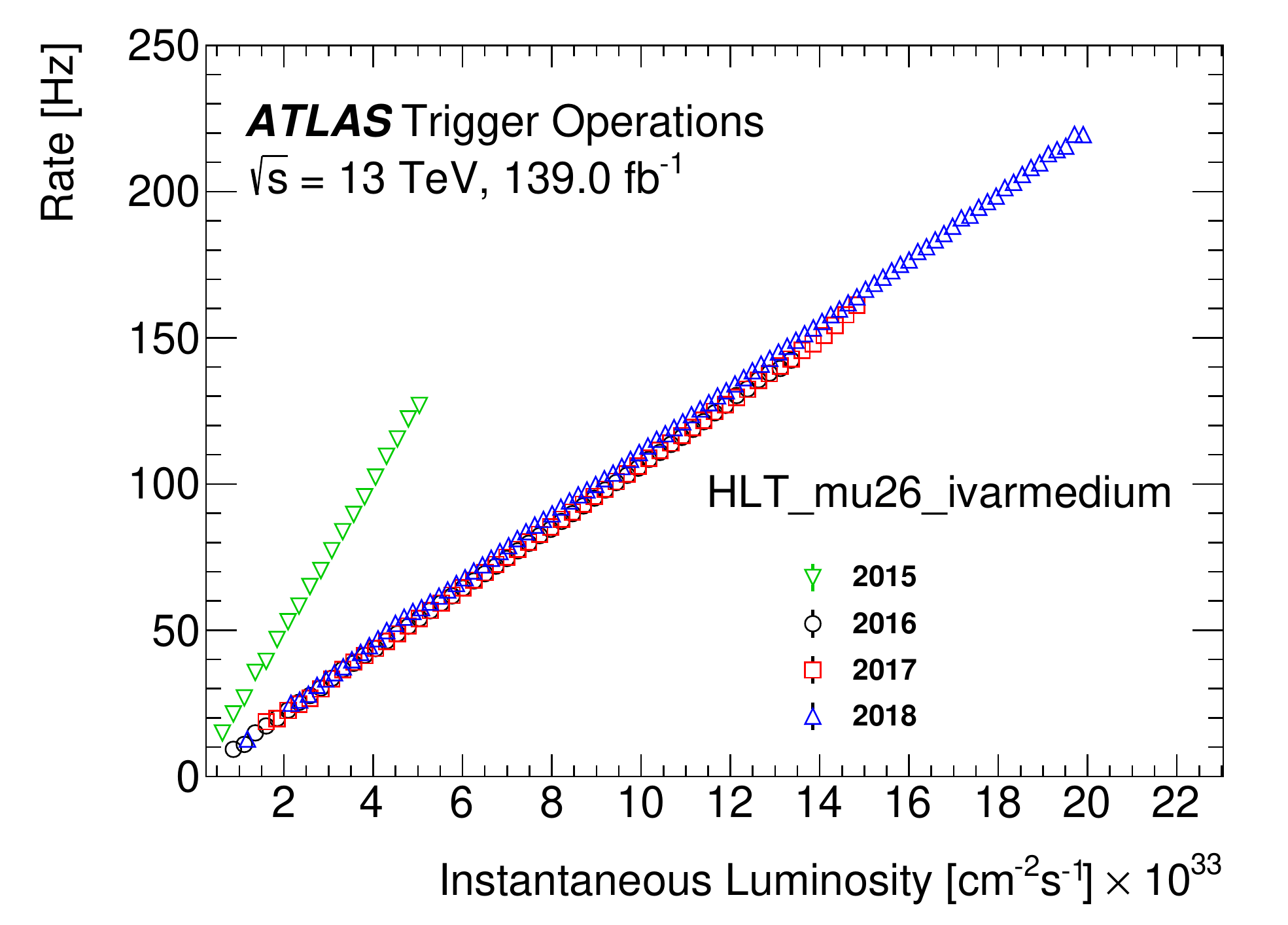}
\includegraphics[width=0.49\textwidth]{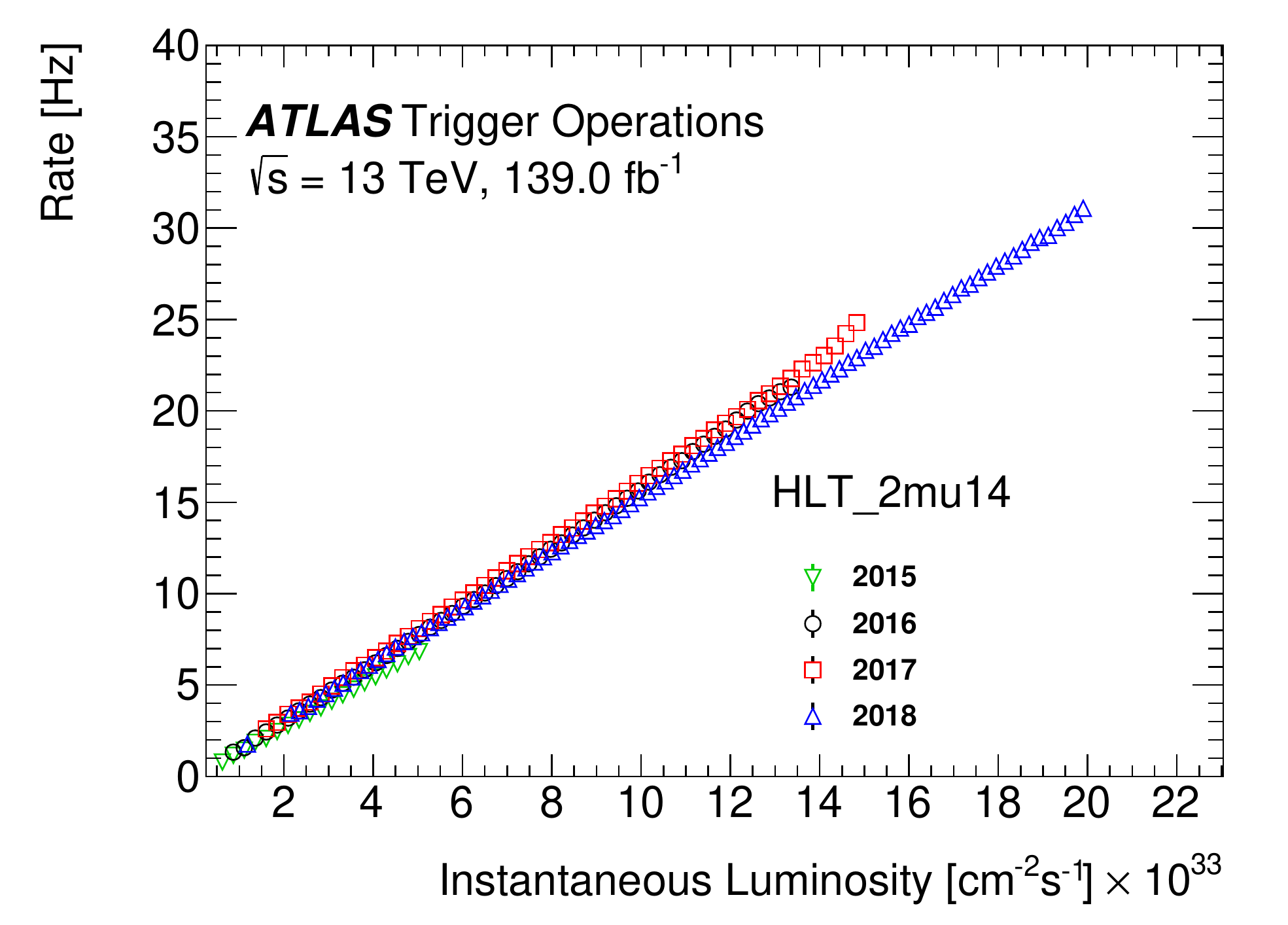}
\caption{HLT trigger rates as a function of the instantaneous luminosity for the lowest unprescaled single-muon and dimuon triggers, HLT\_mu26\_ivarmedium (left) and HLT\_2mu14 (right).
The larger rate in 2015 for the single-muon trigger is due to HLT\_mu20\_iloose\_L1MU15 being the lowest unprescaled trigger.}
\label{fig:rates:HLT}
\end{figure}
 
The efficiency of the primary single-muon triggers is evaluated using the \Zmm \tap method.
The following results are derived for \textit{Medium} quality muons.
Figure \ref{fig:eff_Zmm:HLT_mu26_ivarmedium_OR_HLT_mu50_probe_pt_ratio_2017} shows the efficiency of passing either the HLT\_mu26\_ivarmedium or the HLT\_mu50 trigger in the barrel and endcap regions as a function of the probe muon \pT.
The smaller L1 efficiency in the barrel region translates directly to the HLT efficiency, while the efficiency relative to L1 is close to unity.
The inefficiency observed in the turn-on region is due to the fact that no isolation criteria are applied to the offline selected muon.
A rise in efficiency at $\pT{} \sim 50~\GeV$ is due to the fact that no isolation criterion is applied in the HLT\_mu50 trigger.
The trigger efficiencies in simulation and data are not in complete agreement, particularly in the barrel due to a different RPC efficiency in MC simulation, which is accounted for in analyses by the SFs.
Overall, the \pT dependence of the SFs is small.
The difference in the ratio between data and MC efficiencies is due to a constant L1 efficiency used in the simulation.
\begin{figure}[htbp]
\centering
\includegraphics[width=0.48\textwidth]{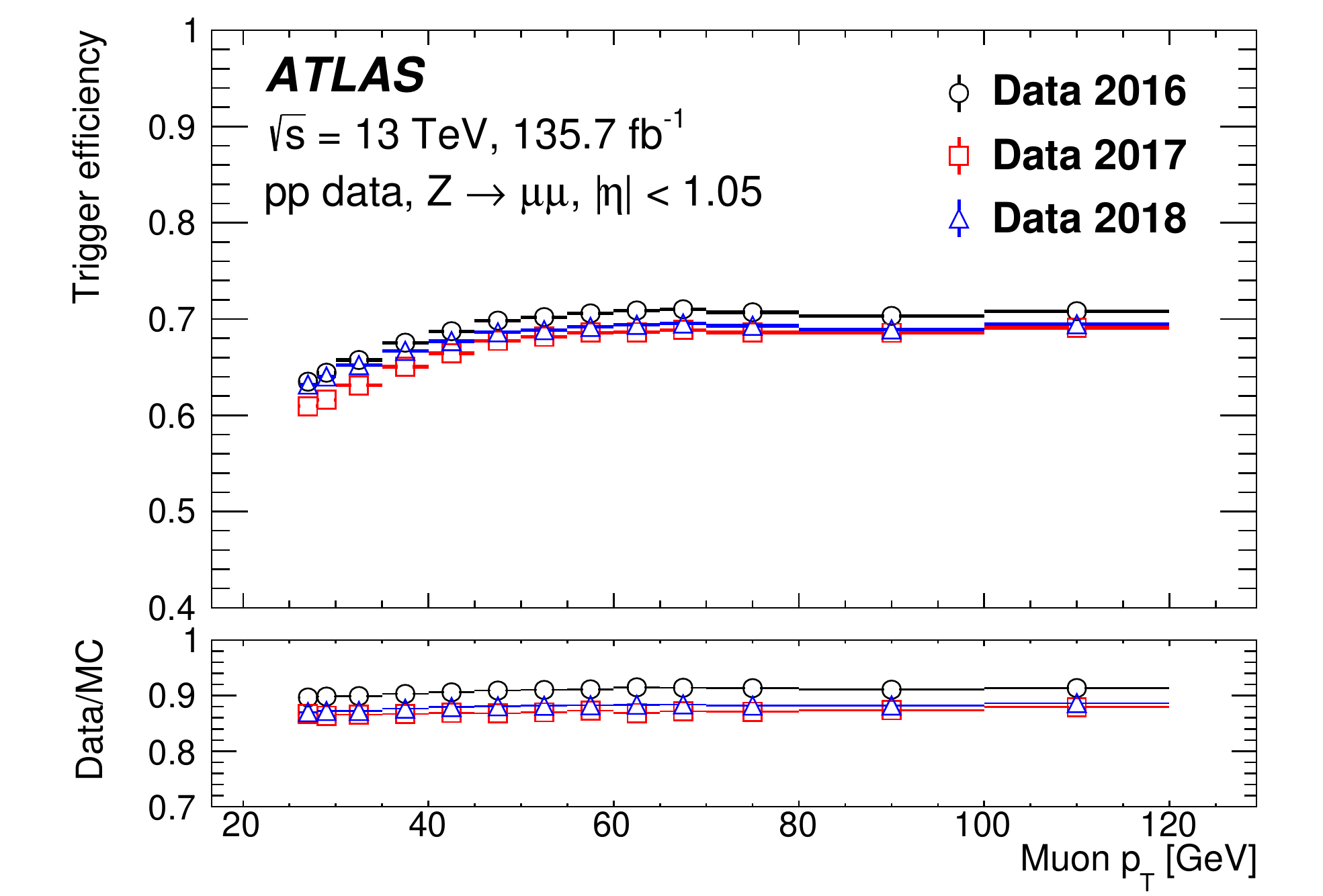}
\includegraphics[width=0.48\textwidth]{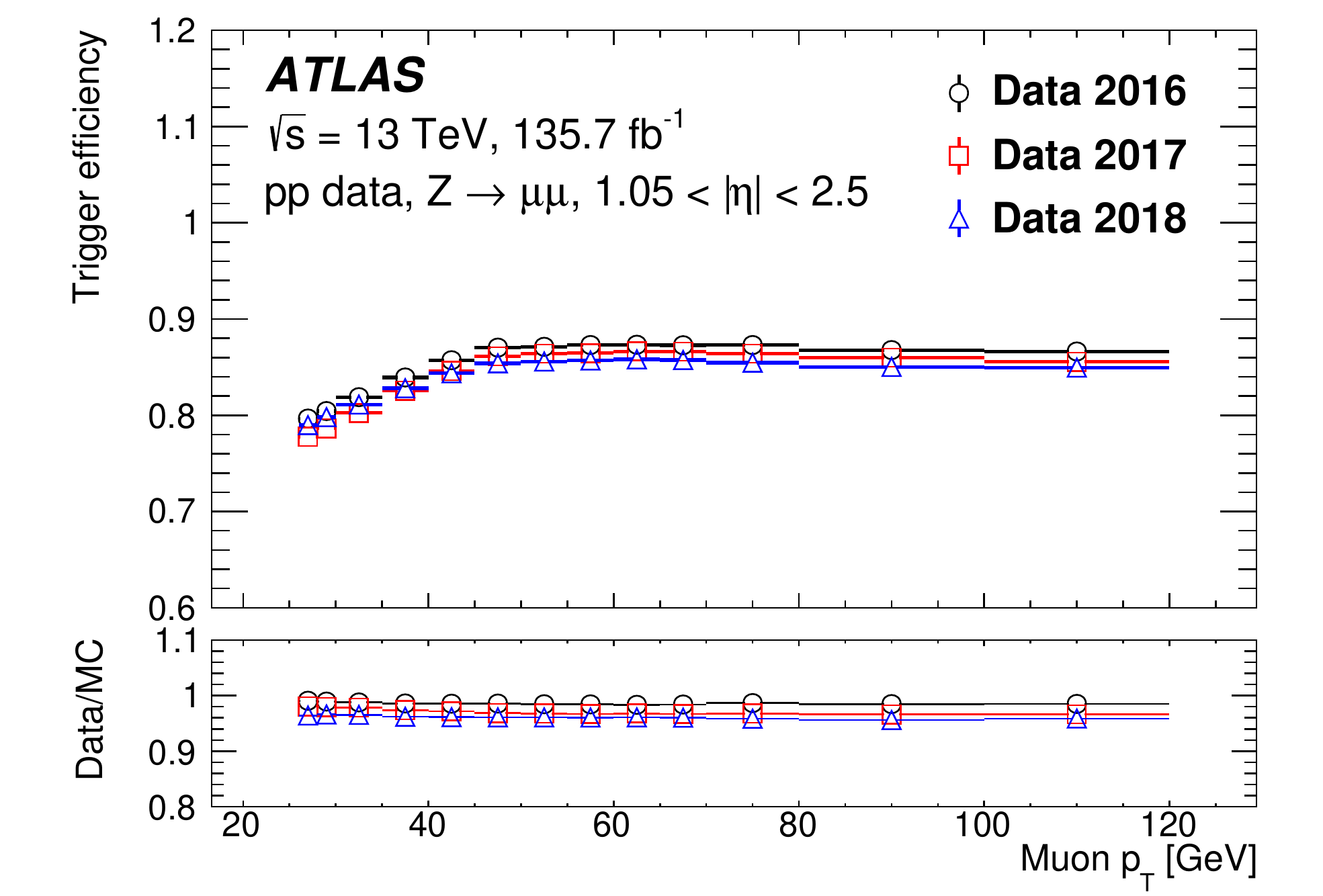}
\caption{Efficiency of passing either the HLT\_mu26\_ivarmedium or the HLT\_mu50 trigger in the barrel (left) and endcaps (right) as a function of the muon \pT, computed using data taken in 2016--2018. The error bars show the statistical uncertainties only.}
\label{fig:eff_Zmm:HLT_mu26_ivarmedium_OR_HLT_mu50_probe_pt_ratio_2017}
\end{figure}
 
Figure~\ref{fig:eff_Zmm:HLT_mu26_ivarmedium_OR_HLT_mu50_probe_eta_phi_ratio_2017} displays in the $\eta$--$\phi$ plane the ratio of data to MC trigger efficiencies of passing either the HLT\_mu26\_ivarmedium or the HLT\_mu50 trigger in the barrel and endcap regions using only 2017 data.
The variations between the ratios in neighbouring bins in the barrel region are due to the gas leaks in some of the RPC chambers discussed above.
To ensure that the probe muon lies in the trigger efficiency plateau, it must have a \pT at least 5\% above the \pT threshold of the trigger.
The binning is chosen to reflect the detector segmentation.
The statistical uncertainties are consistently at or below 5\% in each bin, in both regions.
\begin{figure}[htbp]
\centering
\includegraphics[width=0.48\textwidth]{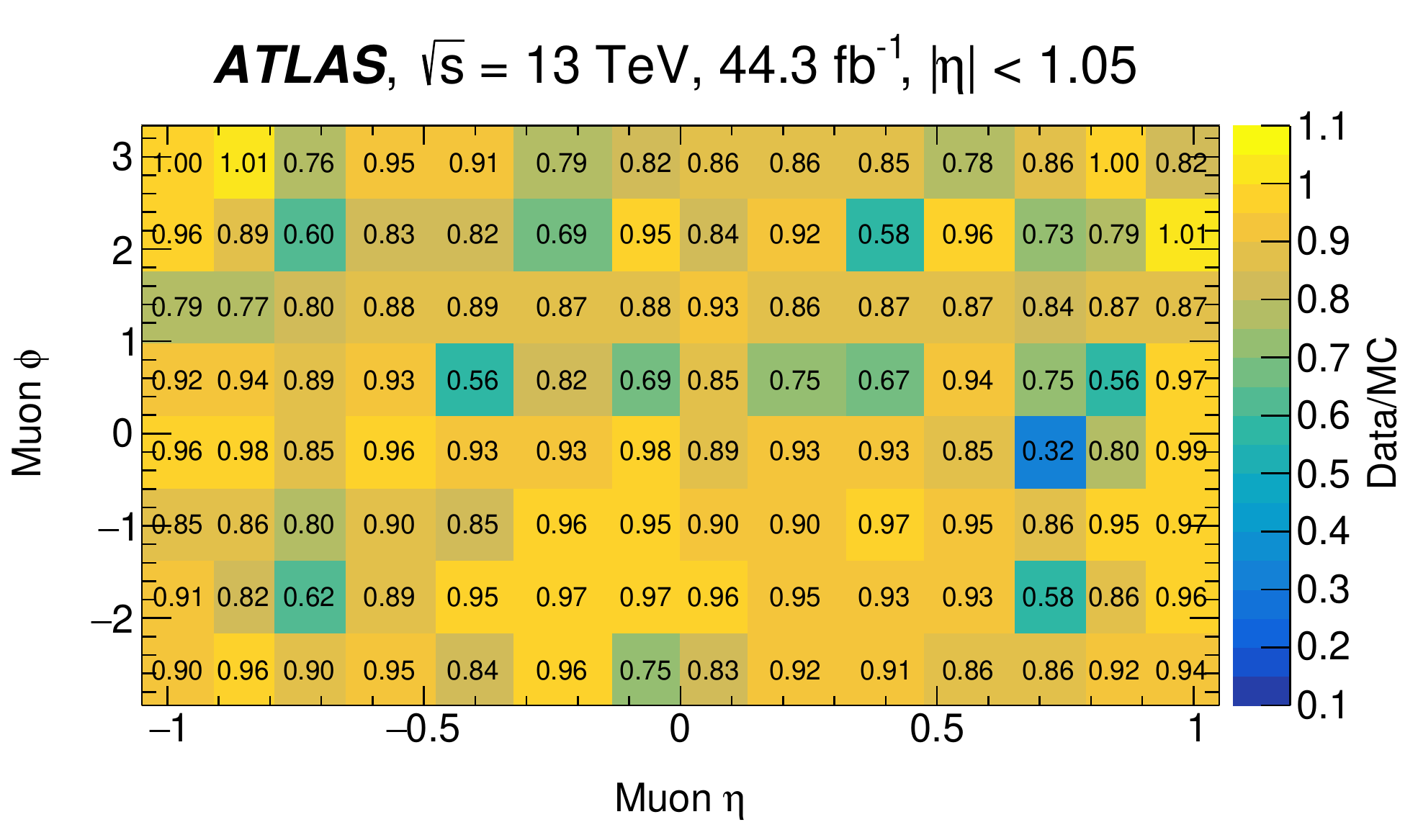}
\includegraphics[width=0.48\textwidth]{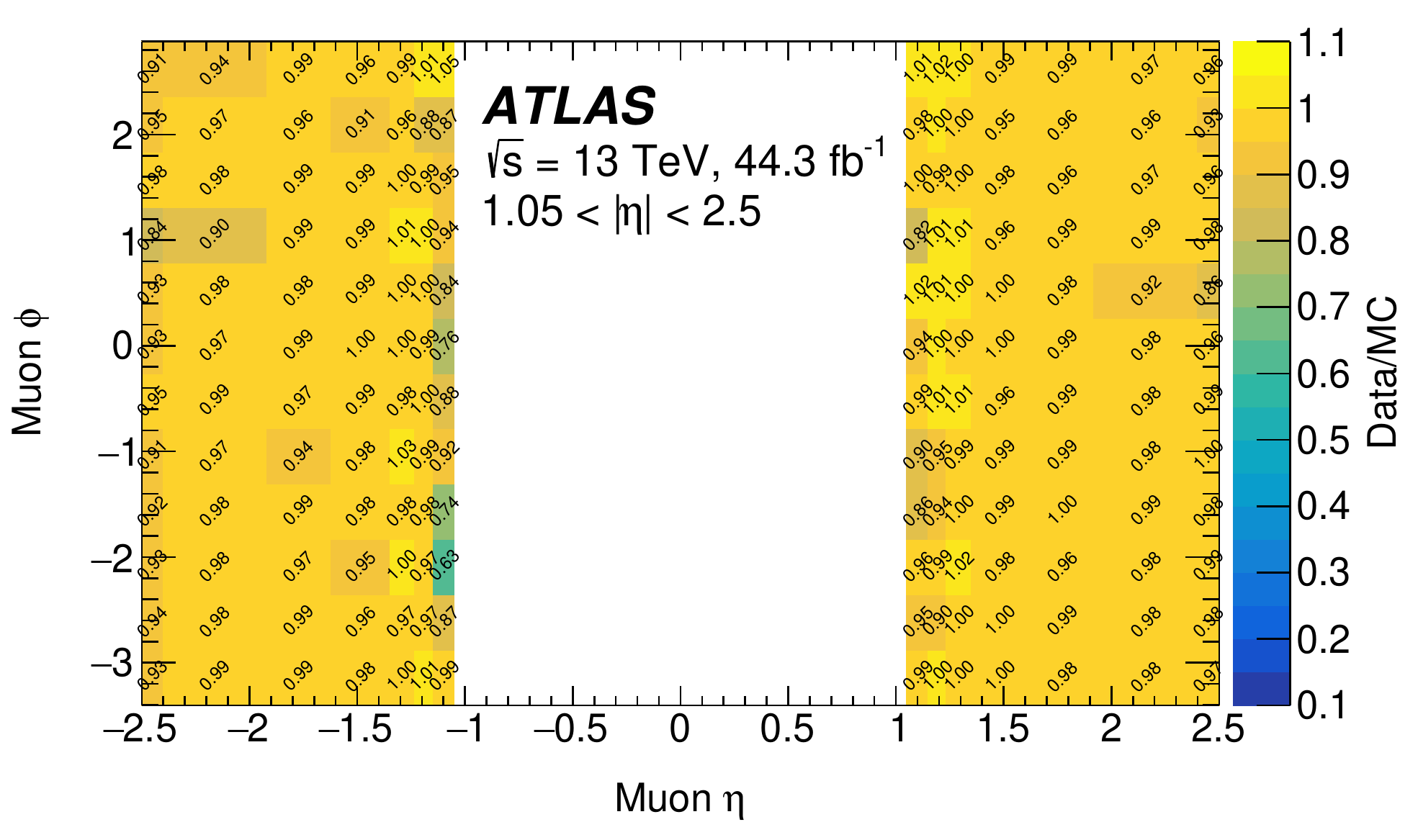}
\caption{Ratio of data to simulation efficiencies of passing either the HLT\_mu26\_ivarmedium or the HLT\_mu50 trigger in the barrel (left) and endcaps (right) in two-dimensional bins of muon $\eta$ and $\phi$, computed using data taken in 2017.}
\label{fig:eff_Zmm:HLT_mu26_ivarmedium_OR_HLT_mu50_probe_eta_phi_ratio_2017}
\end{figure}
 
Figure~\ref{fig:eff_Zmm:HLT_mu26_ivarmedium_average_mu} shows the efficiency of passing the HLT\_mu26\_ivarmedium trigger in the barrel and endcap regions as a function of the average number of interactions per bunch crossing.
An improvement in trigger efficiency for 2018 data is observed at high average numbers of interactions per bunch crossing, which is attributed to the modification of the $\Delta z$ selection criterion described in Section~\ref{sec:MuonTrigger}.
\begin{figure}[htbp]
\centering
\includegraphics[width=0.48\textwidth]{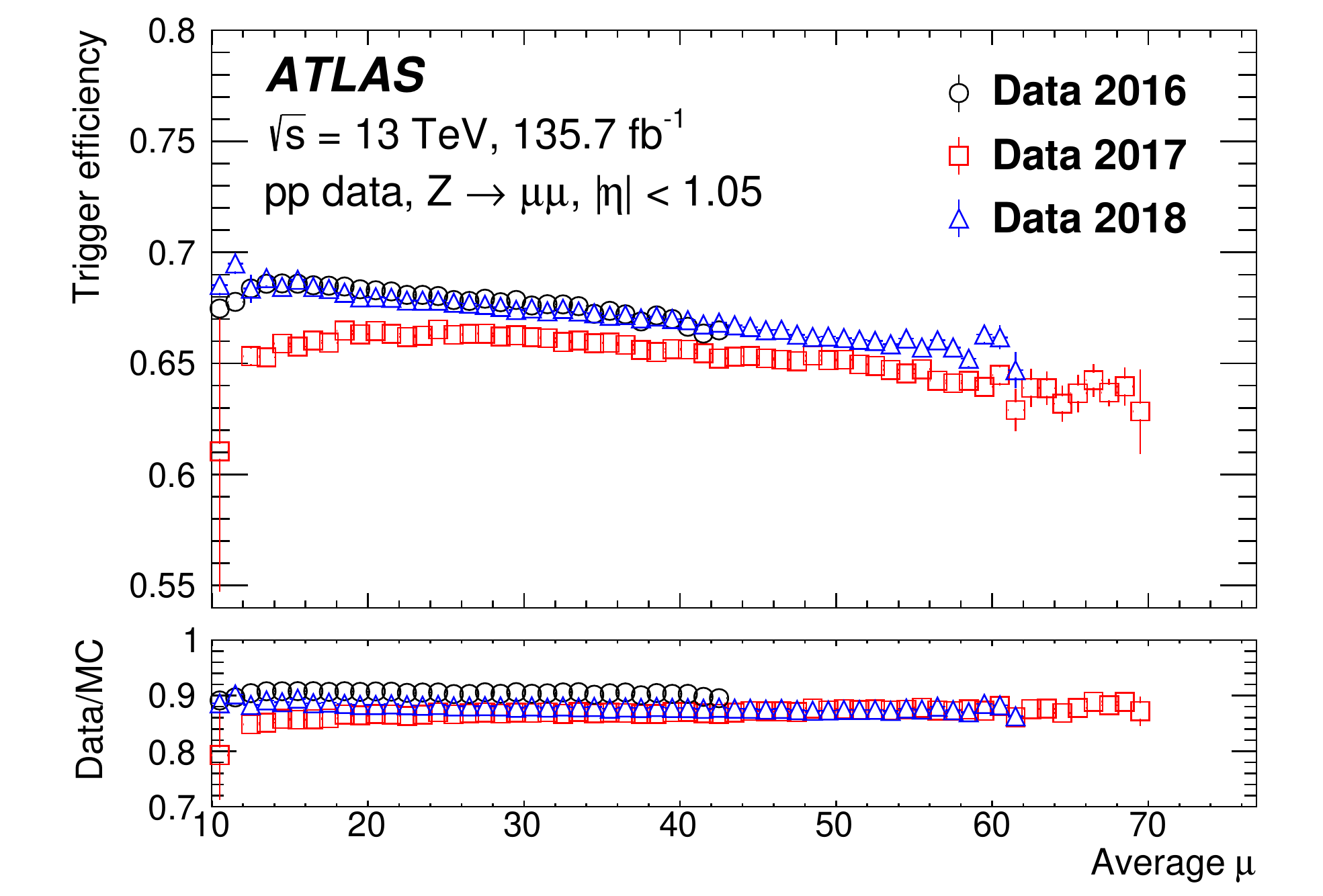}
\includegraphics[width=0.48\textwidth]{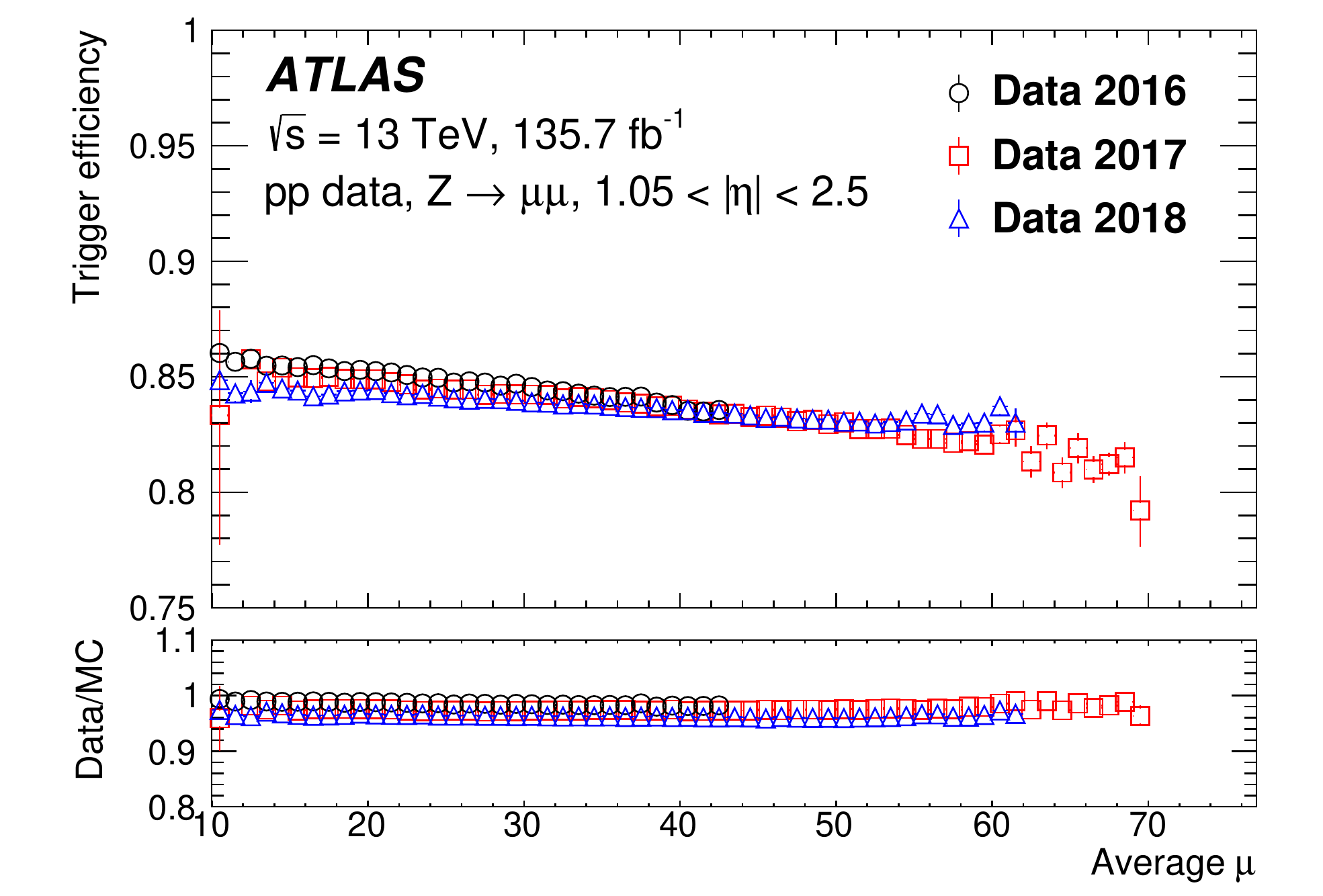}
\caption{Efficiency of passing the HLT\_mu26\_ivarmedium trigger in the barrel (left) and endcaps (right) as a function of the average number of interactions per bunch crossing, computed using data taken in 2016--2018. The error bars show the statistical uncertainties only.}
\label{fig:eff_Zmm:HLT_mu26_ivarmedium_average_mu}
\end{figure}
 
Since the \Zmm \tap analysis is statistically limited at muon \pT above 100~\GeV{}, the same measurement is carried out using \ttbar and \Wjets events.
Figure \ref{fig:effi:highpt:HLT_mu26_ivarmedium_OR_HLT_mu50_2017} shows the efficiency of passing either the HLT\_mu26\_ivarmedium or the HLT\_mu50 trigger in the barrel and endcap regions as a function of the probe muon \pT, for \ttbar and \Wjets events.
Only statistical uncertainties are shown.
The SFs are derived using muons with \pT > 100~\GeV{}.
\begin{figure}[htbp]
\centering
\includegraphics[width=0.45\textwidth]{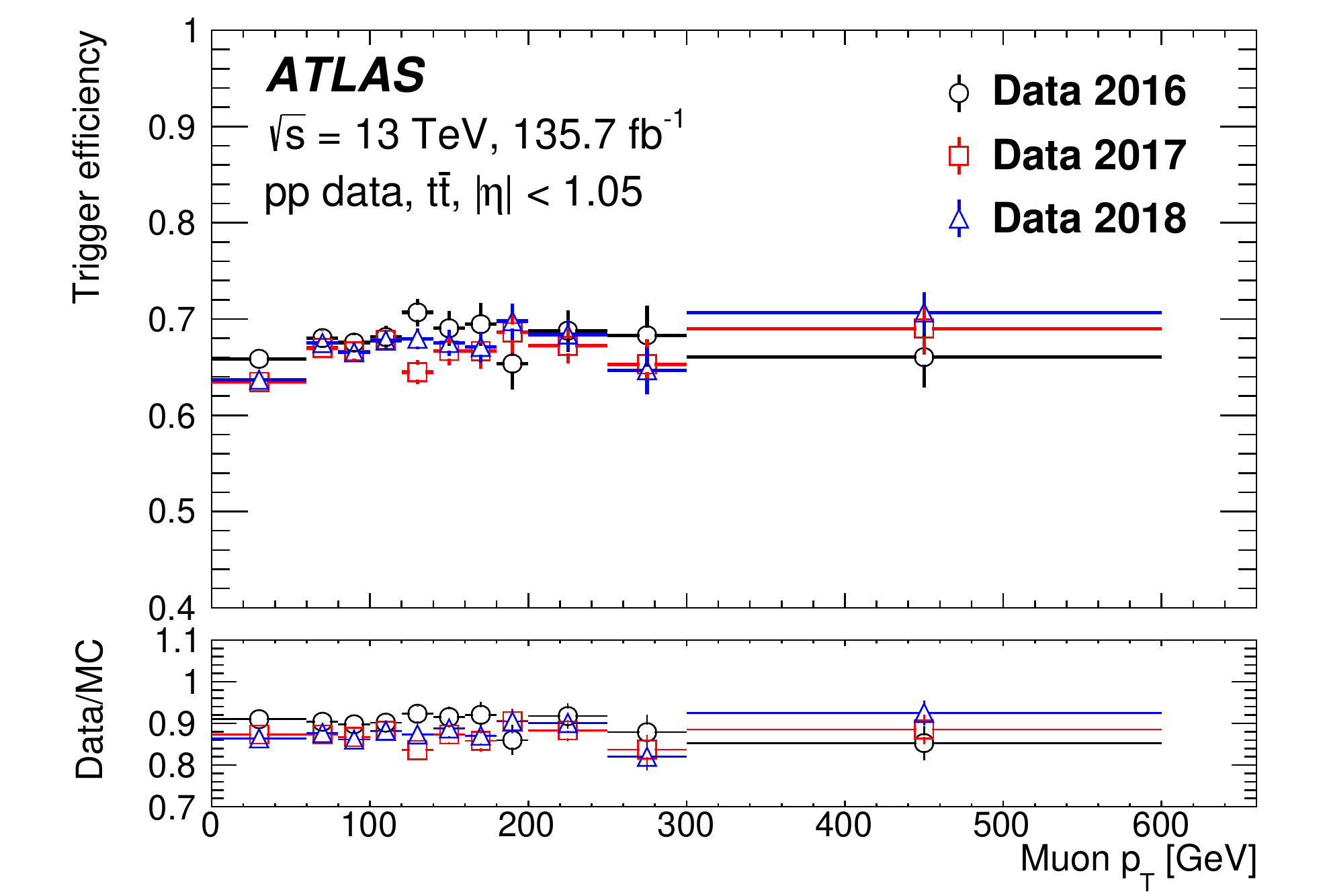}
\includegraphics[width=0.45\textwidth]{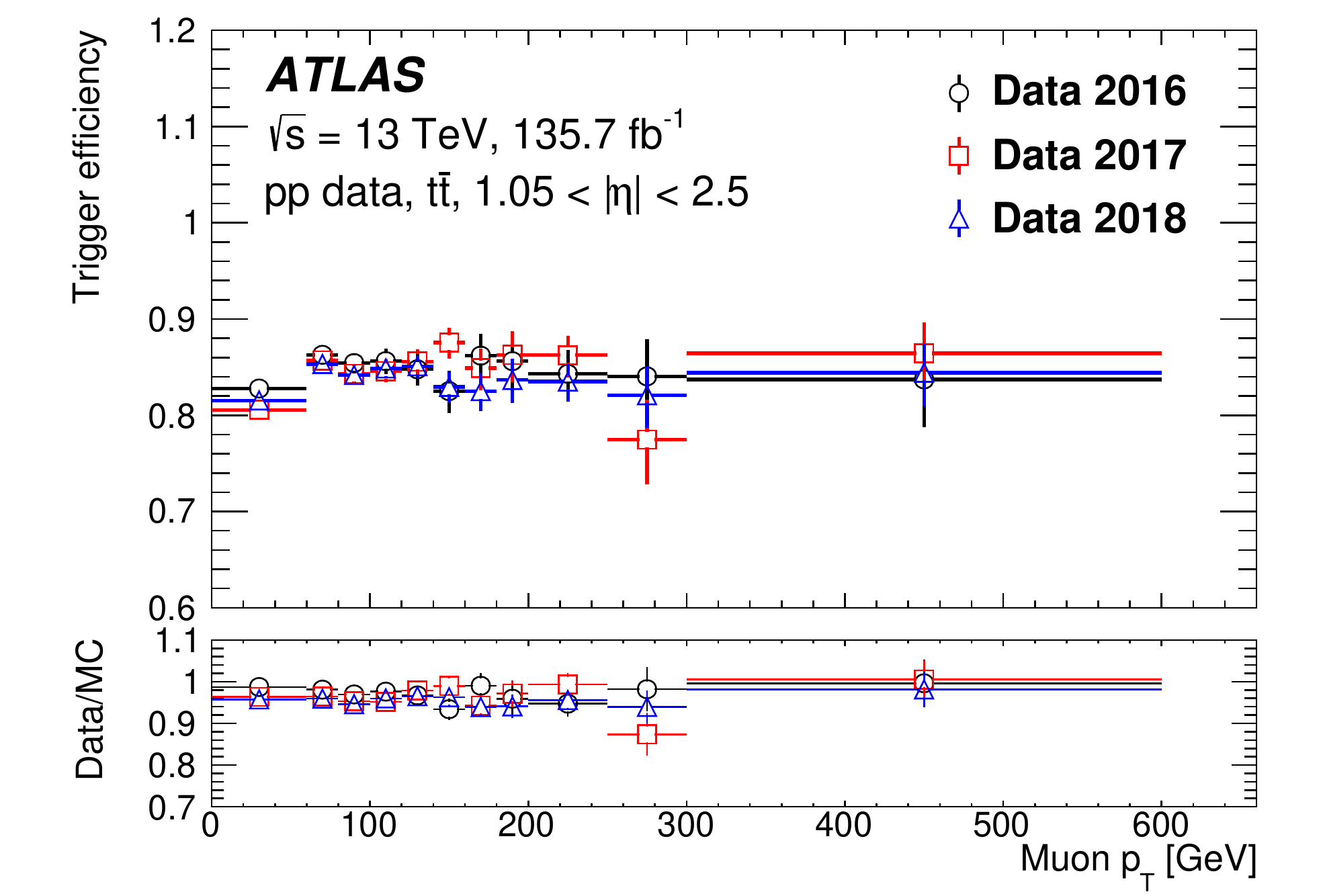}
\includegraphics[width=0.45\textwidth]{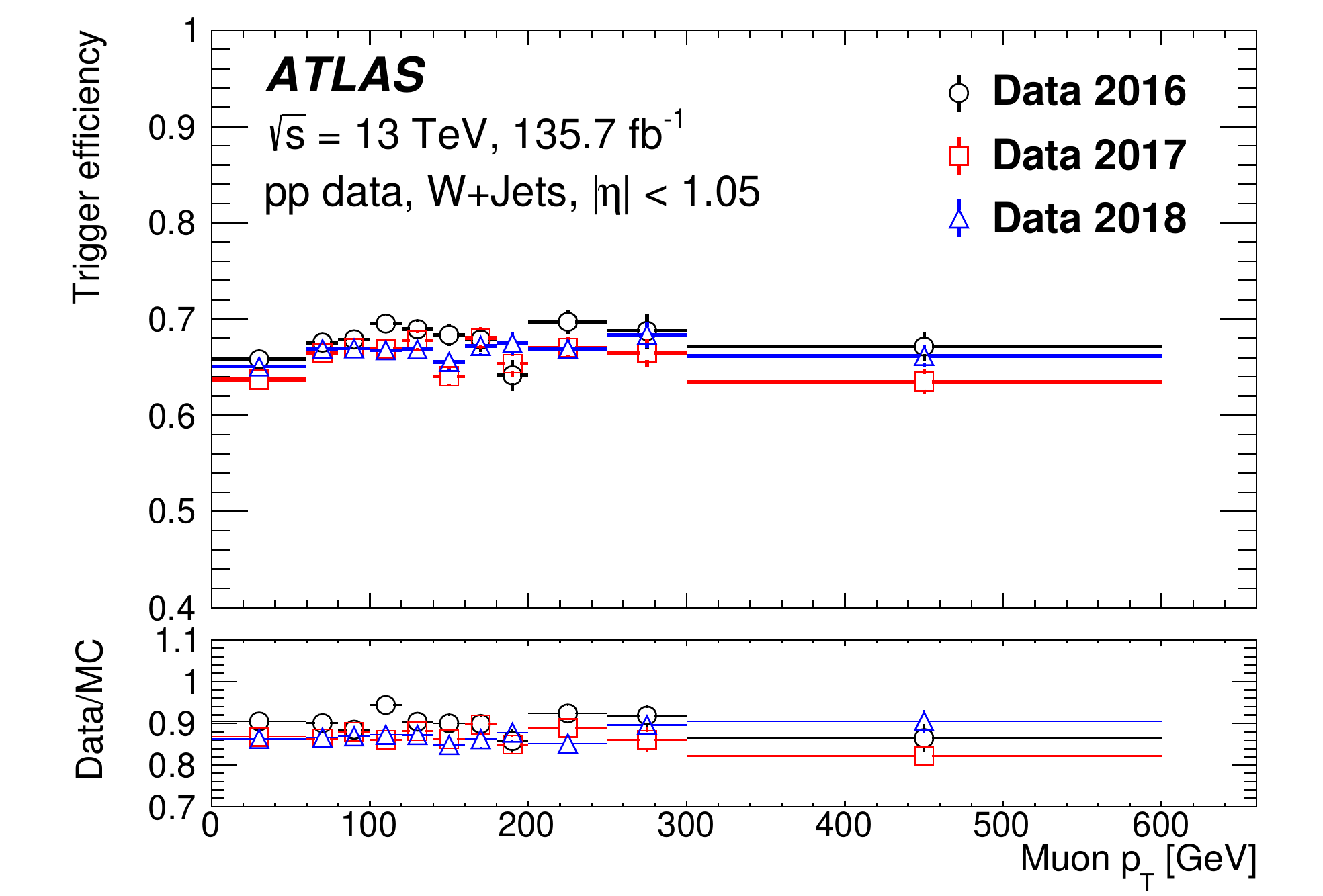}
\includegraphics[width=0.45\textwidth]{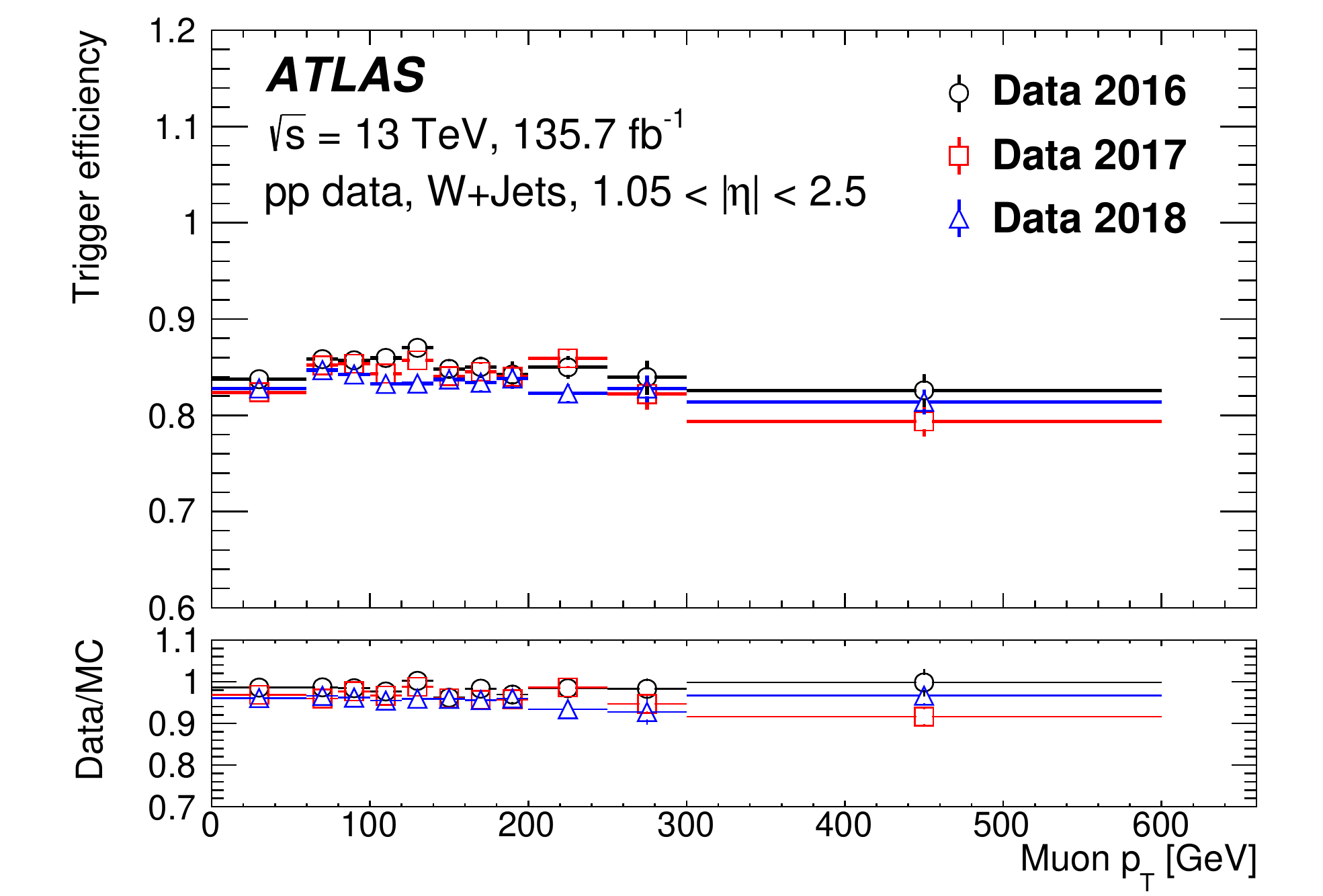}
\caption{Efficiency of passing either the HLT\_mu26\_ivarmedium or the HLT\_mu50 trigger in the barrel (left) and endcaps (right) for \ttbar (top) and \Wjets (bottom) events as a function of the muon \pT, computed using data taken in 2016--2018. The error bars show the statistical uncertainties only.}
\label{fig:effi:highpt:HLT_mu26_ivarmedium_OR_HLT_mu50_2017}
\end{figure}
 
Table \ref{fig:effi:highpt:SF_table_summary} summarises the single-muon trigger SFs derived from the different \tap{} measurements.
The inclusive SFs derived from the \ttbar and \Wjets selections are in good agreement with the results derived for intermediate-\pT muons.
The largest contribution to the systematic uncertainty in the high-\pT measurement arises from the \MET and muon quality WP variations.
\begin{table}[htbp]
\centering
\caption{Summary of the high-\pT muon trigger SF measurements for muons passing either the HLT\_mu50 trigger or the HLT\_mu26\_ivarmedium trigger.
The last two columns give the statistical and systematic uncertainties that compose the total uncertainty.}
\begin{tabular}{lllrrrr}
\toprule
Year                  & Selection & Region  & {SF [\%]}   & {Unc. [\%]} & {Stat. [\%]} & {Syst. [\%]}\\
\midrule
\multirow{6}{*}{2016} & \ttbar    & Barrel  & 90.5        & 1.2         &  0.9         & 0.8\\
& \Wjets    & Barrel  & 90.7        & 1.1         &  0.8         & 0.8\\
& \Zjets    & Barrel  & 90.5        & 0.7         & <0.1         & 0.7\\
& \ttbar    & Endcap  & 96.7        & 2.6         &  0.9         & 2.5\\
& \Wjets    & Endcap  & 98.2        & 0.9         &  0.5         & 0.7\\
& \Zjets    & Endcap  & 98.6        & 0.3         & <0.1         & 0.3\\
\midrule
\multirow{6}{*}{2017} & \ttbar    & Barrel  & 86.9        & 1.0         &  0.8         & 0.6\\
& \Wjets    & Barrel  & 86.8        & 0.7         &  0.6         & 0.4\\
& \Zjets    & Barrel  & 86.8        & 0.4         & <0.1         & 0.4\\
& \ttbar    & Endcap  & 96.7        & 1.1         &  0.8         & 0.7\\
& \Wjets    & Endcap  & 96.6        & 1.1         &  0.5         & 1.0\\
& \Zjets    & Endcap  & 98.6        & 0.5         & <0.1         & 0.5\\
\midrule
\multirow{6}{*}{2018} & \ttbar    & Barrel  & 88.3        & 1.7         &  0.7         & 1.6\\
& \Wjets    & Barrel  & 86.9        & 1.2         &  0.5         & 1.1\\
& \Zjets    & Barrel  & 87.8        & 0.7         & <0.1         & 0.7\\
& \ttbar    & Endcap  & 95.8        & 1.7         &  0.7         & 1.5\\
& \Wjets    & Endcap  & 95.5        & 0.8         &  0.4         & 0.7\\
& \Zjets    & Endcap  & 96.2        & 0.3         & <0.1         & 0.3\\
\bottomrule
\end{tabular}
\label{fig:effi:highpt:SF_table_summary}
\end{table}
 
\subsection{Multi-muon trigger efficiency}
\label{sec:eff:singleleg}
While the primary single-muon triggers serve a wide variety of physics analyses, multi-object triggers are important in many cases where lower thresholds benefit the analysis.
The efficiency of these triggers can be factorised as the product of the single-leg efficiencies.
For analyses focusing on low-momentum muons, such as in the $B$-physics and Light States programme, dimuon and trimuon triggers with thresholds down to 4~\GeV{} are frequently applied.
The corresponding single-leg efficiency is measured by applying the \tap method to \Jpsi events.
Figure~\ref{fig:eff_Jpsi:HLT_mu4} shows the measured efficiency for the barrel and endcap regions as a function of the muon \pT for \textit{Tight} quality muons in 2015 and 2016, for the single-muon trigger HLT\_mu4.
These efficiency measurements provide a crucial input for the BLS programme outlined in Section~\ref{sec:bphys}.
As observed for the single-muon triggers, there are differences between the simulated trigger efficiency and the trigger efficiency in data, particularly in the barrel, where the efficiency is $\sim$10\% lower in data than in simulation.
The \pT dependence of the SFs in the plateau region is found to be small.
\begin{figure}[htbp]
\centering
\includegraphics[width=0.48\textwidth]{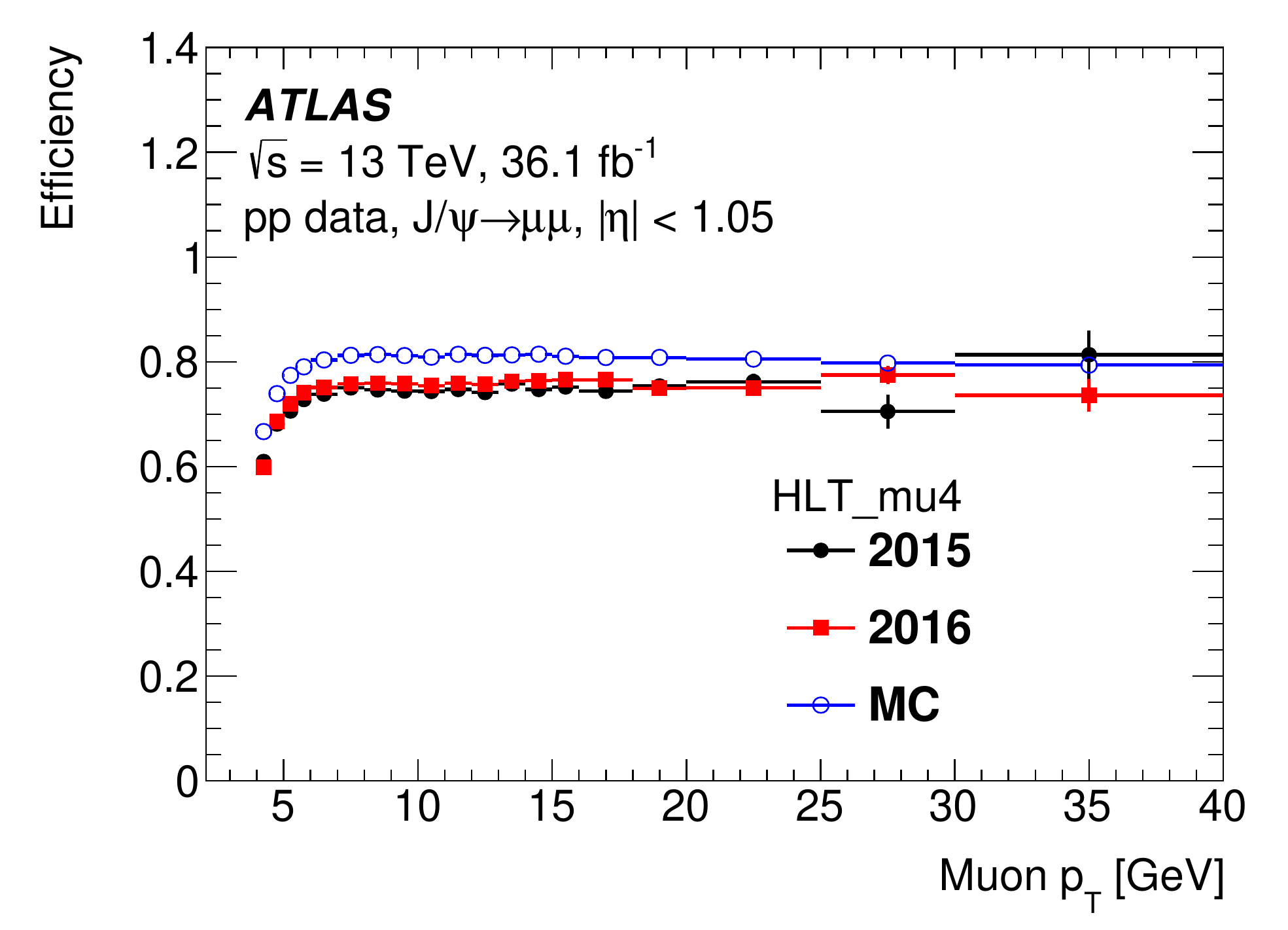}
\includegraphics[width=0.48\textwidth]{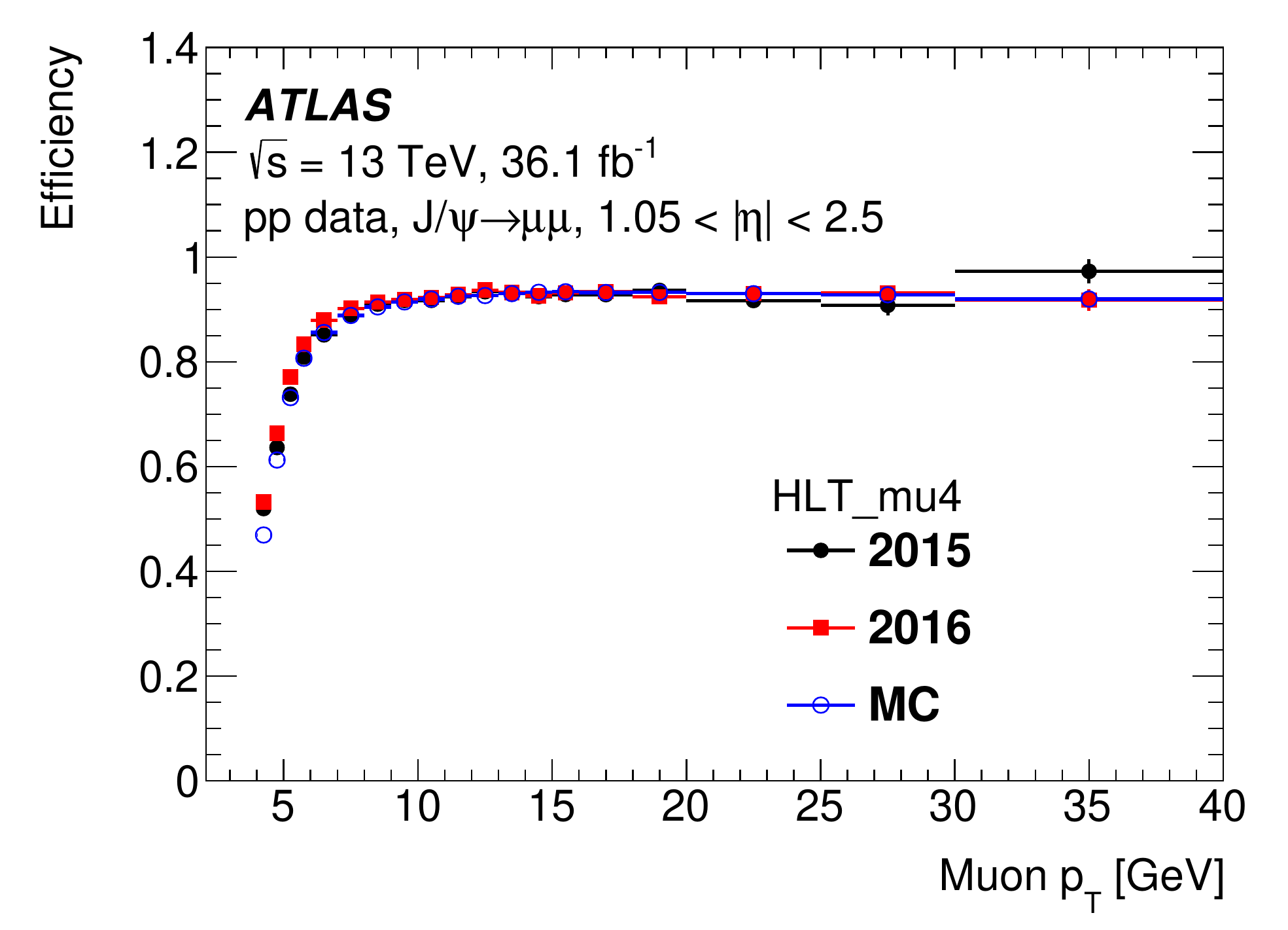}
\caption{Efficiency of the single-muon trigger HLT\_mu4 in the barrel (left) and endcap (right) regions as a function of the muon \pT for \textit{Tight} quality muons, computed using data taken in 2015 and 2016. The full markers correspond to data, while the empty markers represent MC. The error bars show the statistical uncertainties only.}
\label{fig:eff_Jpsi:HLT_mu4}
\end{figure}
 
Measurements of multi-muon trigger efficiencies targeting a higher \pT regime than the previous low-\pT chains are performed by using the \Zmm event selection criteria described above to measure the single-leg efficiencies.
Figure~\ref{fig:eff_Zmm:HLT_mu14} shows the efficiency of HLT\_mu14 in the barrel and endcap regions as a function of the muon \pT for each year of data-taking.
\begin{figure}[htbp]
\centering
\includegraphics[width=0.48\textwidth]{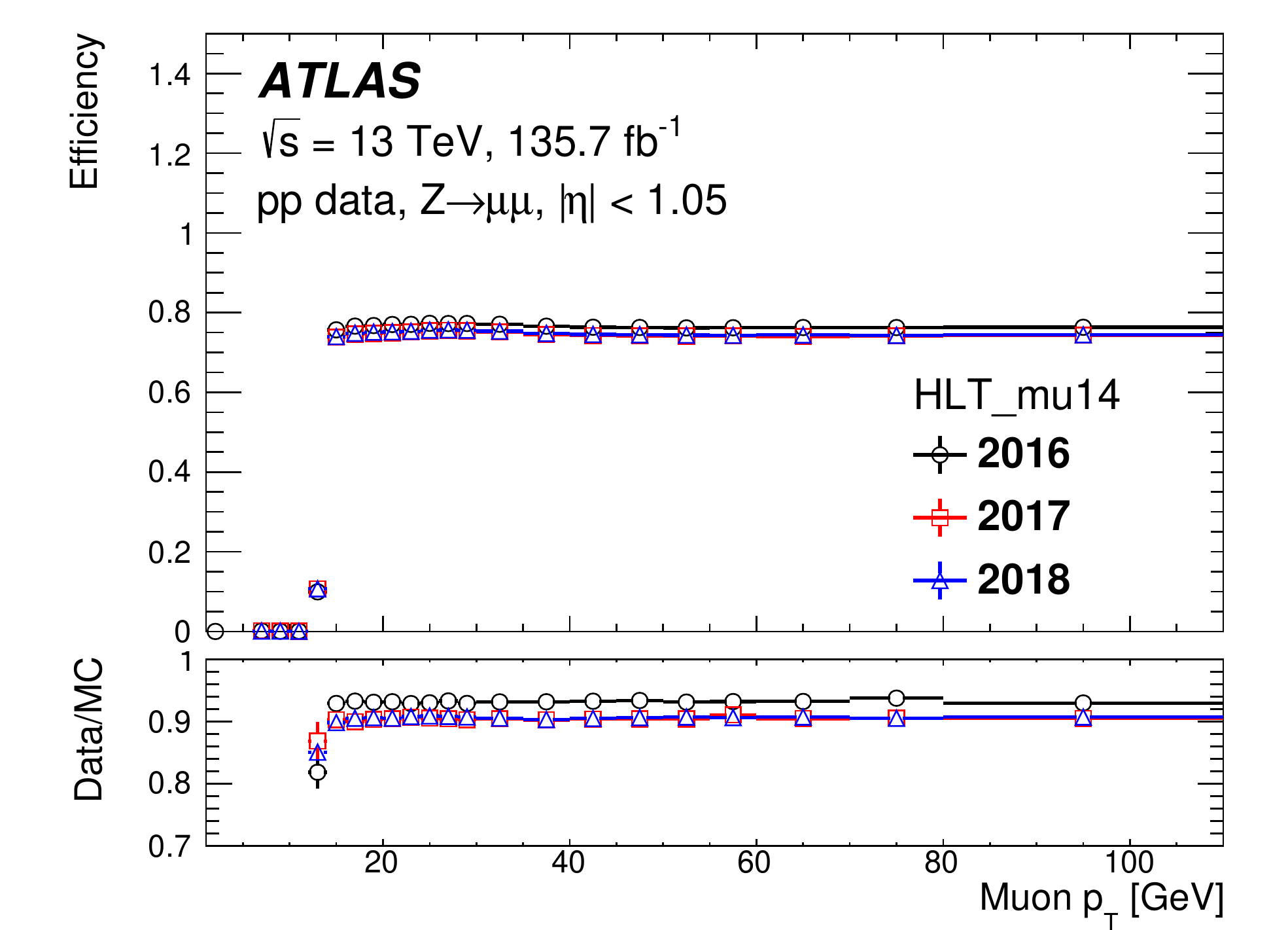}
\includegraphics[width=0.48\textwidth]{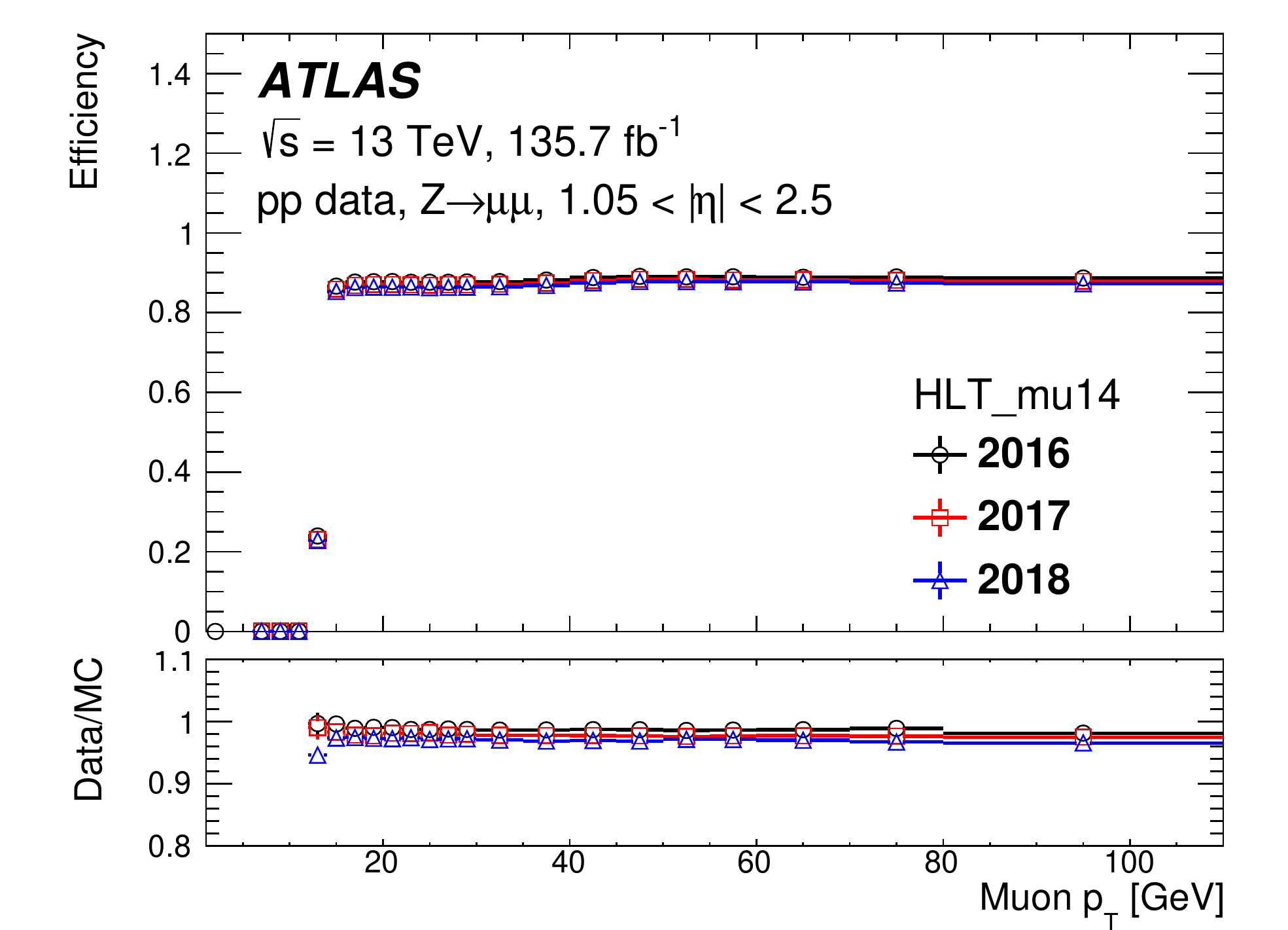}
\caption{Efficiency of the single-muon trigger HLT\_mu14 in the barrel (left) and endcap (right) regions measured in each year of data-taking as a function of the probe muon transverse momentum \pT. The error bars show the statistical uncertainties only.}
\label{fig:eff_Zmm:HLT_mu14}
\end{figure}
 
The efficiency of the FS trigger, HLT\_mu8noL1, is also measured using the same technique and presented for the barrel and endcap regions in Figure~\ref{fig:eff_Zmm:HLT_mu8}.
Compared to the RoI-based trigger efficiency above, the FS efficiency is almost one in both the barrel and endcap regions, because the L1 inefficiencies are circumvented.
\begin{figure}[htbp]
\centering
\includegraphics[width=0.48\textwidth]{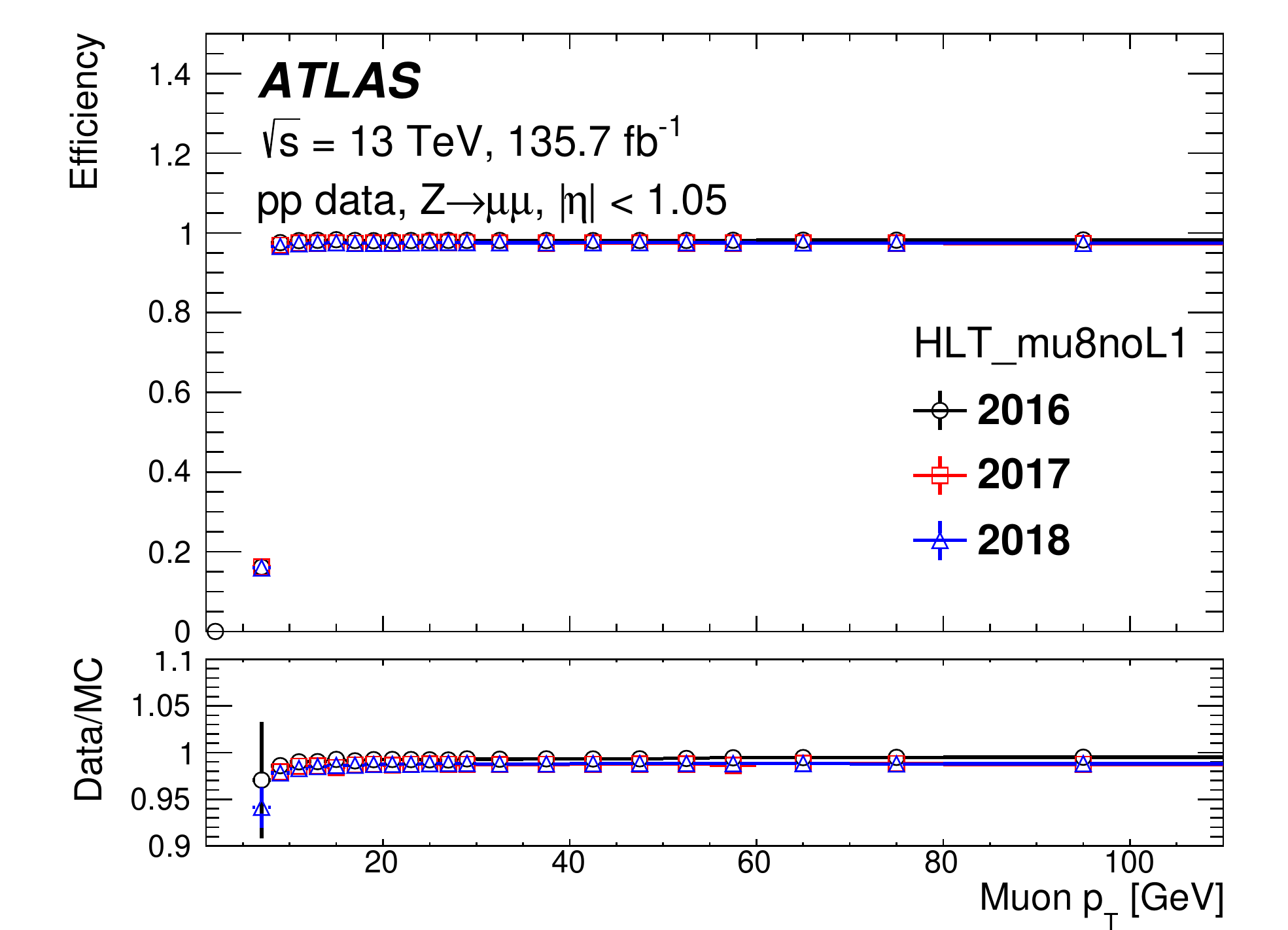}
\includegraphics[width=0.48\textwidth]{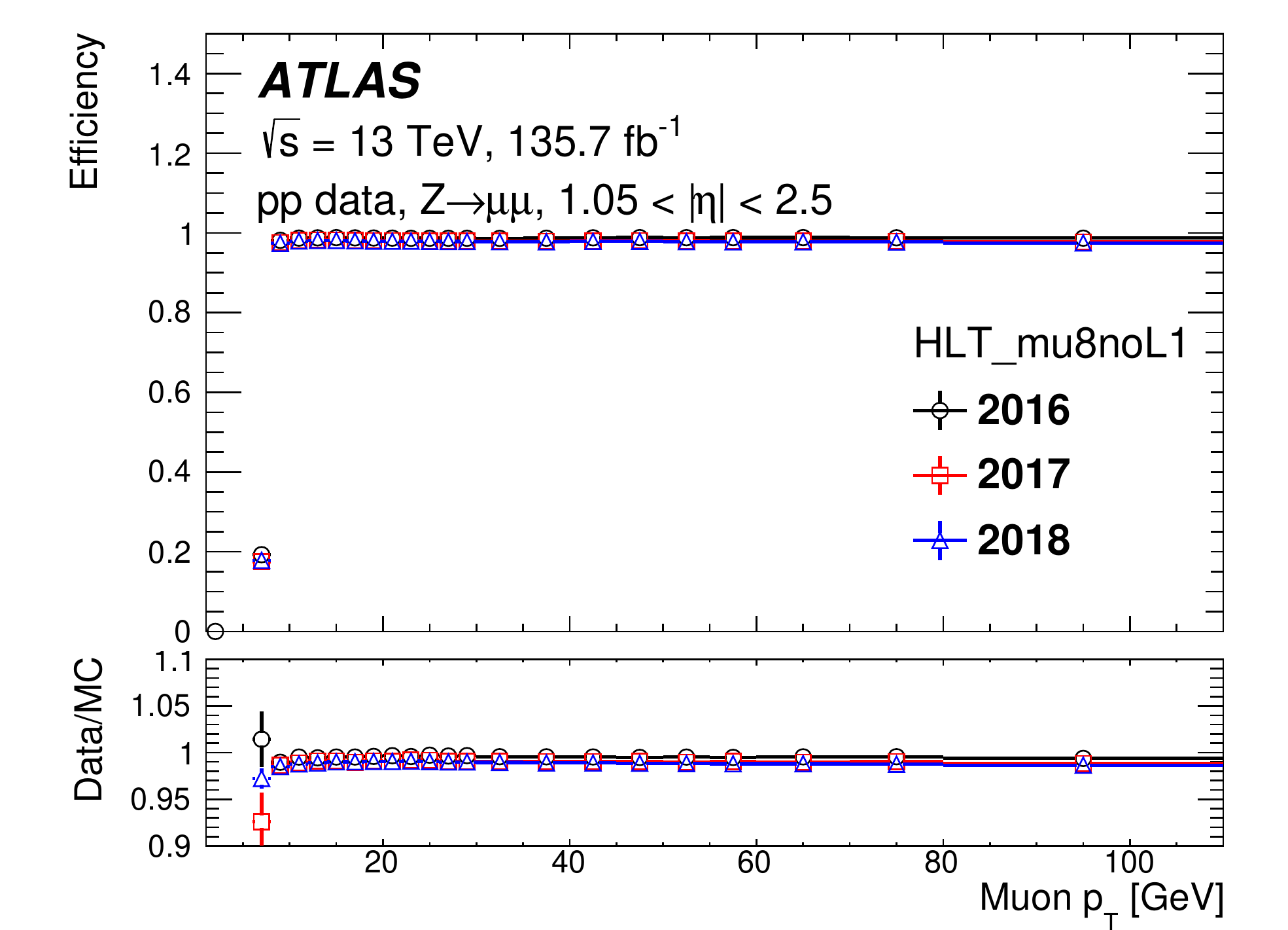}
\caption{Efficiency of the FS single-muon trigger HLT\_mu8noL1 in the barrel (left) and endcap (right) regions measured in each year of data-taking as a function of the probe muon transverse momentum \pT. The error bars show the statistical uncertainties only.}
\label{fig:eff_Zmm:HLT_mu8}
\end{figure}
 
Figure~\ref{fig:eff_Zmm:HLT_mu14:pu} shows that the efficiency of the single-muon trigger HLT\_mu14 has almost no pile-up dependence.
Since no isolation is applied, there is no change in efficiency for events with high pile-up in 2017 and 2018.
\begin{figure}[htbp]
\centering
\includegraphics[width=0.48\textwidth]{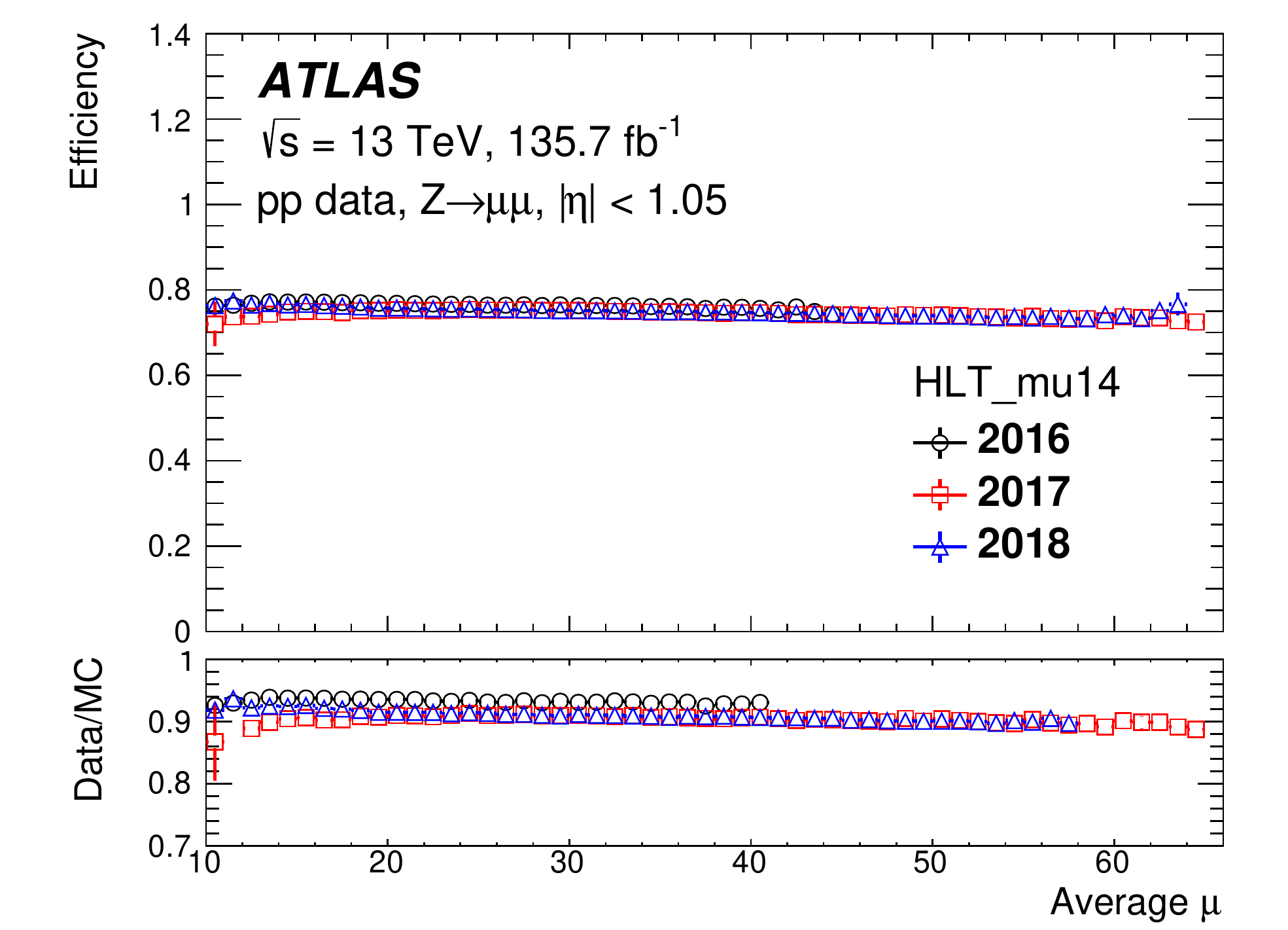}
\includegraphics[width=0.48\textwidth]{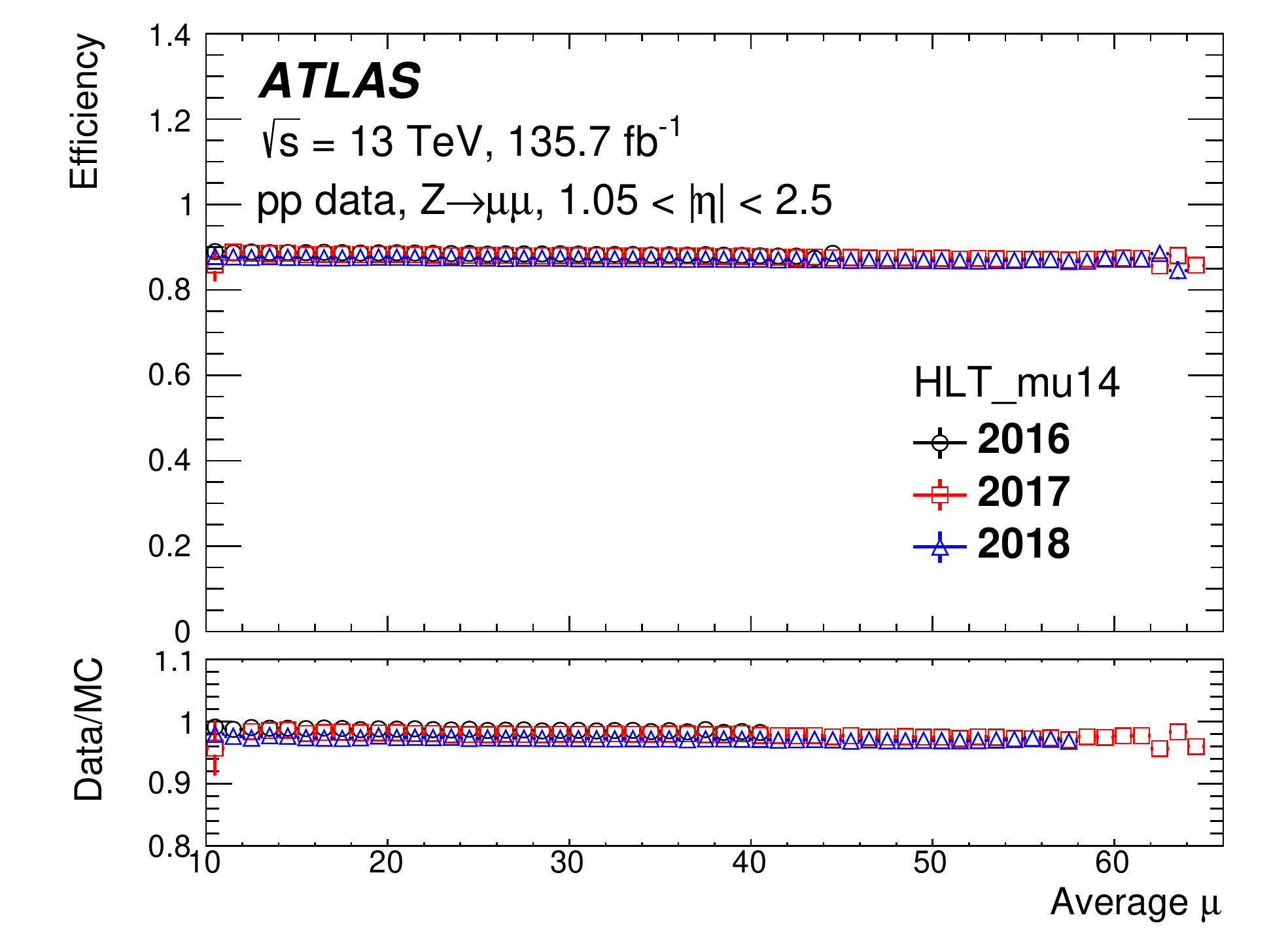}
\caption{Efficiency of the single-muon trigger HLT\_mu14 in the barrel (left) and endcap (right) regions measured in each year of data-taking as a function of the average pile-up <$\mu$>. The error bars show the statistical uncertainties only.}
\label{fig:eff_Zmm:HLT_mu14:pu}
\end{figure}
 
\subsection{High-\texorpdfstring{\pT}{momentum} muons in low-pile-up \texorpdfstring{\pp}{pp} collisions}
This section presents the studies of muon trigger efficiencies and SFs for data from low-pile-up \pp collisions at $\sqrt{s} = 5.02$~\TeV{} collected in November 2015.
The efficiency of the lowest unprescaled single-muon trigger in this data set, HLT\_mu14, is determined using the \tap method in \Zmm events, largely following the procedure described in Section~\ref{sec:eff}, but adapting the tag and probe muon selections to the looser trigger thresholds.
Additionally, the identification criterion of the tag muon is tightened to the \textit{Medium} WP.
A subset of the systematic uncertainties described in Section~\ref{sec:perf} are considered and found to be either negligible or limited by statistical precision and thus are neglected.
The HLT\_mu14 trigger efficiency for \Zmm simulation and measured in data is presented in Figure~\ref{fig_lowmupp:trig-eff-dataMC-pt} as a function of probe muon~\pT.
The trigger efficiency for muons with \pT above 15~\GeV{} is roughly constant, at about 75\% in the barrel region and at about 90\% in the endcap region.
In the endcap region, the simulated efficiency matches the efficiency measured in data, and SFs are close to unity.
However, as discussed above, the efficiency measured in the barrel region in data is lower by 5--10\% than in simulation.
\begin{figure}[htbp]
\centering
\includegraphics[width=0.49\textwidth]{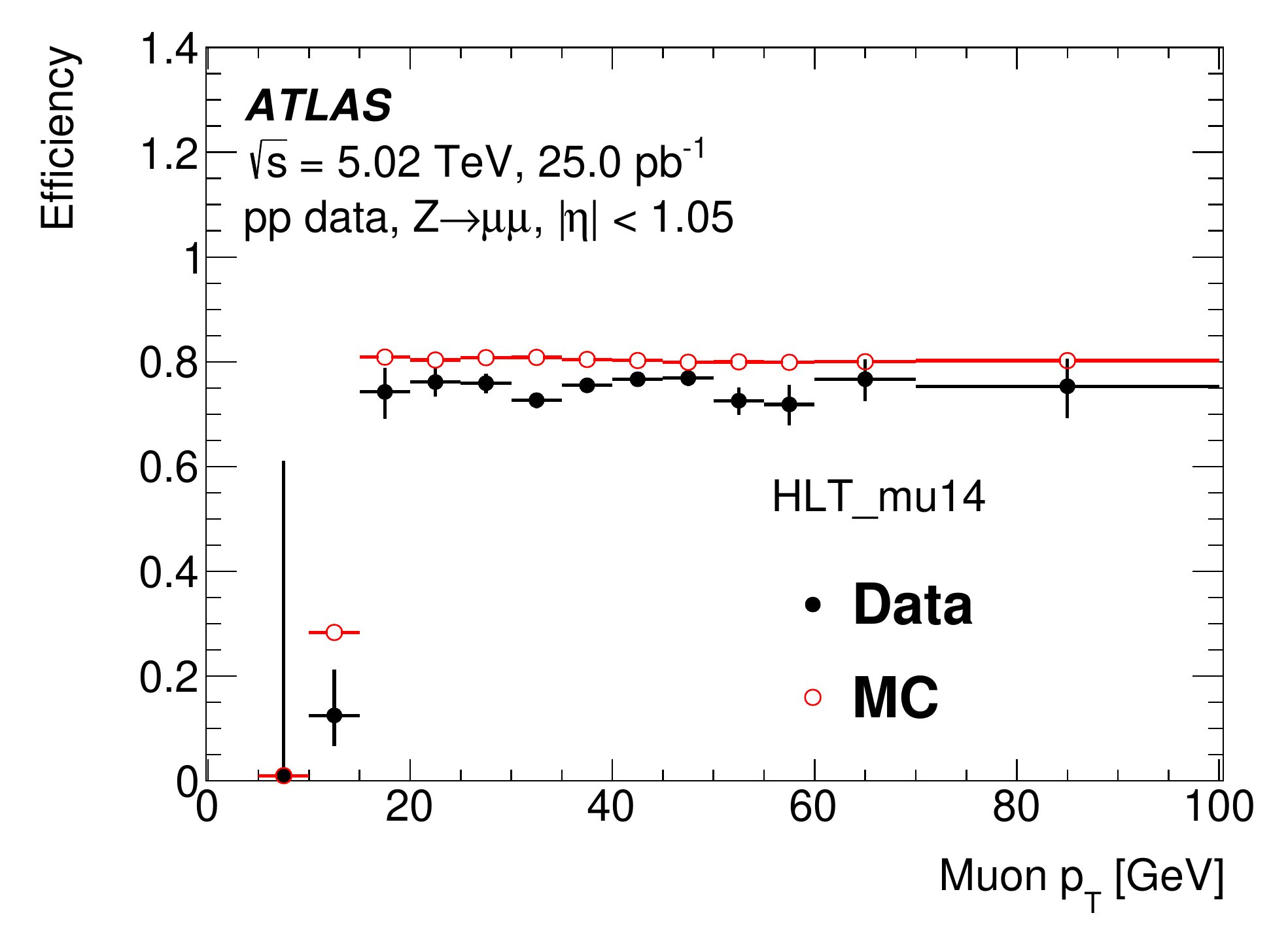}
\includegraphics[width=0.49\textwidth]{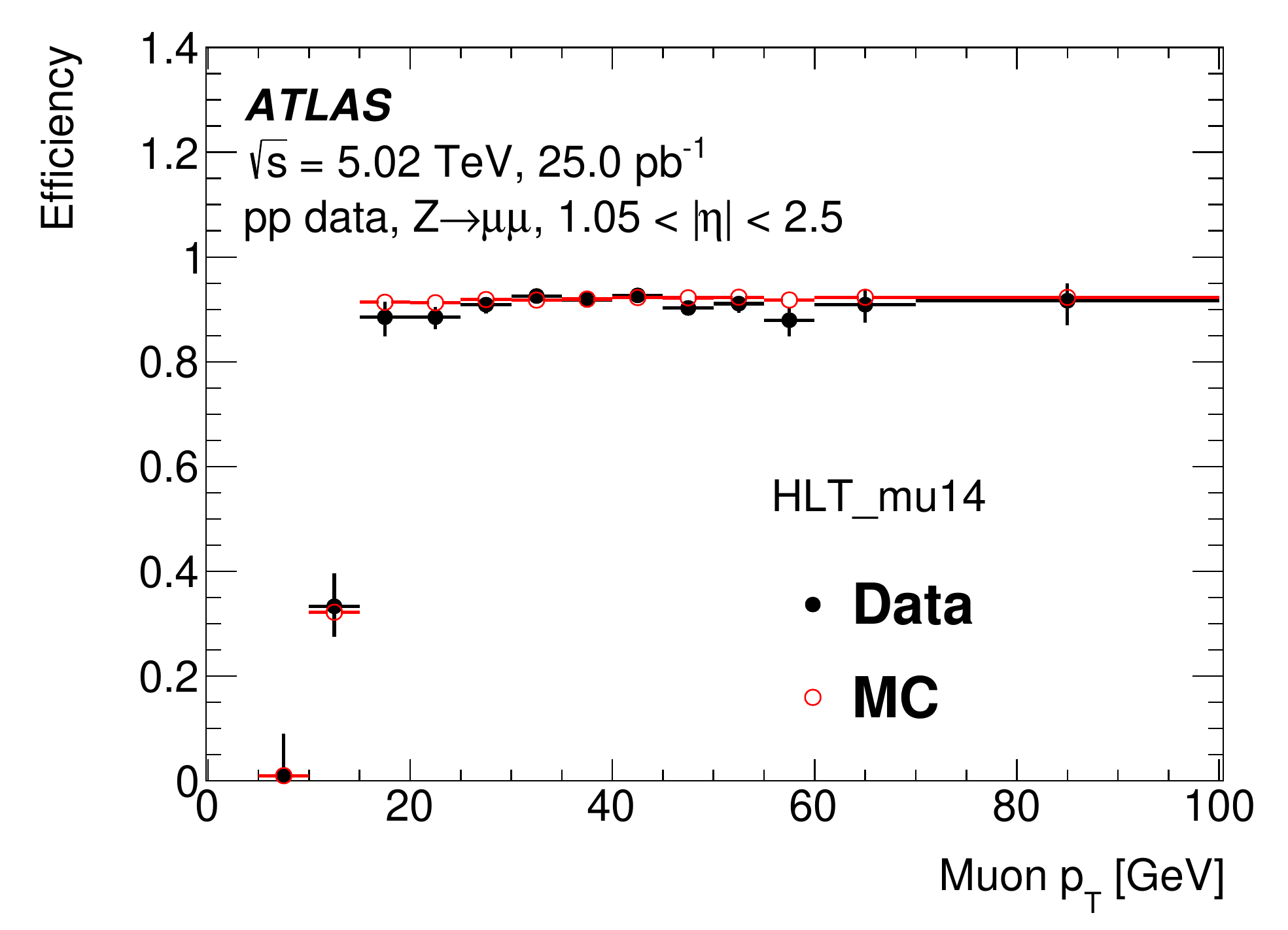}
\caption{Efficiency of the HLT\_mu14 trigger measured as a function of probe muon~\pT in the barrel~(left) and endcap~(right) regions in low-pile-up \pp data (full black circles) and \Zmm simulation (empty red circles). The error bars show the statistical uncertainties only.}
\label{fig_lowmupp:trig-eff-dataMC-pt}
\end{figure}

\section{Muon trigger efficiency in HI data-taking}
 
This section presents single-muon trigger efficiency studies for \PbPb data at $\sqrt{s_\text{NN}}=5.02$~\TeV{} taken in 2015 and 2018.
The trigger efficiency is studied using the \tap method on \Jpsi and \Zmm events described above, but adapting the tag and probe muon selections to the looser trigger thresholds.
A subset of the systematic uncertainties as mentioned above are considered and found to be either negligible or limited by statistical precision.
The performance of the HLT\_mu4 trigger is estimated using the \tap method on \Jpsi events described in Section~\ref{sec:perf} with minor changes.
For this measurement the data taken in 2018 are analysed.
HLT\_mu8 is used as the tag trigger, and the probe muons are required to satisfy the \textit{Tight} quality WP.
Figure~\ref{fig_PbPb:trig-eff-data-pt} (left) shows the trigger efficiencies for HLT\_mu4 in the barrel and endcap regions, as a function of the probe muon \pT.
In the barrel region the efficiencies are about 70\% in the plateau region.
Trigger efficiencies in the endcap are visibly higher, with the plateau at about 90\% and with a more sharply rising turn-on.
No significant difference in the measured efficiencies is observed between central and peripheral collisions.
The performance of the single-muon HLT\_mu8 trigger is analysed using high-\pT muons, which satisfy the \textit{Medium} quality WP, from \Zmm events collected in 2015.
The measurement of the HLT\_mu8 trigger efficiency as a function of the probe muon~\pT is presented in Figure~\ref{fig_PbPb:trig-eff-data-pt} (right).
The trigger efficiency is measured to be around 70\% in the barrel region and around 85\% in the endcap region.
\begin{figure}[htbp]
\centering
\includegraphics[width=0.49\textwidth]{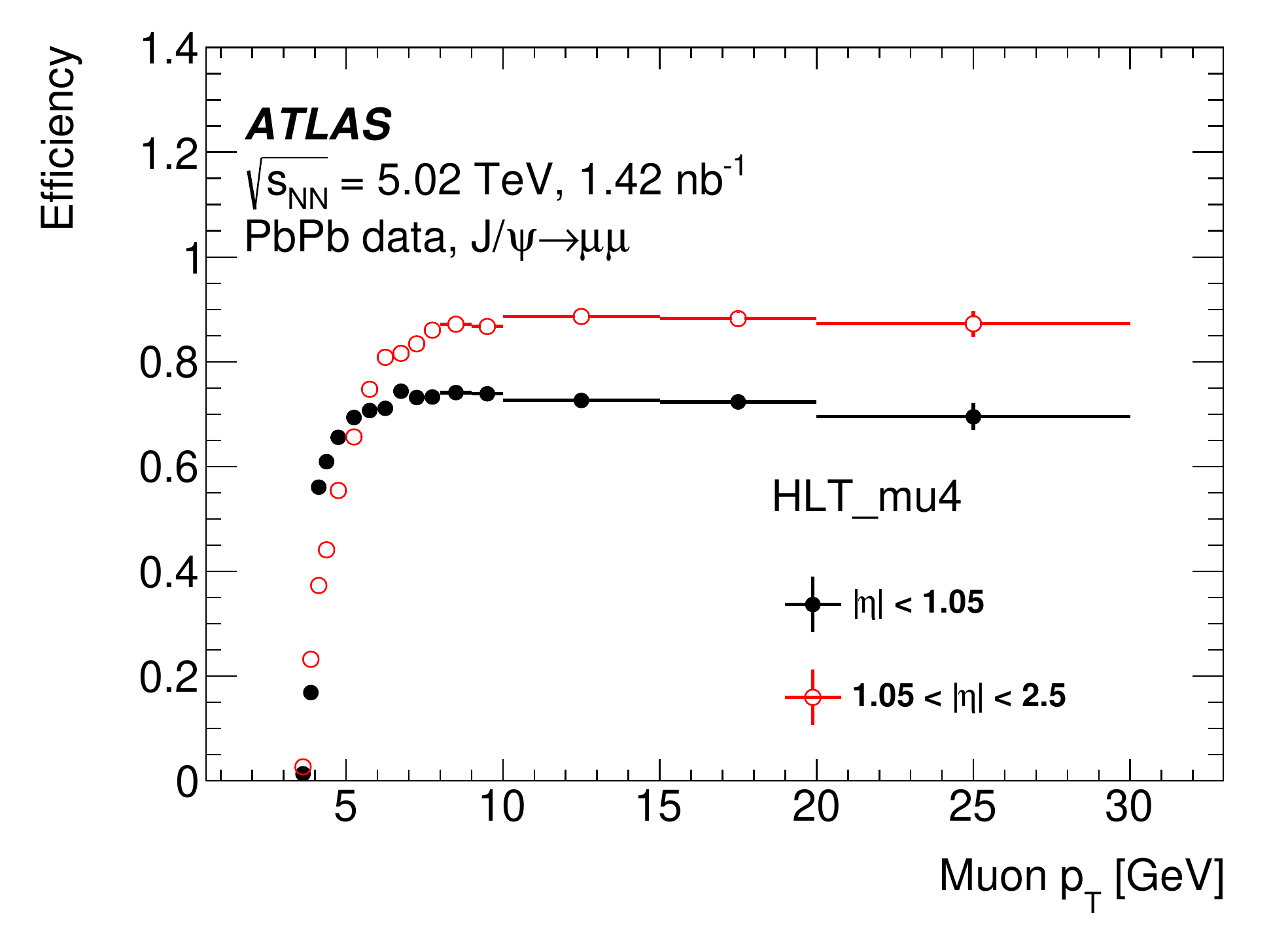}
\includegraphics[width=0.49\textwidth]{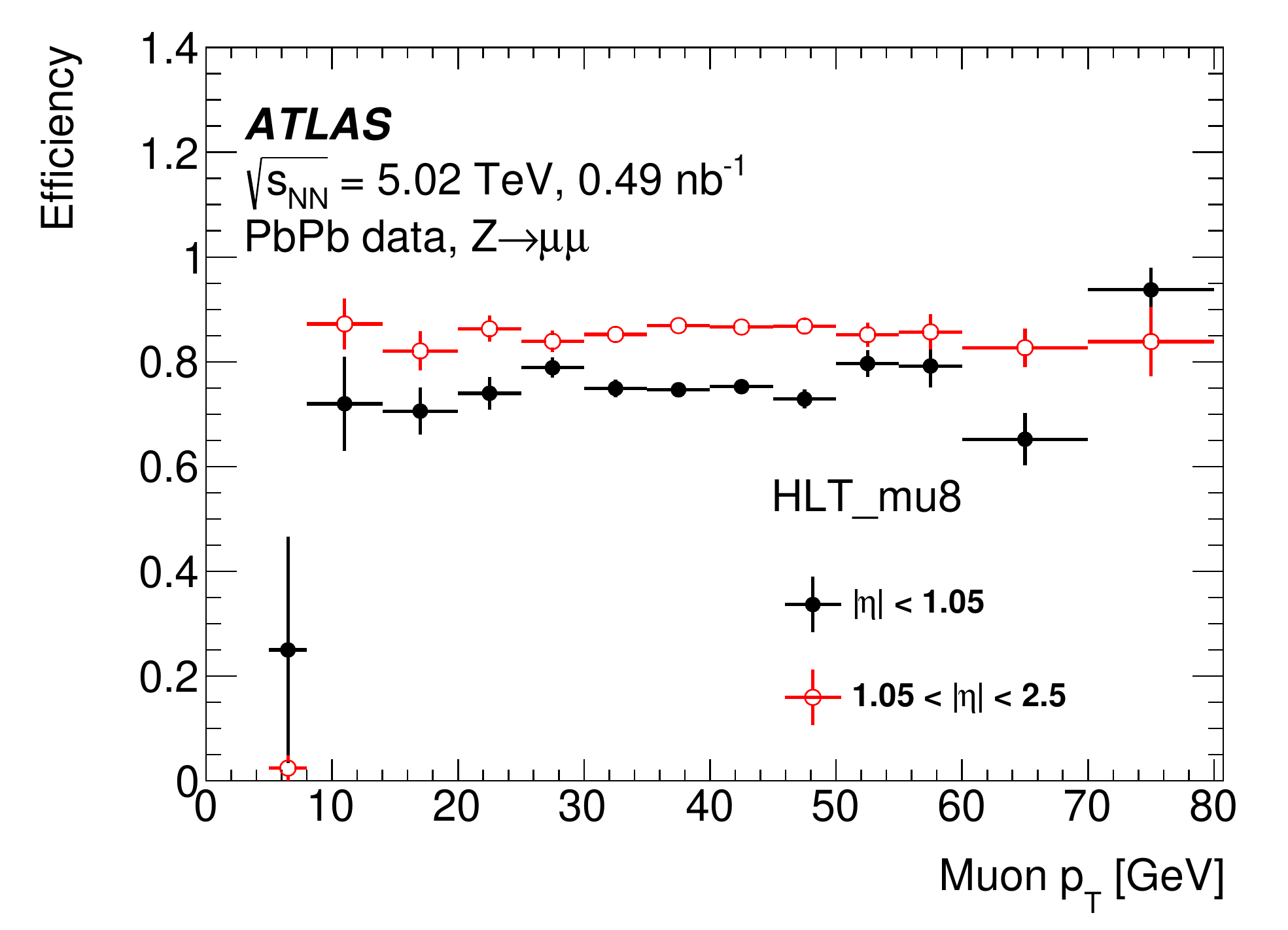}
\caption{Efficiency of the single-muon trigger HLT\_mu4 (left) and HLT\_mu8 (right) measured as a function of muon~\pT in the barrel region (full black circles) and in the endcap region (empty red circles) in \PbPb collision data. The error bars show the statistical uncertainties only.}
\label{fig_PbPb:trig-eff-data-pt}
\end{figure}
 
Trigger efficiency studies of low-\pT and high-\pT muon triggers are performed also for HI collisions in \pPb data taken during November--December 2016.
Both the low- and high-\pT trigger efficiencies are studied using the \tap method described above.
The performance of the low-\pT muon trigger, HLT\_mu4, is analysed using \Jmumu samples, while for the high-\pT muon trigger, HLT\_mu15\_L1MU6, a \Zmm sample is used.
Figure~\ref{fig:pPb:pt} (left) shows the efficiency measurements for the HLT\_mu4 trigger chain as a function of \pT for the barrel and endcap regions.
The maximum efficiency in the barrel region is $\sim$80\%, while in the endcaps it increases to $95\%$.
Figure~\ref{fig:pPb:pt} (right) shows the efficiency for the high-\pT muon trigger as a function of \pT for the barrel and endcap regions.
Overall, the maximum efficiency in the barrel region is $\sim$76\%, while in the endcap regions the efficiency reaches $\sim$90\%.
\begin{figure}[htbp]
\centering
\includegraphics[width=0.49\textwidth]{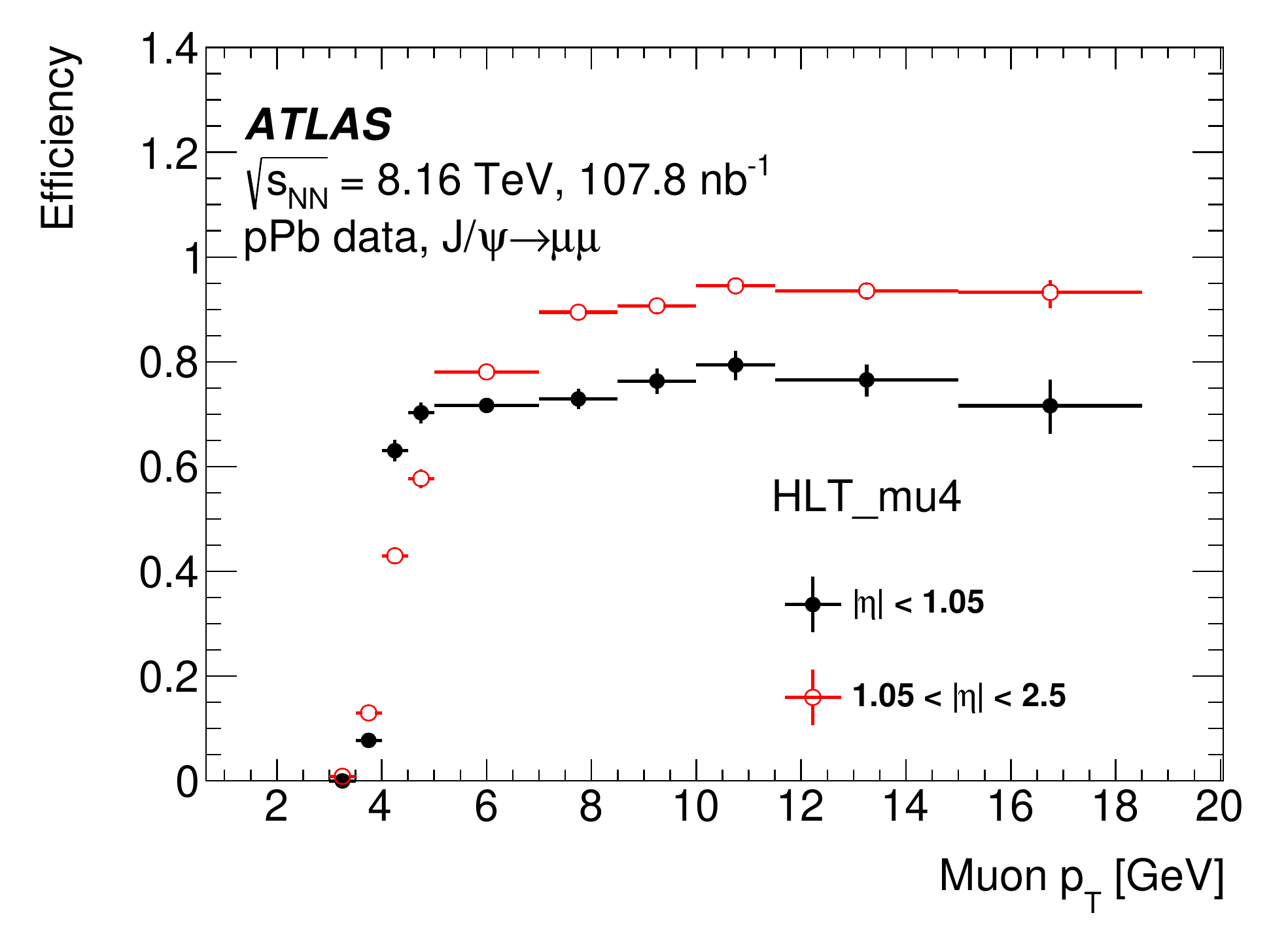}
\includegraphics[width=0.49\textwidth]{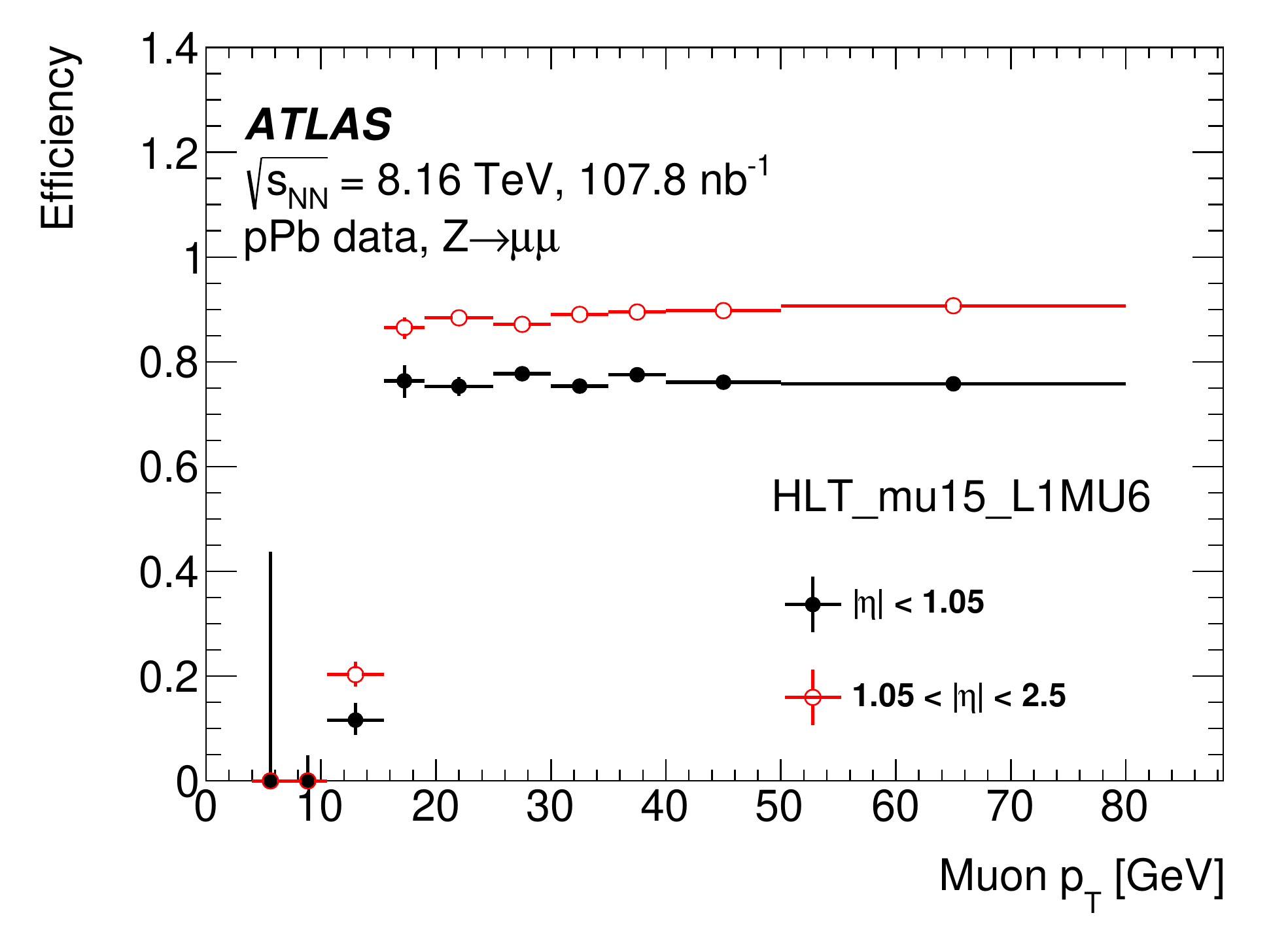}
\caption{Low-\pT (left) and high-\pT (right) trigger efficiency as a function of the muon \pT in the barrel region (full black circles) and in the endcap region (empty red circles) in \pPb collision data. The error bars show the statistical uncertainties only.}
\label{fig:pPb:pt}
\end{figure}

\section{Muon triggers for \textit{B}-physics and Light States programme}
\label{sec:bphys}
 
The ATLAS BLS programme relies on samples of low-\pT leptons, typically muons.
Because of the large instantaneous luminosities delivered by the LHC, both standard single-muon and dimuon triggers are incapable of running with a threshold low enough to provide a sample of data useful for BLS.
To record an appropriately sized sample of data, a set of multi-muon triggers with invariant mass and vertex requirements are used to reduce backgrounds to a level at which trigger rates are low enough to run online.
 
The basic flow of a BLS chain is depicted in Figure~\ref{fig:BphysFlowchart}.
After fast muon reconstruction, ID tracks are reconstructed, and a first vertex reconstruction is performed with the ID tracks.
If this vertex passes the required $\chi^2$ and mass constraints, the precision muon reconstruction proceeds, and combined muons are used in a second vertex reconstruction algorithm.
The trigger is then passed if at least one vertex meets the required $\chi^2$ and mass constraints, which are generally tighter than those in the first-stage vertex reconstruction.
\begin{figure}[htbp]
\centering
\includegraphics[width=0.8\textwidth]{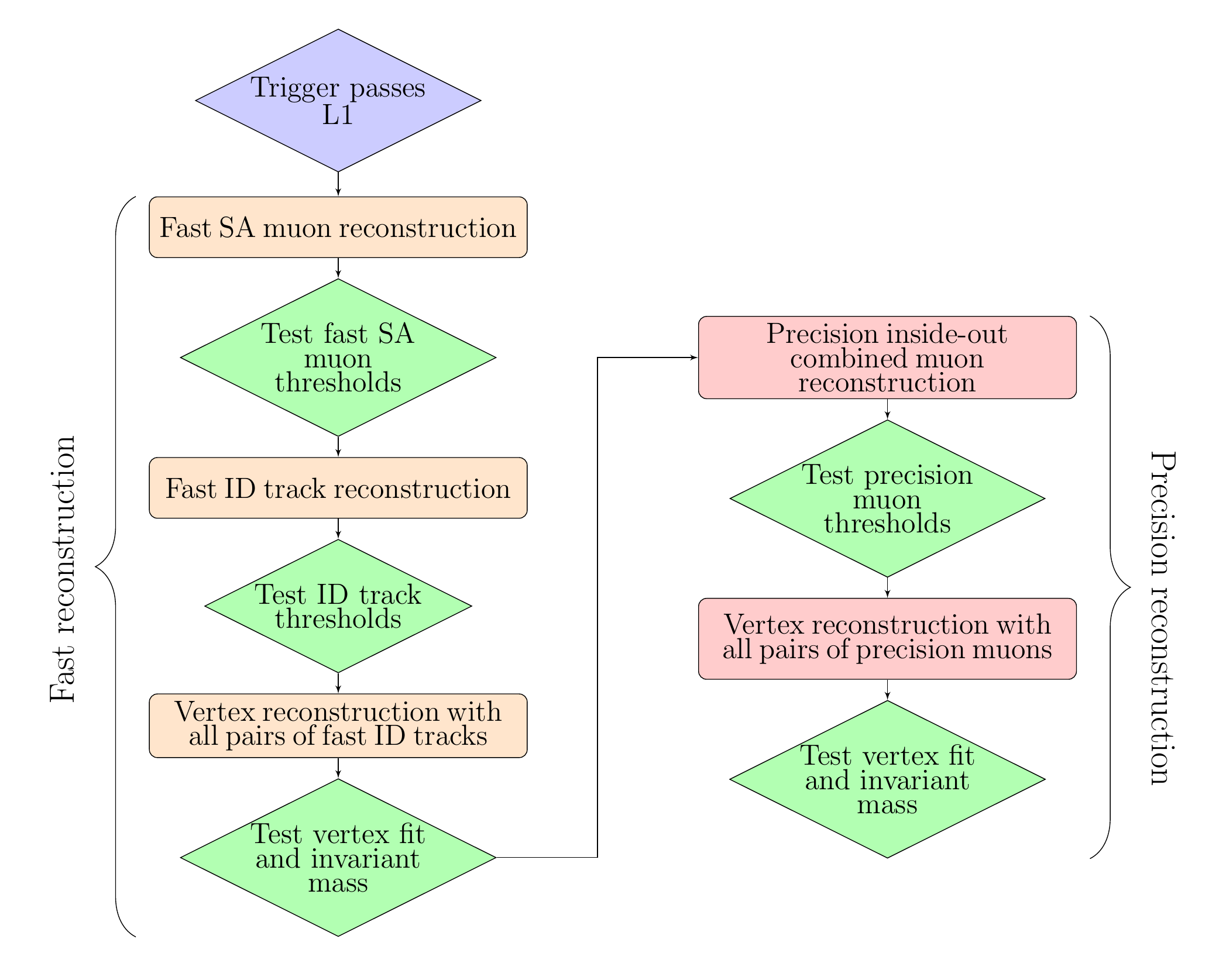}
\caption{General construction of a BLS trigger.
After both fast muon reconstruction and precision muon reconstruction, BLS vertex reconstruction algorithms and hypothesis testing run to see if the muons parameters are consistent with a predefined topology.}
\label{fig:BphysFlowchart}
\end{figure}
 
\subsection{BLS L1 topological trigger}
The backbone of the ATLAS BLS programme is low-\pT muons.
This is reflected in heavy use of dimuon triggers with the lowest possible threshold for both L1 and the HLT.
To cope with the conditions of \RunTwo and to maintain low thresholds, the L1 topological trigger processor~(L1Topo) is heavily used.
It was installed between 2013 and 2014, before the start of \RunTwo.
L1Topo can combine kinematic information from multiple calorimeter and muon trigger objects into topological information about the event, leading to a reduction in rates while keeping the \pT thresholds low.
The L1Topo system was successfully commissioned during \pp collisions in 2016 and has been used to collect data since September 2016.
 
The BLS L1Topo triggers make use of a dimuon invariant mass range cut, where the mass is calculated as $m_{\mu\mu}^2 = 2 p_{\text{T}1} p_{\text{T}2} (\cosh{\Delta\eta} - \cos{\Delta\phi})$, and a cut on the angular distance $\Delta R_{\mu \mu}$ of the two muons.
The L1Topo items used in BLS chains are:
\begin{itemize}
\item L1\_LFV-MU6: requires two muons with \pT > 6~\GeV{} and $\Delta R_{\mu\mu} < 1.5$.
\item L1\_LFV-MU11: requires two muons with \pT > 11 (6)~\GeV{} for the leading (subleading) muon and $\Delta R_{\mu\mu}  < 1.5$.
\item L1\_BPH-2M9-2MU4\_BPH-0DR15-2MU4: requires two muons with \pT > 4~\GeV{} and $2~\GeV{} < m_{\mu\mu} < 9~\GeV{}$ and two muons with \pT > 4~\GeV{} and $\Delta R_{\mu\mu} < 1.5$.
\item L1\_BPH-2M9-MU6MU4\_BPH-2DR15-MU6MU4: requires two muons with \pT > 6 (4)~\GeV{} for the leading (subleading) muon and $2~\GeV{} < m_{\mu\mu} < 9~\GeV{}$ and two muons with \pT > 6 (4)~\GeV{} for the leading (subleading) muon and $0.2 < \Delta R_{\mu\mu}  < 1.5$.
\item L1\_BPH-2M9-2MU6\_BPH-2DR15-2MU6: requires two muons with \pT > 6~\GeV{} and $2~\GeV{} < m_{\mu\mu} < 9~\GeV{}$ and two muons with \pT > 6~\GeV{} and $0.2 < \Delta R_{\mu\mu} < 1.5$.
\item L1\_BPH-8M15-MU6MU4\_BPH-0DR22-MU6MU4-BO: requires two muons with \pT > 6 (4)~\GeV{} for the leading (subleading) muon and $8~\GeV{} < m_{\mu\mu} < 15~\GeV{}$ and two barrel-only (BO) muons ($|\eta_\mu| < 1.05$) with \pT > 6 (4)~\GeV{} for the leading (subleading) muon with $\Delta R_{\mu\mu} < 2.2$.
\item L1\_BPH-8M15-2MU6\_BPH-0DR22-2MU6: requires two muons with \pT > 6~\GeV{} and $8~\GeV{} < m_{\mu\mu} < 15~\GeV{}$ and two muons with \pT > 6~\GeV{} and $\Delta R_{\mu\mu} < 2.2$.
\end{itemize}
These L1Topo triggers all target dimuon topologies, and for triggers with two requirements, the muon pair passing the first requirement is generally the same as the pair that passes the second.
The dimuon mass windows are optimised by taking into account the available L1 muon \pT thresholds.
 
The L1 rate reduction achieved by L1Topo is approximately a factor of four, as can be seen in Figure~\ref{fig:BphysL1TopoRate}, where the comparison of 2MU6 with and without the L1Topo requirement is shown.
The L1 selection requirements result in an approximately 12\% HLT efficiency reduction with respect to offline.
\begin{figure}[htbp]
\centering
\includegraphics[width=0.75\textwidth]{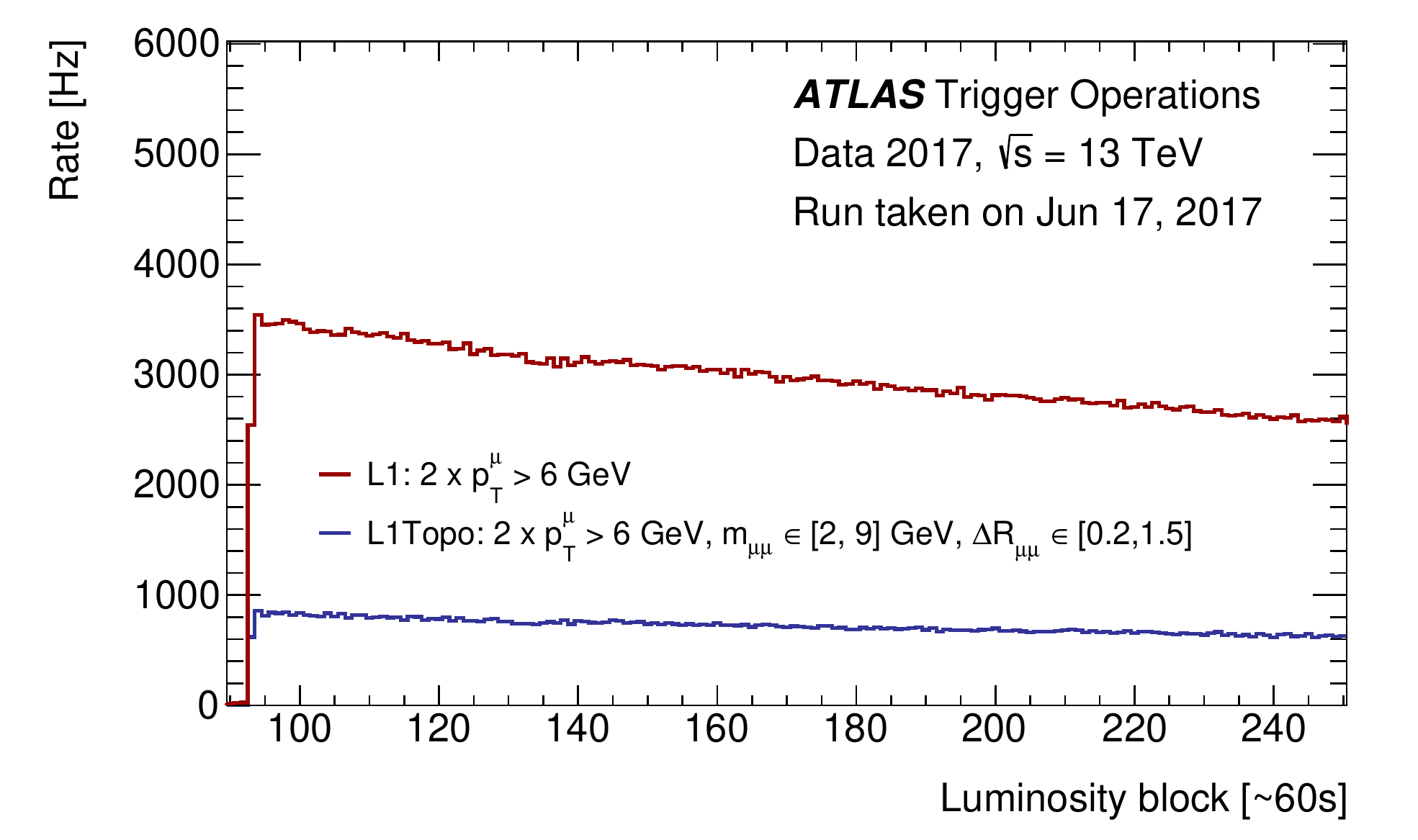}
\caption{The rate of L1 triggers selecting two muons, each with transverse momentum above 6~\GeV{} with (blue) and without (red) the additional L1Topo requirement.
The L1Topo requirement is L1 dimuon invariant mass $2~\GeV{} < m_{\mu\mu} < 9~\GeV{}$, and an angular separation $0.2 < \Delta R < 1.5$ between the muons.
The rate is shown as a function of luminosity block, which on average correspond to a time interval of 60~s, in a run taken in June 2017 with a peak luminosity of $L = 0.79 \times 10^{34}$~cm$^{-2}$s$^{-1}$ and an average pile-up of $\langle \mu \rangle$ = 46.4.
The overall reduction of the rate due to the L1Topo requirement is approximately a factor of four.}
\label{fig:BphysL1TopoRate}
\end{figure}
 
\subsection{BLS HLT algorithms}\label{sec:BLSHLT}
To further reduce the background and the rate of BLS muon trigger chains, several algorithms and selections are applied at the HLT.
Most of the chains start with a multi-track vertex reconstruction algorithm in the fast reconstruction step, followed by vertex hypothesis testing.
The same vertex hypothesis testing algorithm is used in the precision reconstruction step for many BLS topologies.
The following algorithms are used:
\begin{itemize}
\item \textbf{Multi-track vertex reconstruction:} This algorithm takes two or more high-\pT tracks produced by the fast muon reconstruction and applies cuts on their \pT, multiplicity, and sum of charges.
A mass hypothesis is set according to the selected topology and the tracks are passed to a Kalman vertex fitter~\cite{ATL-PHYS-2003-031}.
\item \textbf{Vertex hypothesis testing:} This algorithm selects vertices passing predefined invariant mass and vertex fit $\chi^2$ selections (see Table~\ref{tab:BLS_topologies}).
\end{itemize}
 
A more complicated topological selection happens at the precision step.
Muons passing the combined muon reconstruction and threshold requirements are used for vertex reconstruction.
Additionally, for chains that require ID tracks for vertex reconstruction, ID track reconstruction runs in a $(\Delta\eta\times\Delta\phi) = (1.5\times1.5)$ region centred around each reconstructed precision muon that passes the required \pT thresholds.
The following BLS algorithms are used:
\begin{itemize}
\item \textbf{Dimuon vertex reconstruction and hypothesis testing:} These dimuon algorithms are used for $J/\psi / \Upsilon / B \to \mu\mu$ topologies, as shown in Table~\ref{tab:BLS_topologies}.
The reconstruction algorithm performs vertex fits with pairs of oppositely charged muon tracks after the precision reconstruction step.
Corresponding hypothesis testing requires the reconstructed vertices to meet specified invariant mass and $\chi^2$ requirements.
Additionally, a subset of triggers place pointing requirements on the dimuon vertex by requiring $L_{xy} > 0$, where $L_{xy} = (\Delta\vec{x_{\text{T}}} \cdot \vec{\pT}^{\text{vertex}})/|\vec{\pT}^{\text{vertex}}|$, $\Delta\vec{x_{\text{T}}}$ is the transverse vector between the secondary vertex and the beam spot and $\vec{\pT}^{\text{vertex}}$ is the vector sum of the transverse momenta of the muons forming the vertex.
\item \textbf{Multi-muon + track vertex reconstruction and hypothesis testing:} These algorithms are used for $\mu\mu$+$X$ topologies, e.g., $B^\pm \to J/\psi K^\pm$ (dimuon + one track) or $B_s^0 \to J/\psi \phi$ (dimuon + two tracks).
The vertex reconstruction algorithm is run in both inclusive modes, e.g. $B^\pm \to J/\psi X$, and exclusive modes, e.g. only $B^\pm \to J/\psi K^\pm$.
It starts by building a dimuon candidate, which can be a non-resonant $\mu\mu$ pair, that is then fitted using the Kalman vertex fitter.
ID tracks are then used to make all combinations for desired topologies, defined by number of additional tracks and charge requirements.
The secondary decay is then fitted by combining the ID tracks with the dimuon system, with invariant mass and vertex $\chi^2$ cuts applied to both the $\mu\mu$+$X$ and intermediate $X$ vertices.
\item \textbf{Multi-muon vertex reconstruction and hypothesis testing:} These algorithms fit two to four muons to a common vertex and apply invariant mass and vertex $\chi^2$ cuts.
The algorithm is used for 3- and 4-muon triggers as well as for dimuon triggers targeting $\tau \to 3\mu$ final states.
\item \textbf{Muon+track vertex reconstruction and hypothesis testing:}
The reconstruction algorithms make combinations of one muon and one ID track above a given \pT threshold and fit them to a common vertex.
Hypothesis testing imposes an invariant mass requirement.
\end{itemize}
All vertex fits in the precision reconstruction step are performed using a Kalman vertex fitter.
 
Multi-track vertex reconstruction and multi-muon hypothesis algorithms are also used in BLS Partial Event Building (PEB) triggers.
These were developed in 2017 to provide a large sample of low-\pT muons for offline muon calibration and a minimum-bias sample for quarkonia measurements.
The BLS PEB triggers selectively record raw detector information in a small region centred on muons reconstructed at the precision stage that form $J/\psi$ candidates with an ID track.
Selected events are sent to a special \textit{BphysPEB} stream, described in the following section.
 
The BLS triggers did not use combined muons at the first fast stage because of efficiency losses for closely spaced muons.
The fast SA muon reconstruction sometimes reconstructs two almost identical muon candidates for two closely spaced L1 RoIs, and therefore the subsequent fast combination step cannot resolve them as separate muons.
In 2015, in order to avoid this potential inefficiency the fast muon reconstruction steps were omitted completely and only the precision algorithms were used.
The high CPU demand of this strategy made it unsuitable for the higher instantaneous luminosities delivered by the LHC from 2016 onwards.
Instead of removing the fast reconstruction step, the \pT requirements for fast SA muons were set to the lowest required threshold per chain and the fast combination step was disabled.
The multi-track vertex reconstruction, following the fast muon reconstruction, then makes all checks using ID tracks only.
 
\begin{table}[htbp]
\centering
\small
\caption{BLS topologies selected at the HLT with their fast and precision hypotheses. The muon \pT thresholds used with each topology depend on the rate of each selection, and varied throughout \RunTwo. For topologies with only one listed cut, the same requirements are applied at both the fast and precision stages.\label{tab:BLS_topologies}}
\renewcommand*{\arraystretch}{1.5}
\begin{tabular}{lll}
\toprule
Description         & Fast reco cuts & Precision reco cuts                       \\
\midrule
$2\times\mu$, broad mass    & \multicolumn{2}{c}{1.5 < $m_{\mu\mu}$ < 14~\GeV{}, $\chi^2$ < 20} \\
$2\times\mu$, low mass  & $m_{\mu\mu}$ < 6~\GeV{}, $\chi^2$ < 50                                 & 0.1 < $m_{\mu\mu}$ < 6~\GeV{}, $\chi^2$ < 20 \\
$2\times\mu$, very low mass    & $m_{\mu\mu}$ < 2.7~\GeV{}, $\chi^2$ < 50                               & 0.1 < $m_{\mu\mu}$ < 2.7~\GeV{}, $\chi^2$ < 20 \\
\midrule
$J/\psi\to\mu\mu$    & \multicolumn{2}{c}{2.5 < $m_{\mu\mu}$ < 4.3~\GeV{}, $\chi^2$ < 20} \\
\midrule
$\Upsilon\to\mu\mu$  & \multicolumn{2}{c}{8 < $m_{\mu\mu}$ < 12~\GeV{}, $\chi^2$ < 20} \\
\midrule
$B\to\mu\mu$        & 4 < $m_{\mu\mu}$ < 8.5~\GeV{}, $\chi^2$ < 60                           & 4 < $m_{\mu\mu}$ < 8.5~\GeV{}, $\chi^2$ < 20 \\
$B\to\mu\mu+X$      & 0.1 < $m_{\mu\mu}$ < 6.5~\GeV{}, $\chi^2$ < 20                         & Depends on channel \\
\midrule
\multirow{2}{*}{$\tau\to 3\mu$}  &\multicolumn{2}{c}{ 2-$\mu$ triggers:  $m_{\mu\mu}$ < 2.7~\GeV{}, $\chi^2$ < 50} \\
&\multicolumn{2}{c}{ 3-$\mu$ triggers: $m_{\mu\mu\mu}$ < 2.7~\GeV{}, $\chi^2$ < 50} \\
\midrule
$\phi\to\mu\mu$     & \multicolumn{2}{c}{0.94 < $m_{\mu\mu}$ < 1.1~\GeV{}, $\chi^2$ < 10} \\
\bottomrule
\end{tabular}
\end{table}
 
\subsection{Rates and streaming strategy}
Starting in mid-2016, triggers for dedicated measurements were saved to a stream separate from the rest of the data for physics analyses.
This enabled these raw events to be stored on tape and only reconstructed at a later date, thus reducing the pressure on the Tier-0 system responsible for offline reconstruction.
The rate of this stream as a function of instantaneous luminosity is shown in Figure~\ref{fig:bls_rates_run2} for a representative set of runs from each year.
Averages over the full year are not shown, since the composition of the stream changed during each year.
As the luminosity decreased below predefined thresholds (for 2018, instantaneous luminosities of $1.7\times10^{34}$~cm$^{-2}$s$^{-1}$, $1.5\times10^{34}$~cm$^{-2}$s$^{-1}$, and $1.2\times10^{34}$~cm$^{-2}$s$^{-1}$), lower-threshold triggers were enabled.
These lower-threshold triggers could not be enabled at higher luminosities due to either L1 rate, HLT CPU usage, or HLT output rate limitations, but were enabled as early as resources allowed.
\begin{figure}[htbp]
\centering
\includegraphics[width=0.75\textwidth]{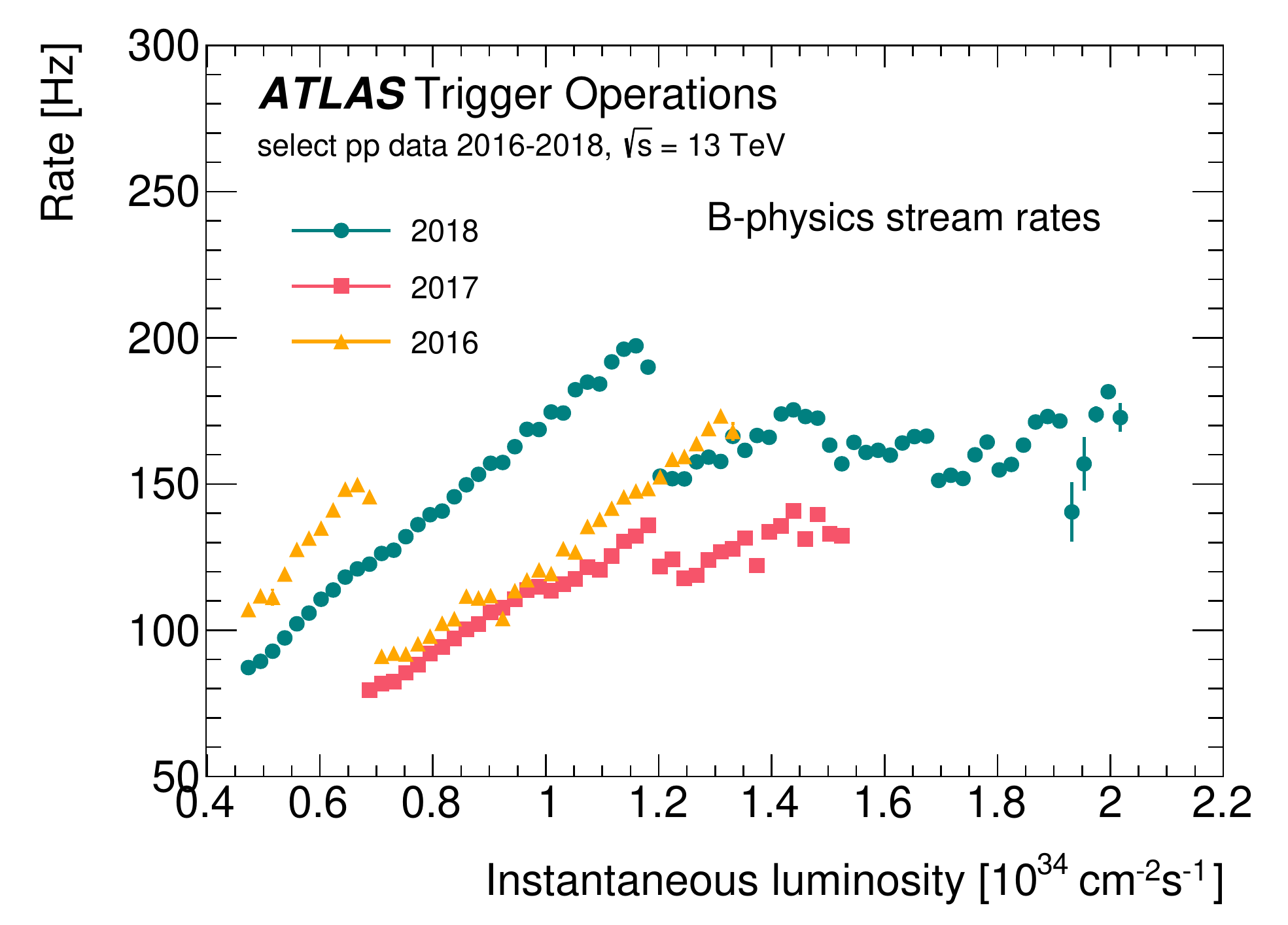}
\caption{Rate of the dedicated BLS stream from representative runs during 2016, 2017, and 2018 \pp 13~\TeV{} data taking. Enabling or lowering the prescale of lower-threshold triggers causes the jumps in the rate at $0.7 \times 10^{34}$~cm$^{-2}$s$^{-1}$ and $0.9 \times 10^{34}$~cm$^{-2}$s$^{-1}$ in 2016;  $1.0 \times 10^{34}$~cm$^{-2}$s$^{-1}$, $1.2 \times 10^{34}$~cm$^{-2}$s$^{-1}$, and $1.5 \times 10^{34}$~cm$^{-2}$s$^{-1}$ in 2017; and $1.2 \times 10^{34}$~cm$^{-2}$s$^{-1}$, $1.5 \times 10^{34}$~cm$^{-2}$s$^{-1}$, and $1.7 \times 10^{34}$~cm$^{-2}$s$^{-1}$ in 2018.}
\label{fig:bls_rates_run2}
\end{figure}
 
An additional PEB stream of events for low-\pT muon calibration and efficiency studies was commissioned in late 2017.
The events in this stream are selected by triggers requiring a first-stage SA low-\pT muon and an ID track with a combined invariant mass $2.5 < m_{\mu\mu} < 4.3$~\GeV{}.
Since a large fraction of $J/\psi$ are boosted,
instead of recording the full event data, events in this stream contain only the raw detector hits in the $(\Delta \eta \times \Delta \phi) = (1.5 \times 1.5)$ region centred on the tag muon without loss of $J/\psi$ candidates.
This reduces the average event size in the stream by a factor of seven, allowing a high rate of calibration data to be recorded without requiring excessive offline storage.
 
Table~\ref{tab:bphys_int_lumi} shows the full integrated luminosities and the integrated luminosities collected by triggers only enabled below specified instantaneous luminosities, the so-called `end-of-fill' strategy, in 2017 and 2018.
During 2015, no end-of-fill strategy was needed due to the low peak instantaneous luminosity.
During 2016, the prescale and streaming strategy changed throughout the year due to the rapidly changing data-taking conditions.
Therefore, integrated luminosities for a single set of instantaneous luminosity ranges are not shown.
The low-\pT muon calibration stream also used an end-of-fill strategy.
\begin{table}[htbp]
\centering
\caption{Integrated luminosity collected in 2017 and 2018 by triggers that were enabled below specified instantaneous luminosities.}
\label{tab:bphys_int_lumi}
\begin{tabular}{lrr}
\toprule
& \multicolumn{2}{c}{Total $\cal{L}$ [fb$^{-1}$]} \\
Enabling $L_{\text{inst}}$ & 2017 & 2018 \\
\midrule
Full run                               & 49 & 63 \\
$1.7 \times 10^{34}$ cm$^{-2}$s$^{-1}$ & -  & 53 \\
$1.5 \times 10^{34}$ cm$^{-2}$s$^{-1}$ & 38 & 43 \\
$1.2 \times 10^{34}$ cm$^{-2}$s$^{-1}$ & 30 & 26 \\
$1.0 \times 10^{34}$ cm$^{-2}$s$^{-1}$ & 20 & -  \\
\bottomrule
\end{tabular}
\end{table}
 
Figures~\ref{fig:rates:L1bphys} and~\ref{fig:rates:HLTbphys} show the 2018 rates of dimuon $J/\psi$ and BLS triggers as a function of the instantaneous luminosity.
The pure dimuon trigger rates are well described by a linear fit with an approximately zero intercept.
For L1 triggers, there is a clear non-linearity caused by an additional dependence on the number of interactions per bunch crossing.
The feature at $L = 0.5\times10^{34}~$cm$^{-2}$s$^{-1}$ is caused by LHC fills where only 600 proton bunches were colliding, but the collisions had a pile-up multiplicity of approximately 60.
This non-linearity is also present in HLT triggers with additional track requirements.
The effect of the end-of-fill strategy can be seen in triggers that have sharp changes in their rates.
\begin{figure}[htbp]
\centering
\includegraphics[width=0.49\textwidth]{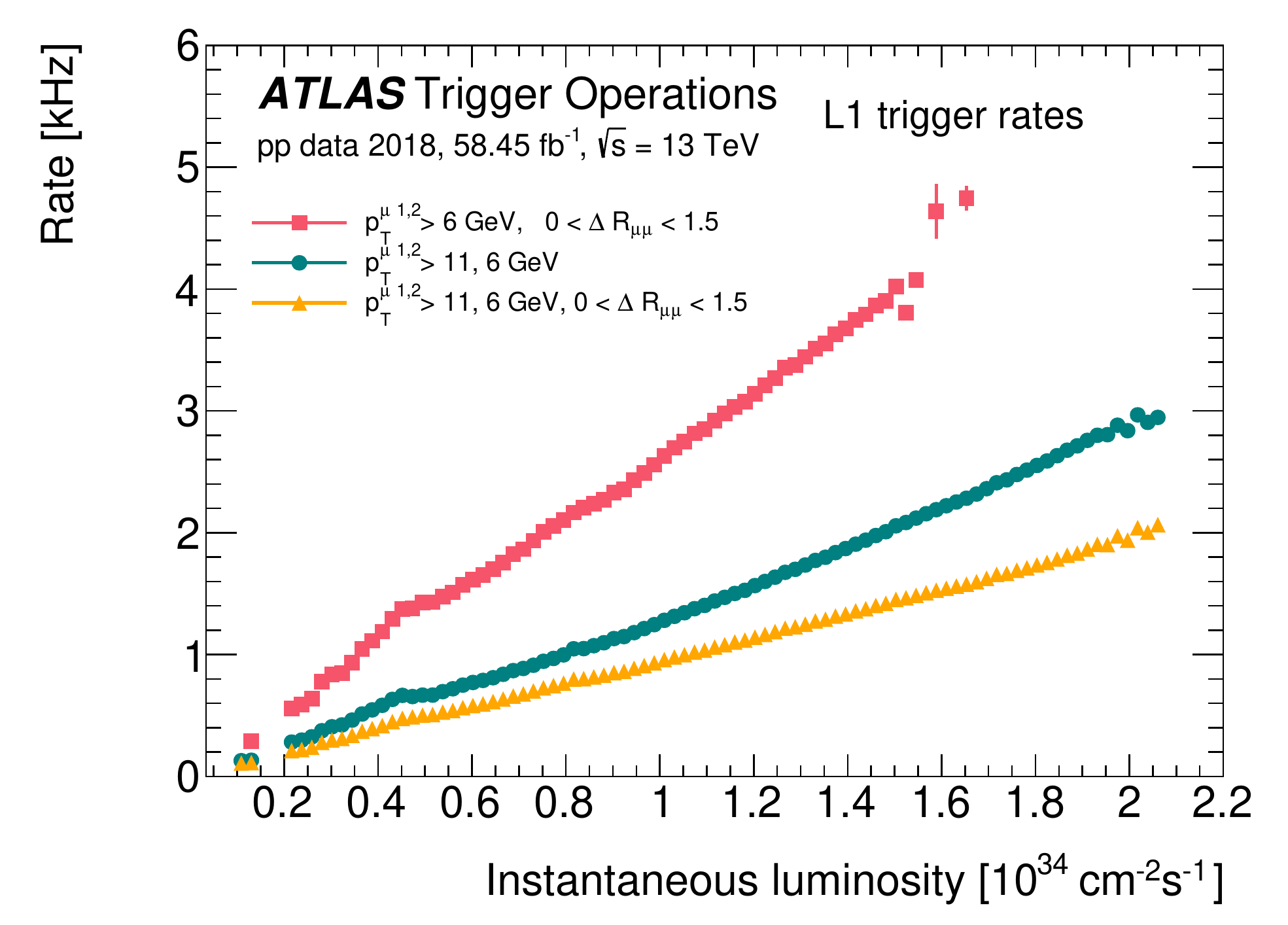}
\includegraphics[width=0.49\textwidth]{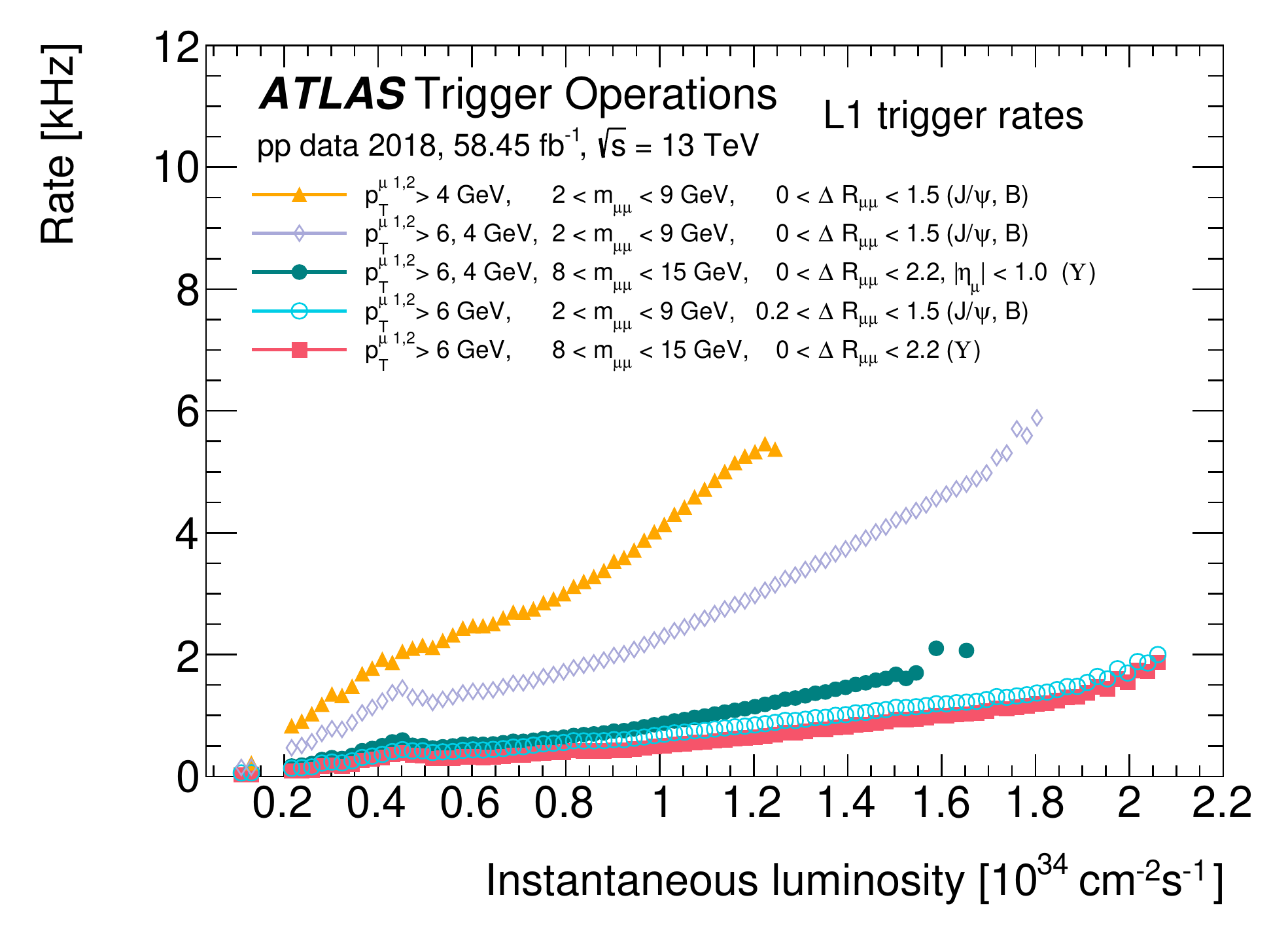}
\caption{L1 trigger rates as a function of the instantaneous luminosity for the unprescaled dimuon triggers used for the BLS programme.
The left plot shows the lowest unprescaled non-L1Topo L1 trigger, L1\_MU11\_2MU6 and two L1Topo triggers with a $\Delta R_{\mu\mu} < 1.5$ requirement.
The right plot shows L1Topo triggers with both the $\Delta R_{\mu\mu}$ and $m_{\mu\mu}$ requirements, targeting specific final states.}
\label{fig:rates:L1bphys}
\end{figure}
\begin{figure}[htbp]
\centering
\includegraphics[width=0.49\textwidth]{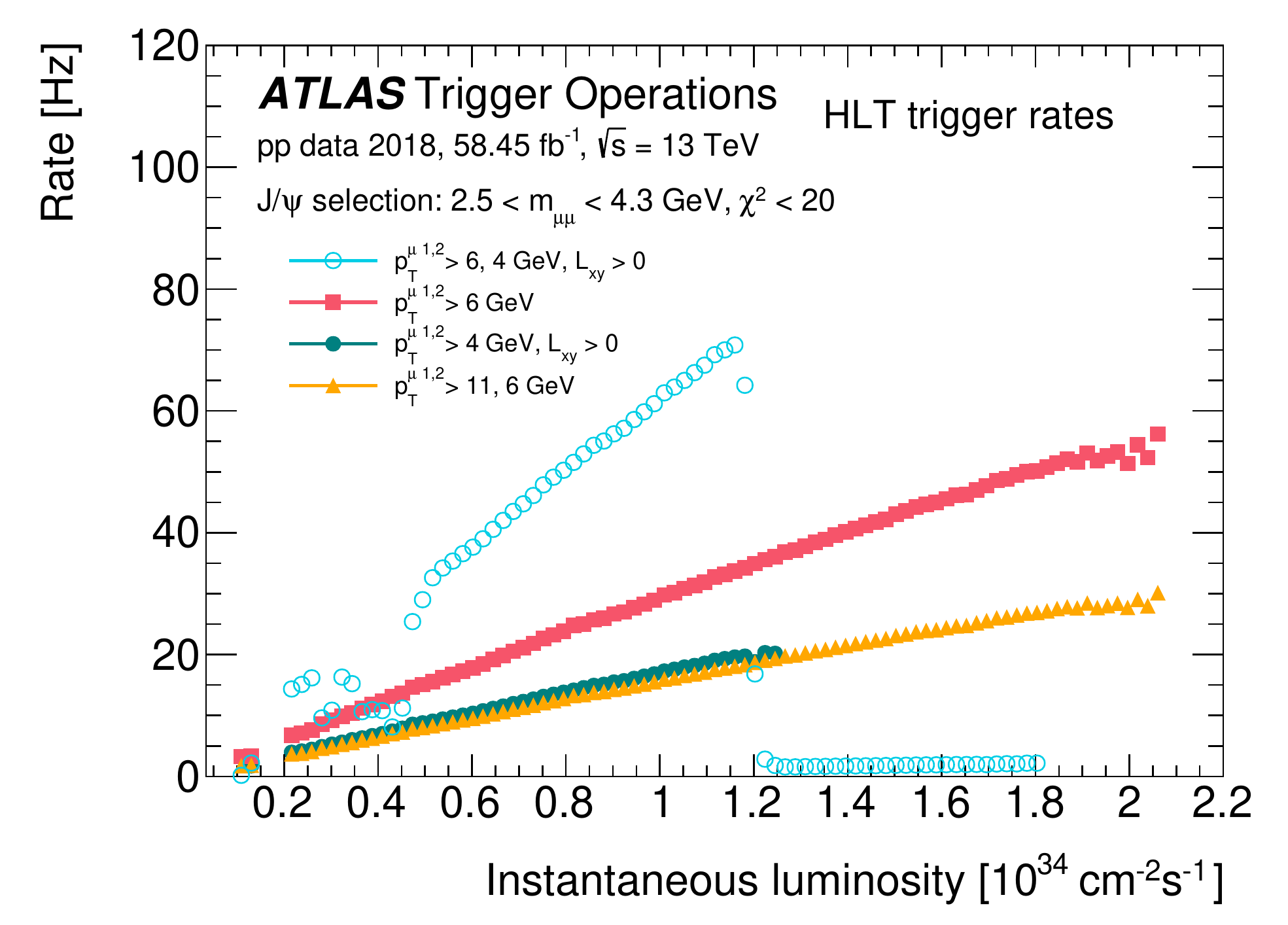}
\includegraphics[width=0.49\textwidth]{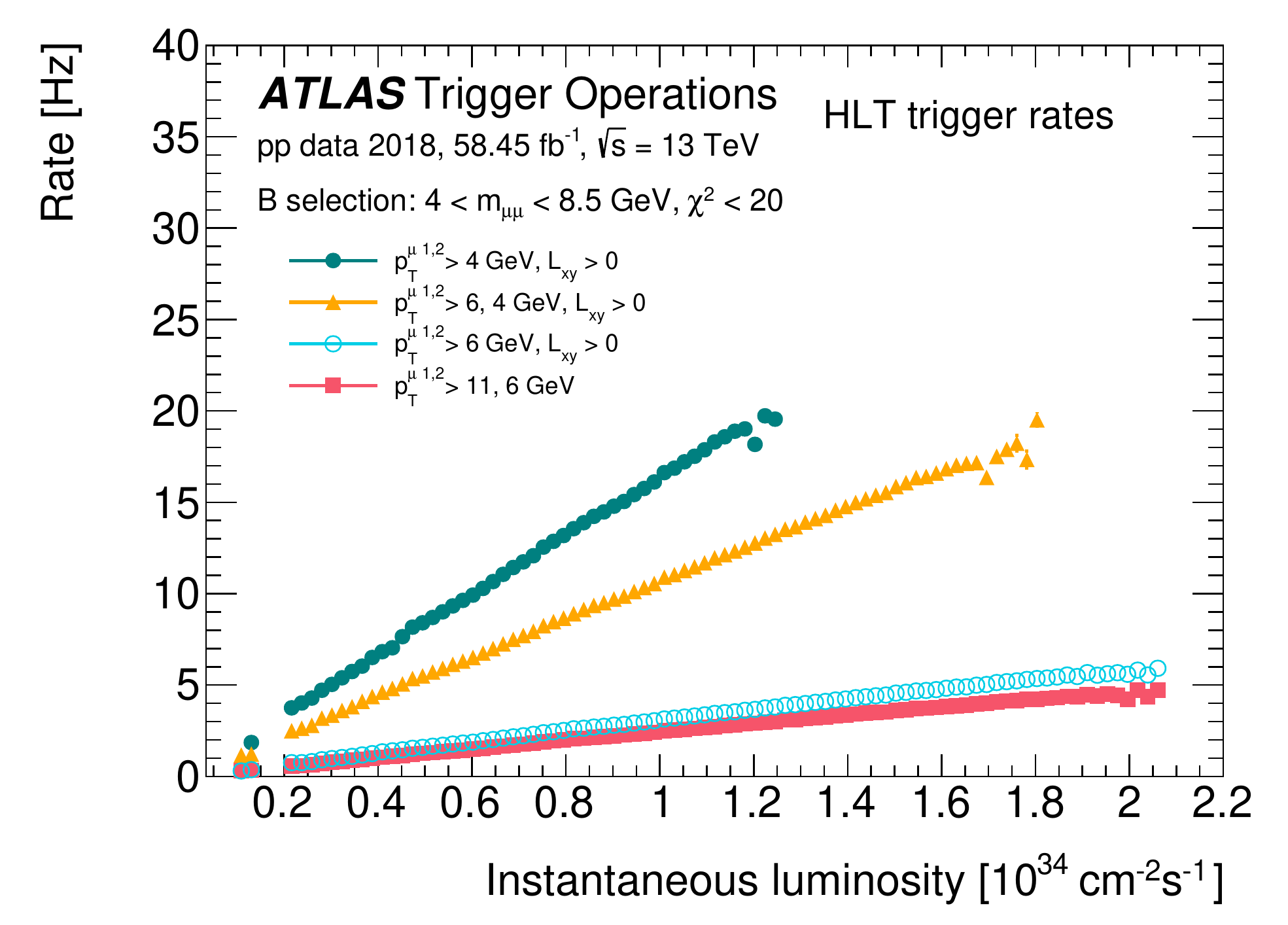}
\caption{HLT trigger rates as a function of the instantaneous luminosity for the lowest unprescaled dimuon triggers targeting $J/\psi$ (left) and $b$-hadron (right) final states.}
\label{fig:rates:HLTbphys}
\end{figure}
 
\subsection{Performance}
Figure~\ref{fig:dimu_mass} shows the dimuon invariant mass spectrum for pairs of muons satisfying \textit{Tight} offline quality criteria~\cite{PERF-2015-10}.
They are fitted to a common vertex, using the ID track parameters, with a $\chi^{2} < 20$ for one degree of freedom.
Muons are also required to have an absolute pseudorapidity less than 2.3, and have a transverse momentum greater than the trigger threshold.
The dimuon triggers require two muons at L1, passing either the low-\pT thresholds of 4~or~6~\GeV{} or the high-\pT threshold of 10~\GeV{} (L1\_MU11).
These thresholds are subsequently confirmed at the HLT.
Invariant mass requirements are applied to restrict events to the charmonium, $b$-hadron and bottomonium  mass ranges.
The yields of events collected for each trigger are cumulative, where overlapping events collected by multiple triggers are retained in the lowest-threshold histogram; hence the integral of events from all histograms is the same as the total event yield.
For certain data-taking periods, trigger prescales were applied, reducing the effective yield of events collected by those triggers.
The inclusive dimuon trigger, which covers the full mass range of interest and is shown as the dashed line in Figure~\ref{fig:dimu_mass}, was unprescaled throughout the data-taking period.
\begin{figure}[htbp]
\centering
\includegraphics[width=0.75\textwidth]{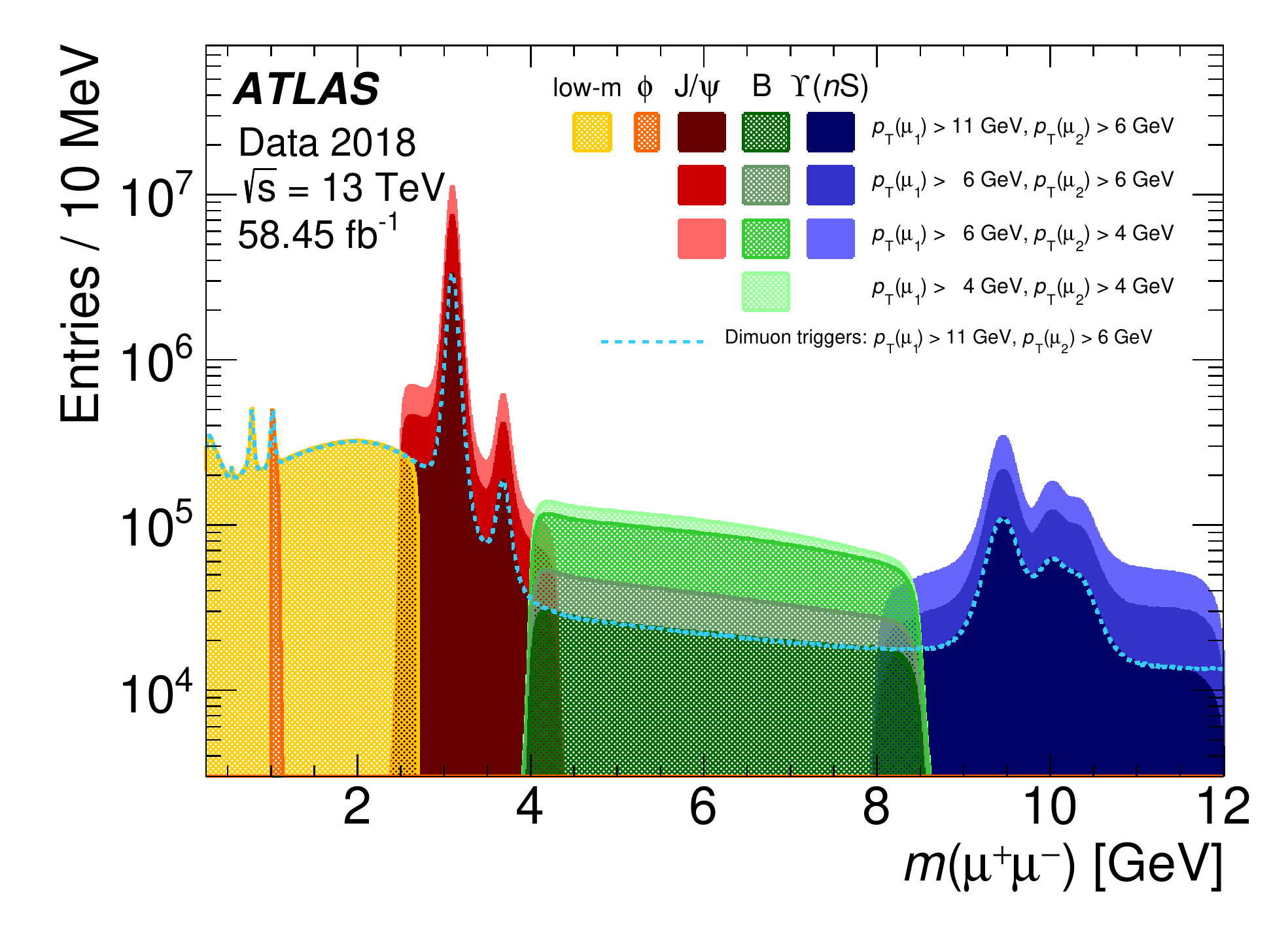}
\caption{Dimuon invariant mass for events passing a selection of BLS triggers. Triggers select specific invariant mass regions at both L1 and the HLT, and, due to vastly differing rates, some thresholds collected a smaller integrated luminosity. Smaller invariant mass regions allow lower-threshold triggers to run unprescaled at higher instantaneous luminosities, thus more signal events can be recorded than is possible with only inclusive triggers. The dashed line represents the events collected by the lowest unprescaled dimuon trigger that is inclusive of the full mass range of interest.}
\label{fig:dimu_mass}
\end{figure}
 
The efficiency of BLS dimuon triggers can be factorised into three terms:
\begin{equation}
\epsilon_\text{dimuon} = \epsilon_\text{muon}(\pT^{\mu_1}, q^{\mu_1}\cdot \eta^{\mu_1}) \times
\epsilon_\text{muon}(\pT^{\mu_2}, q^{\mu_2}\cdot \eta^{\mu_2}) \times
C_{\mu\mu}(\Delta R^{\mu\mu}, |y^{\mu\mu}|),\notag
\end{equation}
where $\epsilon_\text{muon}$ are the efficiencies of single-muon triggers corresponding to the dimuon trigger legs, $C_{\mu\mu}$ is a correction accounting for the effects of the dimuon selection applied by the trigger and $q^{\mu}$ is the muon electric charge.
Single-muon trigger efficiencies for low-\pT muons are discussed in Section~\ref{sec:eff:singleleg}.
The dimuon correction can be represented as a product of two components:
\begin{equation}
C_{\mu\mu}(\Delta R^{\mu\mu}, |y^{\mu\mu}|) =
c_a(|y^{\mu\mu}|) \times c_{\Delta R}(\Delta R^{\mu\mu},|y^{\mu\mu}|).\notag
\end{equation}
Both terms are measured in three regions of dimuon rapidity: barrel ($|y^{\mu\mu}| \leq 1.0$), barrel--endcap overlap ($1.0 < |y^{\mu\mu}| \leq 1.2$), and endcap ($1.2 < |y^{\mu\mu}| \leq 2.3$), to account for the different values of the correction in these regions.
The invariant mass regions selected by each trigger are wide enough such that offline selections are well within the window and there are no residual resolution effects.
 
The asymptotic correction, $c_a$, accounts for the effect of dimuon vertex and opposite-charge requirements applied by BLS algorithms.
The asymptotic correction values are determined using the ratio of \Jmumu decay yields in events selected by the standard dimuon trigger to those selected by a similar dimuon trigger without applying charge or dimuon vertex requirements.
A requirement of  $\Delta R^{\mu\mu} > 0.3$ is imposed on all \Jmumu candidates used to determine the asymptotic correction.
Apart from that, the selection of \Jmumu candidates and extraction of the yields follow the same procedure as described in Section~\ref{sec:eff}.
The correction values are found to be $0.983\pm 0.001$, $0.984\pm 0.003$, and $0.979\pm 0.001$ in barrel, overlap, and endcap regions, respectively.
 
The second component, $c_{\Delta R}$, accounts for the inefficiency in resolving two overlapping muon RoIs by L1 dimuon triggers for closely spaced muons and hence is evaluated as a function of dimuon angular separation, $\Delta R^{\mu\mu}$, in the regions of $|y^{\mu\mu}|$.
Smaller effects, such as inefficiencies for closely spaced muons in the HLT, are also accounted for in this correction.
To evaluate this correction, a sample of dimuon events collected with the single-muon plus track triggers as in Section~\ref{sec:eff} is used.
The $c_{\Delta R}$ is measured using the fraction, $\rho_{\Delta R}$, of \Jmumu events in this sample that are also selected by dimuon triggers requiring each muon to have $\pT > 4$~\GeV{}.
The \Jpsi candidates are selected as described in Section~\ref{sec:eff} with an additional requirement for both muons to have $\pT > 8$~\GeV{} to ensure that single-muon trigger efficiencies reach their plateau values.
Figure~\ref{fig:rho_dr} shows values of $\rho_{\Delta R}$ as a function of $\Delta R^{\mu\mu}$ in three regions of dimuon rapidity.
The higher efficiencies observed in the 2015 data are due to the absence of fast HLT reconstruction algorithms in the trigger sequence, as described in Section~\ref{sec:BLSHLT}.
\begin{figure}[htbp]
\centering
\includegraphics[width=0.48\textwidth]{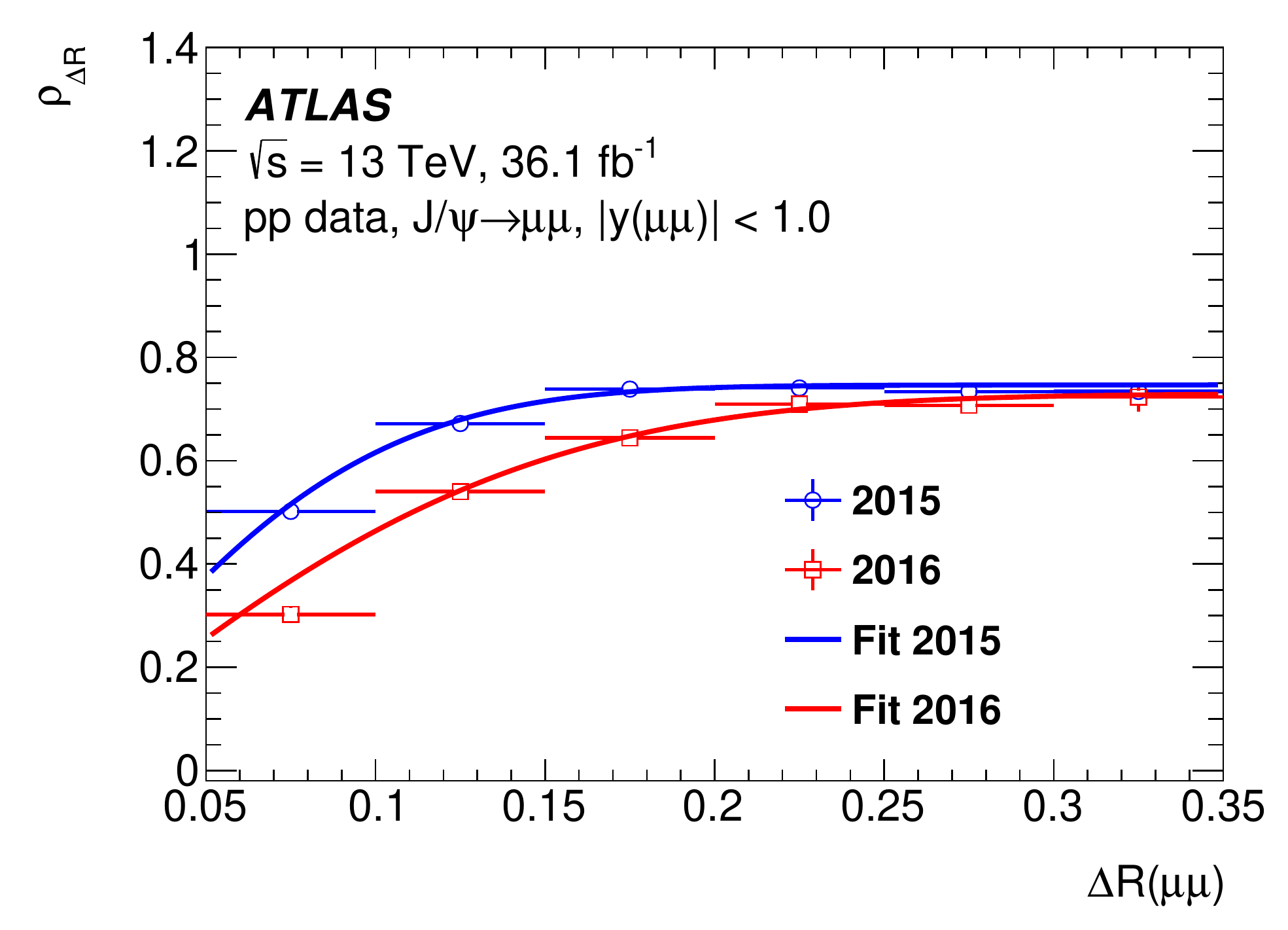}
\includegraphics[width=0.48\textwidth]{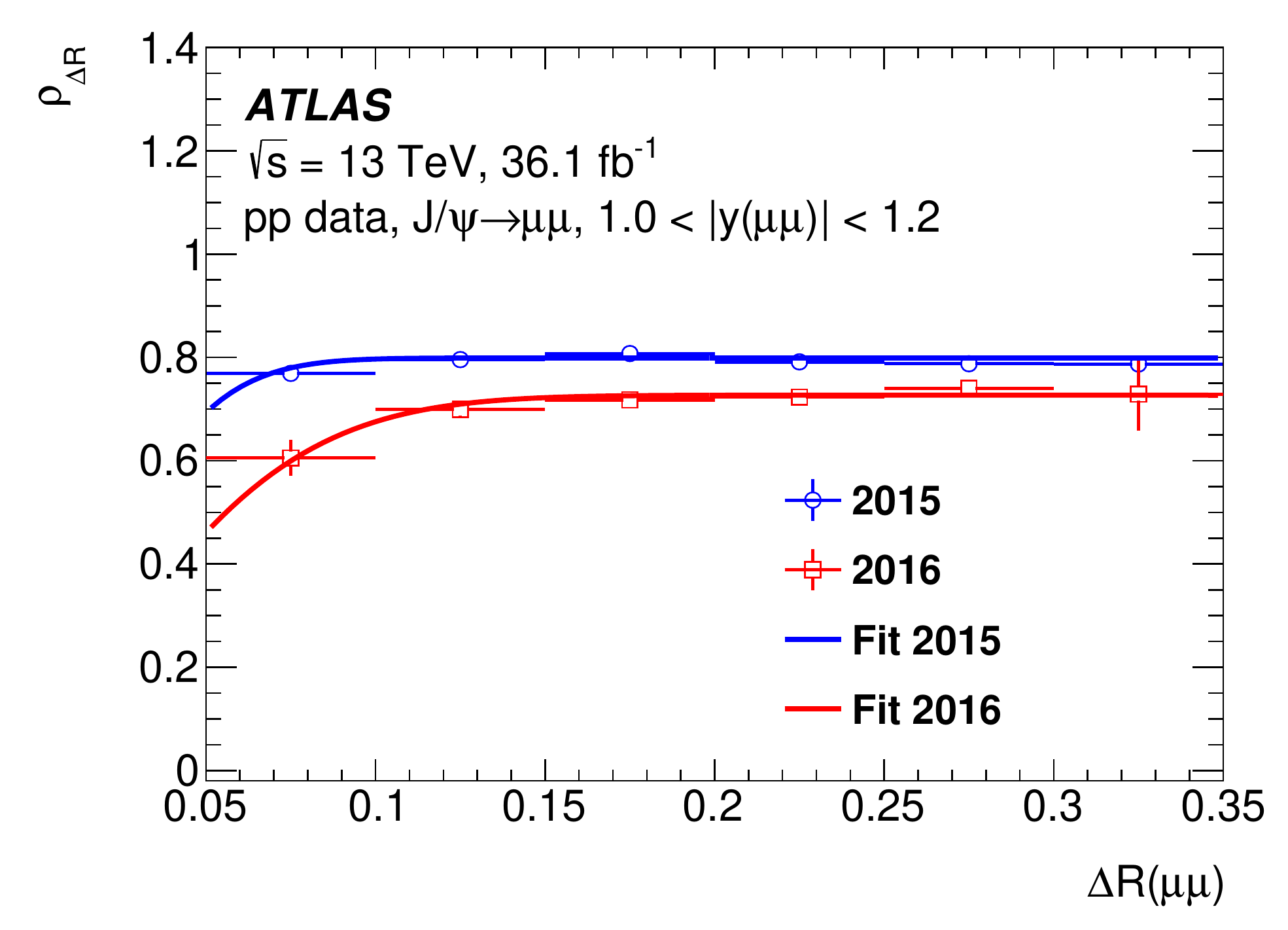}
\includegraphics[width=0.48\textwidth]{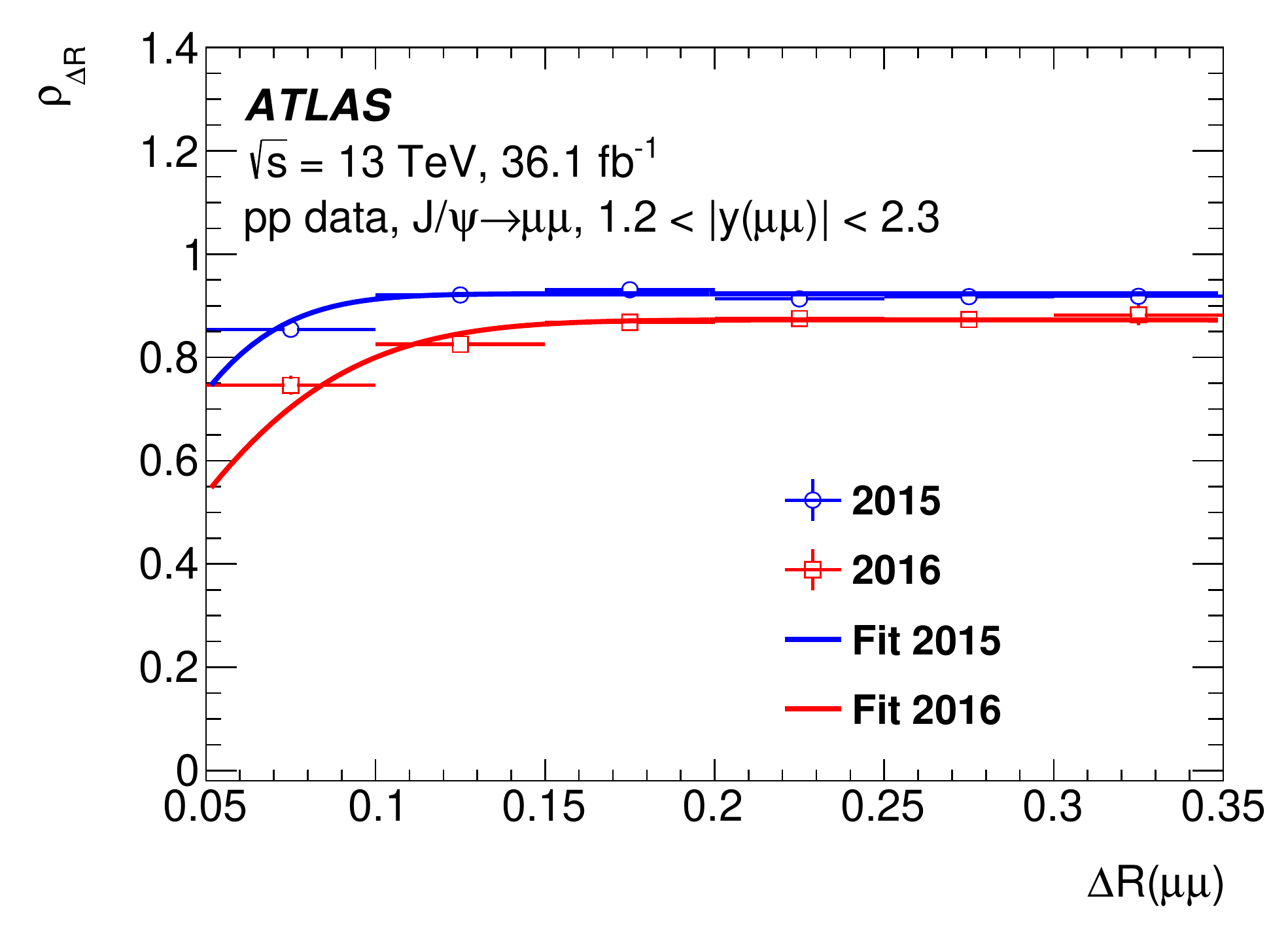}
\caption{Ratio $\rho_{\Delta R}$ of \Jmumu yields in events collected by the dimuon triggers and single-muon plus track triggers and by single-muon plus track triggers only, as a function of the dimuon angular separation $\Delta R^{\mu\mu}$, in data collected in 2015 and 2016 for (top-left) barrel, (top-right) overlap, and (bottom) endcap regions of dimuon rapidity.
Fits with an error function are also shown.}
\label{fig:rho_dr}
\end{figure}
 
The distributions in Figure~\ref{fig:rho_dr} are fitted with a function composed of an error function describing the shape of $\Delta R$ turn-on and a normalisation parameter.
The fitted error function, which approaches unity for $\Delta R \gtrsim 0.3$, corresponds to the $c_{\Delta R}(\Delta R^{\mu\mu})$ correction in each region.
The normalisation may contain contributions from $c_a$ and from the single-muon trigger efficiencies, and is irrelevant for the $c_{\Delta R}$ measurement.
The overall dimuon correction, $C_{\mu\mu}(\Delta R^{\mu\mu}, |y^{\mu\mu}|)$, in each region of $|y^{\mu\mu}|$ is given by a product of the error function extracted from the corresponding fit, and the asymptotic correction value, $c_a(|y^{\mu\mu}|)$, measured in that region.
 
\FloatBarrier

\section{Conclusion}
 
The ATLAS muon trigger has been successfully adapted to the challenging environment at the LHC such that stable and highly efficient data-taking was achieved in \RunTwo.
The transverse momentum threshold for the single-muon trigger was set at 26~\GeV{} for most of the proton--proton data-taking campaign with a well-controlled trigger rate of typically 15~kHz at Level-1 and 180~Hz at the High-Level Trigger for instantaneous luminosities of $1.7\times10^{34}$~cm$^{-2}$s$^{-1}$.
The processing times of the HLT muon trigger algorithms were continuously optimised and are sufficiently short to fit within the computing resource limitations.
The efficiencies are measured with proton--proton and heavy-ion collision data over a wide \pT range (a few~\GeV{} to several hundred~\GeV{}) by using muons from decays of \Jpsi mesons, \PZ and \PW bosons, and top-quark pairs, and show highly uniform and stable performance.
The single-muon trigger efficiency is evaluated to be about 68\% and 85\% in the barrel and endcap region, respectively, considering a kinematic region of $26 < \pT < 100~\GeV{}$.
For lower-threshold multi-muon triggers, increased efficiencies of up to 75\% and 87\% in the barrel and encap region, respectively, are observed due to the loosened requirements at Level-1.
Efficiencies are also measured in heavy-ion collision data and show comparable results.
Low-momentum muon triggers are further combined with vertex and mass selections, providing the backbone of the $B$-physics and Light States programme.
The usage of the topological Level-1 trigger selection is essential to provide well-controlled rates to cope with the conditions during \RunTwo.
Efficiency corrections in addition to the single-leg efficiencies are derived for signatures with closely spaced muons.

\section*{Acknowledgements}
 

We thank CERN for the very successful operation of the LHC, as well as the
support staff from our institutions without whom ATLAS could not be
operated efficiently.
 
We acknowledge the support of ANPCyT, Argentina; YerPhI, Armenia; ARC, Australia; BMWFW and FWF, Austria; ANAS, Azerbaijan; SSTC, Belarus; CNPq and FAPESP, Brazil; NSERC, NRC and CFI, Canada; CERN; CONICYT, Chile; CAS, MOST and NSFC, China; COLCIENCIAS, Colombia; MSMT CR, MPO CR and VSC CR, Czech Republic; DNRF and DNSRC, Denmark; IN2P3-CNRS and CEA-DRF/IRFU, France; SRNSFG, Georgia; BMBF, HGF and MPG, Germany; GSRT, Greece; RGC and Hong Kong SAR, China; ISF and Benoziyo Center, Israel; INFN, Italy; MEXT and JSPS, Japan; CNRST, Morocco; NWO, Netherlands; RCN, Norway; MNiSW and NCN, Poland; FCT, Portugal; MNE/IFA, Romania; MES of Russia and NRC KI, Russia Federation; JINR; MESTD, Serbia; MSSR, Slovakia; ARRS and MIZ\v{S}, Slovenia; DST/NRF, South Africa; MINECO, Spain; SRC and Wallenberg Foundation, Sweden; SERI, SNSF and Cantons of Bern and Geneva, Switzerland; MOST, Taiwan; TAEK, Turkey; STFC, United Kingdom; DOE and NSF, United States of America. In addition, individual groups and members have received support from BCKDF, CANARIE, Compute Canada and CRC, Canada; ERC, ERDF, Horizon 2020, Marie Sk{\l}odowska-Curie Actions and COST, European Union; Investissements d'Avenir Labex, Investissements d'Avenir Idex and ANR, France; DFG and AvH Foundation, Germany; Herakleitos, Thales and Aristeia programmes co-financed by EU-ESF and the Greek NSRF, Greece; BSF-NSF and GIF, Israel; CERCA Programme Generalitat de Catalunya and PROMETEO Programme Generalitat Valenciana, Spain; G\"{o}ran Gustafssons Stiftelse, Sweden; The Royal Society and Leverhulme Trust, United Kingdom.
 
The crucial computing support from all WLCG partners is acknowledged gratefully, in particular from CERN, the ATLAS Tier-1 facilities at TRIUMF (Canada), NDGF (Denmark, Norway, Sweden), CC-IN2P3 (France), KIT/GridKA (Germany), INFN-CNAF (Italy), NL-T1 (Netherlands), PIC (Spain), ASGC (Taiwan), RAL (UK) and BNL (USA), the Tier-2 facilities worldwide and large non-WLCG resource providers. Major contributors of computing resources are listed in Ref.~\cite{ATL-GEN-PUB-2016-002}.
 

\clearpage
 
\printbibliography
 
\clearpage
 
\begin{flushleft}
\hypersetup{urlcolor=black}
{\Large The ATLAS Collaboration}

\bigskip

\AtlasOrcid[0000-0002-6665-4934]{G.~Aad}$^\textrm{\scriptsize 102}$,    
\AtlasOrcid[0000-0002-5888-2734]{B.~Abbott}$^\textrm{\scriptsize 128}$,    
\AtlasOrcid[0000-0002-7248-3203]{D.C.~Abbott}$^\textrm{\scriptsize 103}$,    
\AtlasOrcid[0000-0002-2788-3822]{A.~Abed~Abud}$^\textrm{\scriptsize 36}$,    
\AtlasOrcid[0000-0002-1002-1652]{K.~Abeling}$^\textrm{\scriptsize 53}$,    
\AtlasOrcid[0000-0002-2987-4006]{D.K.~Abhayasinghe}$^\textrm{\scriptsize 94}$,    
\AtlasOrcid[0000-0002-8496-9294]{S.H.~Abidi}$^\textrm{\scriptsize 167}$,    
\AtlasOrcid[0000-0002-8279-9324]{O.S.~AbouZeid}$^\textrm{\scriptsize 40}$,    
\AtlasOrcid{N.L.~Abraham}$^\textrm{\scriptsize 156}$,    
\AtlasOrcid[0000-0001-5329-6640]{H.~Abramowicz}$^\textrm{\scriptsize 161}$
% ...
\bigskip
\\

% ...
$^{36}$CERN, Geneva; Switzerland.\\
% ...
$^{40}$Niels Bohr Institute, University of Copenhagen, Copenhagen; Denmark.\\
% ...
$^{53}$II. Physikalisches Institut, Georg-August-Universit\"{a}t G\"ottingen, G\"ottingen; Germany.\\
% ...
$^{94}$Department of Physics, Royal Holloway University of London, Egham; United Kingdom.\\
% ...
$^{102}$CPPM, Aix-Marseille Universit\'e, CNRS/IN2P3, Marseille; France.\\
$^{103}$Department of Physics, University of Massachusetts, Amherst MA; United States of America.\\
% ...
$^{128}$Homer L. Dodge Department of Physics and Astronomy, University of Oklahoma, Norman OK; United States of America.\\
% ...
$^{156}$Department of Physics and Astronomy, University of Sussex, Brighton; United Kingdom.\\
% ...
$^{161}$Raymond and Beverly Sackler School of Physics and Astronomy, Tel Aviv University, Tel Aviv; Israel.\\
% ...
$^{167}$Department of Physics, University of Toronto, Toronto ON; Canada.\\
% ...

\end{flushleft}

 
\end{document}